\documentclass[onecollarge]{svjour2}       
\smartqed  

\usepackage[dvipdfmx]{graphicx}

\usepackage{amsmath, amssymb}
\usepackage{amsfonts}
\usepackage{multicol}
\usepackage{braket}

\usepackage{ascmac}
\usepackage{fancybox}
\usepackage{comment}

\usepackage{cite}

\usepackage{multirow}

\newcommand{\eq}[1]{\begin{equation} #1 \end{equation}}
\newcommand{\eqa}[2]{\begin{equation} #1 \label{#2} \end{equation}}
\newcommand{\balign}[1]{\begin{align} #1 \end{align}}
\newcommand{\bcases}[1]{\begin{cases} #1 \end{cases}}

\newcommand{\mx}[1]{\begin{pmatrix}#1 \end{pmatrix}}

\newcommand{\fn}{\footnote}

\newcommand{\todayd}{\the\year/\the\month/\the\day}

\newcommand{\bib}{\bibitem}

\newcommand{\up}{\uparrow}

\newcommand{\lr}{\leftrightarrow}

\newcommand{\lb}{\label}
\newcommand{\nt}{\notag}

\newcommand{\eref}[1]{Eq.~\eqref{#1}}
\newcommand{\tref}[1]{Theorem~\ref{t:#1}}
\newcommand{\lref}[1]{Lemma~\ref{l:#1}}

\newcommand{\sref}[1]{Sec.~\ref{s:#1}}
\newcommand{\cref}[1]{Chap.~\ref{c:#1}}

\newcommand{\bel}{\begin{easylist}}
\newcommand{\eel}{\end{easylist}}
\newcommand{\bi}[1]{\begin{itemize} #1 \end{itemize}}
\newcommand{\be}[1]{\begin{enumerate} #1 \end{enumerate}}

\def \({\left(}
\def \){\right)}

\def \[{\left[}
\def \]{\right]}
\newcommand{\abs}[1]{\left|#1\right|}

\newcommand{\sumtwo}[2]%
{\mathop{\sum_{#1}}_{#2}}
\newcommand{\sumthree}[3]%
{\mathop{\mathop{\sum_{#1}}_{#2}}_{#3}}
\newcommand{\sumfour}[4]%
{\mathop{\mathop{\mathop{\sum_{#1}}_{#2}}_{#3}}_{#4}} 
\newcommand{\prodtwo}[2]%
{\mathop{\prod_{#1}}_{#2}}
\newcommand{\mintwo}[2]%
{\mathop{\min_{#1}}_{#2}}
\newcommand{\maxtwo}[2]%
{\mathop{\max_{#1}}_{#2}}
\newcommand{\maxthree}[3]%
{\mathop{\mathop{\max_{#1}}_{#2}}_{#3}}
\newcommand{\limtwo}[2]%
{\mathop{\lim_{#1}}_{#2}}
\newcommand{\suptwo}[2]%
{\mathop{\sup_{#1}}_{#2}}
\newcommand{\supthree}[3]%
{\mathop{\mathop{\sup_{#1}}_{#2}}_{#3}}
\newcommand{\supfour}[4]%
{\mathop{\mathop{\mathop{\sup_{#1}}_{#2}}_{#3}}_{#4}} 
\newcommand{\inftwo}[2]%
{\mathop{\inf_{#1}}_{#2}}
\newcommand{\infthree}[3]%
{\mathop{\mathop{\inf_{#1}}_{#2}}_{#3}}
\newcommand{\inffour}[4]%
{\mathop{\mathop{\mathop{\inf_{#1}}_{#2}}_{#3}}_{#4}} 

\newcommand\calG{{\cal G}}

\newcommand\calN{{\cal N}}

\newcommand\calP{{\cal P}}

\newcommand\calS{{\cal S}}

\newcommand{\bsA}{\boldsymbol{A}}
\newcommand{\bsB}{\boldsymbol{B}}
\newcommand{\bsC}{\boldsymbol{C}}
\newcommand{\bsD}{\boldsymbol{D}}

\newcommand{\bbC}{\mathbb{C}}

\newcommand{\bbN}{\mathbb{N}}

\newcommand{\bbR}{\mathbb{R}}

\newcommand{\ba}[2]{
\begin{array}{#1}
#2
\end{array}
}

\def\x{\overline{X}}
\def\y{\overline{Y}}
\def\z{\overline{Z}}
\def\a{\overline{A}}

\def\w{\overline{W}}

\def\tx{\widetilde{X}}
\def\ty{\widetilde{Y}}
\def\tz{\widetilde{Z}}

\def\brzy{\overset{\ zy}{|}}

\def\hrzy{\underset{\ zy}{\up}}

\def\tb{\widetilde{B}}
\def\tL{\widetilde{L}}
\def\tR{\widetilde{R}}

\newcommand{\qh}{[Q, H] }
\def\i{{\rm i}}

\newcommand{\lplus}{\overset{\to}{+}}
\newcommand{\rplus}{\overset{\leftarrow}{+}}
\newcommand{\threeskip}{\ \ }

\def\rnum#1{\resizebox{0.5em}{\height}{\expandafter{\romannumeral #1}}}
\def\Rnum#1{\resizebox{0.5em}{\height}{\uppercase\expandafter{\romannumeral #1}}}

\newcommand{\Wc}{W^{\rm c}}
\newcommand{\Wpc}{{W'}^{\rm c}}
\newcommand{\lXi}{\overset{\leftarrow}{\Xi}}
\newcommand{\rXi}{\overset{\to}{\Xi}}
\newcommand{\lOmega}{\overset{\leftarrow}{\Omega}}
\newcommand{\rOmega}{\overset{\to}{\Omega}}

\newcommand{\bthm}[1]{\begin{theorem} #1 \end{theorem}}
\newcommand{\blm}[1]{\begin{lemma} #1 \end{lemma}}
\newcommand{\bdf}[1]{\begin{definition} #1 \end{definition}}

\makeatletter
\newcommand{\subsubsubsection}{\@startsection{paragraph}{4}{\z@}%
  {-3.25ex plus -1ex minus -0.2ex}
 {1.5ex plus 0.2ex}
  {\reset@font\itshape\normalsize}
}
\makeatother
\begin{document}

\title{Complete classification of integrability and non-integrability of S=1/2 spin chains with symmetric next-nearest-neighbor interaction}


\author{Naoto Shiraishi }


\institute{Naoto Shiraishi  \at
              Graduate School of arts and sciences, University of Tokyo, Komaba 3-8-1, Meguro-ku, Tokyo, Japan \\
             \email{shiraishi@phys.c.u-tokyo.ac.jp}           
          }

\date{Received: date / Accepted: date}

\maketitle

\begin{abstract}
We study S=1/2 quantum spin chains with shift-invariant and inversion-symmetric next-nearest-neighbor interaction, also known as zigzag spin chains.
We completely classify the integrability and non-integrability of the above class of spin systems.
We prove that in this class there are only two integrable models, a classical model and a model solvable by the Bethe ansatz, and all the remaining systems are non-integrable.
Our classification theorem confirms that within this class of spin chains, there is no missing integrable model.
This theorem also implies the absence of intermediate models with a finite number of local conserved quantities.

\keywords{Integrable systems \and non-integrability \and local integral of motion \and zigzag spin chain}
\end{abstract}

\setcounter{tocdepth}{4}
\tableofcontents

\section{Introduction}\label{intro}

A quantum integrable system is one of the important subjects in mathematical physics.
Integrable systems have exact solutions of eigenenergies, energy eigenstates, and more detailed quantities such as correlation functions~\cite{JM, Bax, Tak}.
The quantum inverse scattering method (for locally interacting integrable systems) systematically provides an infinite sequence of local conserved quantities, with which we can compute the above quantities~\cite{Fad, KBI}.
Many other approaches to constructing local conserved quantities including the method of the Boost operator are also intensively studied~\cite{DG, FF, GM94, GM95-1, NF20}
On the other hand, integrable systems are also known to show several anomalous behaviors, including the absence of thermalization~\cite{Caz, RDYO, Lan, EF}, violation of the linear response theory~\cite{Kub, SF, SK}, anomalous transport~\cite{Zot}, and the Poisson distribution of level statistics~\cite{Haabook}, which are distinct properties from generic quantum many-body systems.
Thus, generic systems are strongly expected to be non-integrable.

In spite of the ubiquitousness of non-integrable systems, few studies in mathematical physics have addressed quantum non-integrability.
One reason for this unsatisfactory situation is that proving negative propositions (no solution or no local conserved quantity) is usually a hard task compared to existential propositions.
Most of the existing studies on non-integrability adopt approaches with fixing a solution method (e.g., Bethe ansatz and Yang-Baxter equation) and proving the unsolvability within this method~\cite{Ken92, BY94, Hie92, LPR19, Lee20, Mai24, LP24}.
However, this approach cannot exclude the possibility that a system is unsolvable by a method in consideration but is solvable by another method.
We here remark on a notable attempt by Grabowski and Mathieu~\cite{GM95-2}.
On the basis of examinations on known integrable models, they conjectured that a shift-invariant Hamiltonian with nearest-neighbor interaction is non-integrable if this Hamiltonian does not have a nontrivial 3-local conserved quantity (a conserved quantity consisting of operators with supports on three contiguous sites).
They also showed, through an exhaustive search, that the $S=1/2$ XYZ model with $z$ magnetic field and some $S=1/2$ zigzag spin chains have no 3-local conserved quantity.
Provided their conjecture, they also conjectured that these models are non-integrable.
The generalization of Grabowski-Mathieu conjecture to systems with next-nearest-neighbor interaction is investigated in Ref.~\cite{GP21}.

Recently, a mathematical technique proving non-integrability has been proposed~\cite{Shi19}, and various quantum spin systems have been rigorously proven to be non-integrable.
Here and in the remainder of this paper, we use the word {\it non-integrable} in the sense that a system has no nontrivial local conserved quantities~\cite{GM95-2, CM11, GE16}.
The non-integrability was first proven in the $S=1/2$ XYZ chain with $z$ magnetic field~\cite{Shi19}, which solves the Grabowski-Mathieu conjecture for this model in the affirmative.
After this work, the mixed field Ising chain~\cite{Chi24-1}, the next-nearest-neighbor Heisenberg model~\cite{Shi24}, and the PXP model~\cite{PL24-1} were proven to be non-integrable.
Aiming further generalizations, high-dimensional systems~\cite{ST24, Chi24-2} and $S=1$ spin chains~\cite{PL24-2, HYC24} have also been addressed.
In addition, recently the integrability and non-integrability of all possible $S=1/2$ spin chains with shift-invariant and inversion-symmetric nearest-neighbor interaction is completely classified~\cite{YCS24-1, YCS24-2}.
This classification theorem confirms that known integrable models are the complete list of integrable systems and there is no missing integrable models in this class.
This theorem also establishes the fact that all models in this class have an infinite number of local conserved quantities or no nontrivial local conserved quantity, and there is no intermediate model with a finite number of nontrivial local conserved quantities.

We notice that most of the aforementioned studies treat nearest-neighbor spin systems (on a square lattice in high dimensional cases), and systems beyond nearest-neighbor interaction have not yet been fully addressed.
The exceptions are on two specific models, the next-nearest-neighbor Heisenberg model~\cite{Shi24} and the PXP model~\cite{PL24-1}, while a general and comprehensive characterization of systems with next-nearest-neighbor interaction is elusive.
We note that spin chains with next-nearest-neighbor interaction are also called {\it zigzag spin chains} which have been intensively studied in the field of condensed matter physics~\cite{Ham88, Bat09, BS12, SH24}.

In this paper, we establish the classification theorem of the integrability and non-integrability of general $S=1/2$ spin chains with shift-invariant and inversion-symmetric next-nearest-neighbor interaction.
This theorem confirms that in this class there are only two integrable models, a classical model and a model solvable by the Bethe ansatz, and all other models are rigorously proven to be non-integrable.
The Grabowski-Mathieu conjecture on shift-invariant zigzag spin chains is again solved in the affirmative in this class.
In addition, an intermediate model with a finite number of nontrivial local conserved quantities is also excluded in this class.

Our proof is inspired by Refs.~\cite{YCS24-1, YCS24-2}, though the proof is much more complicated and longer than them.
This hardness stems from the fact that we have two interaction matrices, the next-nearest-neighbor interaction matrix and the nearest-neighbor interaction matrix, and thus in general we cannot diagonalize these two matrices simultaneously, leading to nonzero off-diagonal elements.
The presence of off-diagonal elements in the interaction Hamiltonian makes the proof complicated.

This paper is organized as follows.
In \sref{problem}, we present our main result, the classification theorem, and the basic proof idea.
We also present some symbols used in this proof.
The remainder of this paper is devoted to proving this classification theorem.
Our Hamiltonian is divided into three cases; rank 3, rank 2, and rank 1.
The non-integrability of rank 3 and rank 2 is proven in \sref{rank3} and \sref{rank2}, respectively.
The rank 1 case is further divided into three cases; case A, case B1, and case B2.
We first provide some general statements valid for all these three cases in \sref{rank1-gen}, and then treat these three cases in \sref{rank1-A}, \sref{rank1-B1}, and \sref{rank1-B2}, respectively.
We note that case B1 has an integrable case, and thus we further divide cases into integrable and (several) non-integrable ones.

\section{Main result}\lb{s:problem}

\subsection{Main claim}\lb{s:claim}

In this paper, we treat a general shift-invariant and inversion-symmetric $S=1/2$ spin chain with next-nearest-neighbor interaction with the periodic boundary condition, whose Hamiltonian takes the form of
\eqa{
H=\sum_{i=1}^L \sum_{\alpha,\beta\in \{X,Y,Z\}} J^2_{\alpha\beta}\sigma^\alpha_i \sigma^\beta_{i+2}+\sum_{i=1}^L \sum_{\alpha,\beta\in \{X,Y,Z\}} J^1_{\alpha\beta}\sigma^\alpha_i \sigma^\beta_{i+1}+\sum_{i=1}^L \sum_{\alpha\in \{X,Y,Z\}} h_{\alpha}\sigma^\alpha_i 
}{gen-form}
with $J^2_{\alpha\beta}=J^2_{\beta\alpha}$ and $J^1_{\alpha\beta}=J^1_{\beta\alpha}$.
Here, $\sigma^X$, $\sigma^Y$, and $\sigma^Z$ represent the Pauli matrices, and we identify sites $L+1$ and $L+2$ to sites 1 and 2, meaning the periodic boundary condition.
In this paper, we denote by $X$, $Y$, and $Z$ the Pauli matrices $\sigma^X$, $\sigma^Y$, and $\sigma^Z$.
By regarding even sites and odd sites as two parallel chains, this Hamiltonian is equivalent to spin systems on the zigzag chain.

Our goal of this paper is to classify the integrability and non-integrability of all models described as \eref{gen-form} rigorously.
Before going to our main claim, we first remark on a special case of \eref{gen-form}, where integrability and non-integrability have already been fully classified.
If one of $3\times 3$ interaction matrices $J^2$ or $J^1$ is a zero matrix, then the system is reduced to a nearest-neighbor interaction system with the same symmetry, which has been analyzed in Refs.~\cite{YCS24-1, YCS24-2}.
Thus, it suffices to treat only the case that both $J^2$ and $J^1$ are not zero matrices.

We first define the notion of the locality of operators.

\bdf{
An operator $C$ is a {\it $k$-support operator} if its minimum contiguous support is among $k$ sites.
In the case without confusion, the sum of $k$-support operators is also called simply a $k$-support operator.
}

Let us see several examples.
We denote by the subscript of an operator the site it acts.
Then, $X_4Y_5Z_6$ is a 3-support operator and $X_2Y_5$ is a 4-support operator, since their contiguous supports are $\{4,5,6\}$ and $\{2,3,4,5\}$, respectively.
An example of the sum of $3$-support operators is $\sum_i X_iY_{i+1}Z_{i+2}$, which we also call a 3-support operator.

We denote by $\calP^l$ a set of sequences of $l$ operators with $A\in \{ X,Y,Z,I\}$ such that the first and the last operators, $A^1$ and $A^l$, are one of the Pauli operators ($X$, $Y$, or $Z$), not an identity operator $I$, while other operators $A^2,\cdots ,A^{l-1}$ are one of $\{ X,Y,Z,I\}$.
We express such an operator sequence starting from site $i$ to site $i+l-1$ by a shorthand symbol $\bsA^l_i:=A^1_iA^2_{i+1}\cdots A^l_{i+l-1}$.
Using these symbols, a candidate of a shift-invariant local conserved quantity up to $k$-support operators can be expressed as
\eqa{
Q=\sum_{l=1}^{k} \sum_{\bsA^l\in \calP^l} \sum_{i=1}^L q_{\bsA^l}\bsA^l_i
}{Qform}
with coefficients $q_{\bsA^l}\in \bbR$.
The sum of $\bsA^l$ runs over all possible $9\times 4^{l-2}$ sequences in $\calP^l$.
Since the Pauli matrices and the identity span the space of $2\times2$ Hermitian matrices, the above form \eqref{Qform} covers all possible shift invariant quantities whose contiguous support of summand is less than or equal to $k$.

\bdf{
An operator $Q$ in the form \eqref{Qform} is a {\it $k$-support conserved quantity} if (i) $Q$ is conserved (i.e., $[Q,H]=0$), and (ii) one of $q_{\bsA ^k}$ is nonzero.
}

Conventionally, a {\it local} conserved quantity refers to a $k$-support conserved quantity with $k=O(1)$ with respect to the system size $L$.
Our main theorem excludes a much larger class of conserved quantities, including some $k=O(L)$ cases.

\bthm{\lb{t:main}
Consider a $S=1/2$ spin chain with Hamiltonian \eqref{gen-form} with $J^2_{\alpha\beta}=J^2_{\beta\alpha}$ and $J^1_{\alpha\beta}=J^1_{\beta\alpha}$.
Assume that both $J^2$ and $J^1$ are not zero matrices.

Then, this Hamiltonian has no $k$-support conserved quantity with $4\leq k\leq L/2$, except for the case that a Hamiltonian can be mapped onto one of the following two Hamiltonians by a global spin rotation:
\bi{
\item \underline{classical}: Hamiltonians with only $J^2_{ZZ}$, $J^1_{ZZ}$, and $h_Z$ may take nonzero values.
\item \underline{Bethe solvable}: Hamiltonians with only $J^2_{ZZ}$, $J^1_{XZ}=J^1_{ZX}$, and $h_Y$ may take nonzero values.
}
}

We shall explain the global spin rotation soon later.
Here the upper bound of $k$, $L/2$, is almost optimal since the square of the Hamiltonian, $H^2$, is a $L/2+3$-local conserved quantity.
This result partially solves the conjectures raised by Gombor and Pozsgay~\cite{GP21} in the affirmative for systems where the support size of its local Hamiltonian is equal to 2.

The rest of this paper is devoted to proving this classification theorem of $S=1/2$ symmetric next-nearest-neighbor spin chains.
Although the basic proof idea itself is similar to Refs.~\cite{YCS24-1, YCS24-2} and Ref.~\cite{Shi24}, there are many cases to think and some of these cases are uneasy to treat, which makes the proof longer and more complicated than existing papers.

\subsection{Proof strategy}\lb{s:proof-idea}

Our proof strategy is very similar to the classification of integrability and non-integrability of nearest-neighbor interaction spin chains~\cite{YCS24-1, YCS24-2}.
A pedagogical review of the basic idea of proof techniques is also presented in Ref.~\cite{Shi24}.

Since the next-nearest interaction coefficient matrix $J^2$ is a real symmetric matrix, $J^2$ is diagonalized by applying a proper $3\time 3$ orthogonal matrix $R$ to a Pauli matrix vector $\mx{X\\ Y\\ Z}$.
Denoting the newly obtained Pauli matrix vector $\mx{X'\\ Y'\\ Z'}=R\mx{X\\ Y\\ Z}$ simply by $\mx{X\\ Y\\ Z}$, any Hamiltonian written in \eref{gen-form} is reduced to the following {\it standard form}:
\eqa{
H=\sum_{i=1}^L (J^2_X X_iX_{i+2}+J^2_YY_iY_{i+2}+J^2_ZZ_iZ_{i+2}) +\sum_{i=1}^L \sum_{\alpha,\beta\in \{X,Y,Z\}} J^1_{\alpha\beta}\sigma^\alpha_i \sigma^\beta_{i+1}+\sum_{i=1}^L \sum_{\alpha\in \{X,Y,Z\}} h_{\alpha}\sigma^\alpha_i ,
}{standard-form}
where we abbreviated $J^2_{XX}$, $J^2_{YY}$, and $J^2_{ZZ}$ as $J^2_X$, $J^2_Y$, and $J^2_Z$, respectively.

We adopt highly different proof approaches depending on how many elements in $J^2_X$, $J^2_Y$, and $J^2_Z$ are nonzero.
We call the number of nonzero elements {\it rank} and provide proofs for the cases of rank 3, rank 2, and rank 1 separately.
(Precisely, we further divide the rank 1 case into many cases and treat them one by one.)

We notice that the commutator of a $k$-support operator $Q$ and $H$ is an at most $k+2$-support operator, which guarantees the expansion as
\eqa{
[Q,H]=\sum_{l=1}^{k+2}\sum_{\bsB^l\in \calP^l} \sum_{i=1}^L r_{\bsB^l} \bsB^l_i.
}{QH}
Since the left-hand side can be expressed in terms of a linear sum of $q_{\bsA}$ by inserting \eref{Qform}, and the conservation of $Q$ implies $r_{\bsB^l}=0$ for any $\bsB^l$, by comparing both sides of \eref{QH} we obtain many constraints (linear relations) on $q_{\bsA}$.
Our goal is to show that these linear relations do not have solutions except for $q_{\bsA ^k}=0$ for all $\bsA^k$, which means that $Q$ cannot be a $k$-support conserved quantity.

In all cases, our proof consists of two steps.
In step 1, employing mainly $r_{\bsB^{k+2}}=r_{\bsB^{k+1}}=0$ (sometimes $\bsB^{k}$ is also employed), we show that $\bsA^k$ in a specific form may have nonzero coefficients, and all the other ones have zero coefficients.
In addition, we show that most of the remaining coefficients are linearly connected.
This means that it suffices to show the coefficients of one or a few remaining operators zero for the proof of the absence of $k$-local conserved quantity.
In step 2, employing conditions for shorter supports ($\bsB^l$ with $l=k$ and sometimes that with $l=k-1, k-2$) we demonstrate that the remaining coefficients are zero.

\subsection{Symbols and terms (1)}\lb{s:symbol1}


We promise that an aligned Pauli operator as $XY$ means an operator where $X$ acts on a site and $Y$ acts on the next site (i.e., $X_i Y_{i+1}$).
If we intend to express a product of Pauli operators on the same site, we use a dot symbol $\cdot$ as $X\cdot Y$, whose rule of the product of Pauli matrices is a conventional one.
For completeness, we present the rule of products of Pauli matrices below:
\balign{
X\cdot X=Y\cdot Y=Z\cdot Z=&I, \\
X\cdot Y=-Y\cdot X=&iZ, \\
Y\cdot Z=-Z\cdot Y=&iX, \\
Z\cdot X=-X\cdot Z=&iY.
}
A commutator of two different Pauli matrices $A,B \in \{ X,Y,Z\}$ ($A\neq B$) satisfies
\eqa{
[A, B]=2A\cdot B.
}{commutator-product}
We frequently express a Pauli matrix in $\{X,Y\}$ as $W$, and a Pauli matrix in $\{X,Y,Z\}$ as $P$.
We also use $*$ to express an unknown operator.

For our later use, it is convenient to define {\it divestment} of a phase factor and {\it signless product} of Pauli matrices.
Using the symbol $|\cdot |$, we define the divestment of a phase factor of Pauli matrices as
\eq{
|aX|=|a|X, \hspace{10pt}
|aY|=|a|Y, \hspace{10pt}
|aZ|=|a|Z
}
with $a\in \bbC$.
The signless product of Pauli matrices is a product with divestment;
\balign{
|X\cdot Y|=|Y\cdot X|=&Z, \\
|Y\cdot Z|=|Z\cdot Y|=&X, \\
|Z\cdot X|=|X\cdot Z|=&Y.
}
The lost phase factor is given by the {\it sign factor} $\sigma(A, B)=\pm 1$ for $A,B\in \{ X,Y,Z\}$ ($A\neq B$)  so that
\eq{
A\cdot B=i\sigma(A, B)|A\cdot B|.
}
A concrete expression is
\balign{
\sigma(X,Y)=\sigma(Y,Z)=\sigma(Z,X)=&1, \\
\sigma(Y,X)=\sigma(Z,Y)=\sigma(X,Z)=&-1.
}

For a product of $l$ operators $\bsA=A_i^1A_{i+1}^2\cdots A_{i+l-1}^l$, the divestment is defined as
\eq{
|\bsA|=|A_i^1||A_{i+1}^2|\cdots |A_{i+l-1}^l|.
}
If $A^i\in \{X,Y,Z\}$ and $A^i\neq A^{i+1}$ are satisfied for all $i$, its sign factor is defined recursively as
\balign{
\sigma(A^1,A^2,\ldots, A^l)=\sigma(A^1,A^2)\sigma(A^2,A^3,\ldots , A^l)=&\sigma(A^1,A^2)\sigma(A^2,A^3)\sigma(A^3,\ldots , A^l) \nt \\
=&\sigma(A^1,A^2)\sigma(A^2,A^3)\cdots \sigma(A^{l-1},A^l). \lb{def-sigma-gen}
}
For example, we have $\sigma(X,Y,X,Z)=\sigma(X,Y)\sigma(Y,X)\sigma(X,Z)=1$.

\bigskip

We next introduce some concepts to describe commutators.
When a commutation relation $[\bsA, \bsC]=c\bsD$ holds with a number coefficient $c$, we say that the operator $\bsD$ is {\it generated} by the commutator of $\bsA$ and $\bsC$.
In our proof, we examine commutators generating a given operator and derive a relation of coefficients of operators.

Consider a candidate of conserved quantity $Q$ with $k=4$.
We take 6-support operator $X_iI_{i+1}Y_{i+2}X_{i+3}I_{i+4}Y_{i+5}$ in $\qh$ as an example.
This operator is generated by the following two commutators:
\balign{
-\i [X_iI_{i+1}Y_{i+2}Z_{i+3}, Y_{i+3}I_{i+4}Y_{i+5}]&=-2X_iI_{i+1}Y_{i+2}X_{i+3}I_{i+4}Y_{i+5}, \lb{com-ex1} \\
-\i [Z_{i+2}X_{i+3}I_{i+4}Y_{i+5}, X_iI_{i+1}X_{i+2}]&=2X_iI_{i+1}Y_{i+2}X_{i+3}I_{i+4}Y_{i+5}. \lb{com-ex2}
}
In case without confusion, we drop subscripts of operators for brevity.
The operator $XIYXIY$ in $[Q,H]$ is generated only by the above two commutators.
In such a case, we say that $XIYZ$ and $ZXIY$ form a {\it pair}.
Then, the condition $r_{XIYXIY}=0$, which comes from $\qh=0$, implies the following relation
\eq{
-J^2_Yq_{XIYZ}+J^2_Xq_{ZXIY}=0.
}
The precise relation of the above is $-J^2_Yq_{(XIYZ)_i}+J^2_Xq_{(ZXIY)_{i+2}}=0$.
Employing such linear relations, we specify a possible form of $q_{\bsA}$ and finally show that they are zero.

To capture these relations intuitively, we visualize commutators in a signless form by a {\it column expression}\fn{
In previous literature, the column expression usually describes commutators without divestment.
In this case, the column expression also contains the sign $\pm$.
}
similarly to the column addition.
For example, two commutation relations \eqref{com-ex1} and \eqref{com-ex2} are visualized as
\eq{
\begin{array}{rcccccc}
&X&I&Y&Z&& \\
&&&&Y&I&Y \\ \hline
&X&I&Y&X&I&Y
\end{array}
\ \ \ \ \
\begin{array}{rcccccc}
&&&Z&X&I&Y \\
&X&I&X&&& \\ \hline
&X&I&Y&X&I&Y
\end{array}. \nt 
}
Here, two arguments of the commutator are written above the horizontal line, and the result of the commutator with divestment is written below the horizontal line.
The horizontal positions in this visualization represent the spatial positions of spin operators.

\bigskip

We next introduce useful symbols $\a_i :=A_iA_{i+1}$ and $\widetilde{A}_i:=A_iA_{i+2}$, which we call as {\it doubling operator} and {\it extended doubling operator}, respectively.
The next-nearest-neighbor interactions in the Hamiltonian \eqref{standard-form} are three extended doubling operators, $\tx=X_iX_{i+2}=XIX$, $\ty=YIY$, and $\tz=ZIZ$.
The nearest-neighbor interactions in the Hamiltonian \eqref{standard-form} contain three doubling operators, $\x=XX$, $\y=YY$, and $\z=ZZ$, and off-diagonal interactions such as $XZ$.

When we align doubling operators and extended doubling operators, we promise that a neighboring doubling operator has its support with a single-site shift and a neighboring extended doubling operator has its support with a two-site shift.
For example, we can express
\eqa{
\x\y\x _i=|(X_iX_{i+1})(Y_{i+1}Y_{i+2})(X_{i+2}X_{i+3})| =X_i|X_{i+1}\cdot Y_{i+1}||Y_{i+2}\cdot X_{i+2}|X_{i+3} =X_i Z_{i+1} Z_{i+2}X_{i+3}.
}{doubling-product-def}
and
\eq{
\tx\ty\z_i=|(X_iX_{i+2})(Y_{i+2}Y_{i+4})(Z_{i+4}Z_{i+5})|=X_iZ_{i+2}X_{i+4}Z_{i+5}.
}
We require that doubling operators and extended doubling operators with the same symbol cannot be next to each other (e.g., $\x\x\z$ and $\tz\y\ty\x$ are not allowed).
If an operator is expressed by products of only doubling operators, we call this operator as {\it doubling-product operator}.
Similarly, if an operator is expressed by products of only extended doubling operators, we call this operator as {\it extended-doubling-product operator}.
If an operator is expressed by products of both doubling operators and extended doubling operators, we call this operator as {\it generalized doubling-product operator}.

To see the usefulness of (generalized) doubling-product operators, we also express a doubling-product operator $\overline{ABC\cdots D}$ as
\eqa{
\ba{cccccc}{
A&A&&&& \\
&B&B&&& \\
&&C&C&& \\
&&&\ddots&& \\
&&&&D&D \\ \hline \hline
}.
}{ladder}
Here, the double horizontal line means the multiplication of all the operators with divestment (removing the phase factor) from top to bottom, which we use for both doubling-product and non-doubling-product operators (e.g., off-diagonal operator $XZ$).
Keep in mind not to confuse a single horizontal line, which represents a commutation relation.
For example, $XZZYZ=\x\y\x\z$ is expressed as
\eq{
XZZYZ=
\ba{ccccc}{
X&X&&& \\
&Y&Y&& \\
&&X&X& \\
&&&Z&Z \\ \hline \hline
X&Z&Z&Y&Z
}.
}

Single and double horizontal lines are sometimes used at the same time as
\eqa{
\ba{ccccccc}{
A&A&&&&& \\
&B&B&&&& \\
&&C&C&&& \\
&&&\ddots&&& \\
&&&&D&D& \\ \hline \hline
&&&&&E&E \\ \hline
},
}{ex-ABCD-E}
which represents the commutator $[\overline{ABC\cdots D}, \overline{E}]$.
Here we abbreviated the last row (the resulting operator of these commutators) for brevity.
An example of this expression is
\eq{
\ba{cccccc}{
X&Z&Z&Y&Z& \\
&&&&X&X \\ \hline
X&Z&Z&Y&Y&X
}
=
\ba{cccccc}{
X&X&&&& \\
&Y&Y&&& \\
&&X&X&& \\
&&&Z&Z& \\ \hline \hline
&&&&X&X \\ \hline
X&Z&Z&Y&Y&X
}.
}

This type of expression also works for extended-doubling-product operators and generalized doubling-product operators.
For example, operators $\tx\ty\tx\tz$ and $\tx\y\z$ read respectively
\eq{
\tx\ty\tx\tz=\ba{ccccccccc}{
X&I&X&&&&&& \\
&&Y&I&Y&&&& \\
&&&&X&I&X&& \\
&&&&&&Z&I&Z \\ \hline \hline
}
=XIZIZIYIZ
}
and
\eq{
\tx\y\z=
\ba{ccccc}{
X&I&X&& \\
&&Y&Y& \\
&&&Z&Z \\ \hline \hline
}
=XIZXZ.
}
Note that this expression can be easily extended to operators which are not doubling-product operators.
For example, operator $\tz(XZ)\y$ reads
\eq{
\tz(XZ)\y=
\ba{ccccc}{
Z&I&Z&& \\
&&X&Z& \\
&&&Y&Y \\ \hline \hline
}
=ZIYXY.
}


\section{Rank 3}\lb{s:rank3}

Now we start proving our main classification theorem.

We first treat the case of rank 3, where the Hamiltonian is expressed as
\balign{
H=&\sum_i \mx{X_{i+2}&Y_{i+2}&Z_{i+2}}\mx{J_X^2 && \\ &J_Y^2& \\ &&J_Z^2}\mx{X_i\\ Y_i\\ Z_i}+\sum_i \mx{X_{i+1}&Y_{i+1}&Z_{i+1}}\mx{J_{XX}^1 &J_{XY}^1&J_{XZ}^1 \\ J_{XY}^1&J_{YY}^1&J_{YZ}^1 \\ J_{XZ}^1&J_{YZ}^1&J_{ZZ}^1}\mx{X_i\\ Y_i\\ Z_i} \nt \\
&+\sum_i \mx{h_X&h_Y&h_Z}\mx{X_i\\ Y_i\\ Z_i} \lb{rank3-standard}
}
with nonzero $J_X^2$, $J_Y^2$, and $J_Z^2$, and $J^1\neq O$.
In this section, we prove that the above Hamiltonian has no $k$-local conserved quantity with $4\leq k\leq L/2$.

Note in passing that the following proof is similar to the proof presented in Ref.~\cite{Shi24}.

\subsection{Restricting possible forms of $k$-support operators}\lb{s:NNN-step1}

A commutator of a $k$-support operator and the Hamiltonian can generate at most $k+2$-support operators.
Therefore, we first consider the case that the commutator generates $k+2$-support operators.
A $k+2$-support operator in $\qh$ is generated only by a commutator such that the next-nearest-neighbor interaction term ($XIX$, $YIY$, and $ZIZ$) acts on the left end or right end of a $k$-support operator.
The following two types of commutators serve as examples:
\eq{
\ba{ccccccc}{
A^1&A^2&\cdots&A^{k-1}&A^k&& \\
&&&&X&I&X \\ \hline
A^1&A^2&\cdots&A^{k-1}&*&I &X
}
, \ \
\ba{ccccccc}{
&&A^1&A^2&\cdots&A^{k-1}&A^k \\
Y&I &Y&&&& \\ \hline
Y&I &*&A^2&\cdots&A^{k-1}&A^k
}.
}

From this, we find an important restriction on the possible form of operators which may have nonzero coefficients.
First, $A^1\neq A^3$ and $A^2=I$ are necessary for an operator $\bsA$ to have a nonzero coefficient.
To confirm this fact, we first take $\bsA=XYZ\cdots Z$ as an example of $A^2\neq I$.
We consider commutators generating $XYZ\cdots YIX$ as
\eq{
\ba{ccccccc}{
X&Y&Z&\cdots&Z&& \\
&&&&X&I&X \\ \hline
X&Y&Z&\cdots&Y&I&X
}.
}
It is easy to confirm that this commutator is the unique commutator generating $XYZ\cdots YIX$.
In fact, in order to generate  $XYZ\cdots YIX$ by adding a 3-support operator from left, the 3-support operator should take the form of $XY*$ as
\eq{
\ba{ccccccc}{
&&*&\cdots&Y&I&X \\
X&Y&*&&&& \\ \hline
X&Y&Z&\cdots&Y&I&X
},
}
while our Hamiltonian \eqref{standard-form} does not have a term in the form of $XY*$.
From this, we find
\eq{
J^2_X q_{XYZ\cdots Z}=0,
}
which means $q_{XYZ\cdots Z}=0$.

We next take $\bsA=XIX\cdots Z$ as an example of $A^1=A^3$.
We consider commutators generating $XIX\cdots YIX$ as
\eq{
\ba{ccccccc}{
X&I&X&\cdots&Z&& \\
&&&&X&I&X \\ \hline
X&I&X&\cdots&Y&I&X
}.
}
It is easy to confirm that this commutator is the unique commutator generating $XIX\cdots YIX$.
In fact, in order to generate  $XIX\cdots YIX$ by adding a 3-support operator from left, the added 3-support operator should be $XIX$ and the commutator reads
\eq{
\ba{ccccccc}{
&&*&\cdots&Y&I&X \\
X&I&X&&&&\\ \hline
X&I&X&\cdots&Y&I&X
},
}
while no $*$ satisfies the relation $|[*, X]|=X$.
From this, we find
\eq{
J^2_X q_{XIX\cdots Z}=0,
}
which means $q_{XIX\cdots Z}=0$.
In general, if two leftmost operators are not one of $XI\cdots$, $YI\cdots$, or $ZI\cdots$, or two rightmost operators are not one of $\cdots IX$, $\cdots IY$, or $\cdots IZ$, then there exists a $k+2$-support operator which is uniquely generated by this operator, implying zero coefficient.

Due to the inversion symmetry, the above arguments also hold for the right end of $\bsA$.
If a $k$-support operator $\bsA$ satisfies $A^1\neq A^3$, $A^{k-2}\neq A^k$, and $A^2=A^{k-1}=I$, operator $\bsA$ forms a {\it pair} with another $k$-support operator.
For example, $k$-support operator $XIY\cdots ZIX$ forms a pair with $ZIYIY\cdots Y$ by considering commutators generating $k+2$-support operator $ZIYIY\cdots ZIX$ as
\eq{
\ba{cccccccc}{
&&X&I&Y\cdots&Z&I&X \\
Z&I&Z&&&&& \\ \hline
Z&I&Y&I&Y\cdots&Z&I&X
}
\hspace{15pt}
\ba{cccccccc}{
Z&I&Y&I&Y\cdots&Y&& \\
&&&&&X&I&X \\ \hline
Z&I&Y&I&Y\cdots&Z&I&X
},
}
which leads to a linear relation of coefficients:
\eq{
-J^2_Zq_{XIY\cdots ZIX}-J^2_X q_{ZIYIY\cdots Y}=0.
}
This relation suggests that $q_{ZIYIY\cdots Y}=0$ directly implies $q_{XIY\cdots ZIX}=0$.
In general, if two operators $\bsA$ and $\bsA'$ form a pair, then $q_{\bsA'}=0$ directly implies $q_{\bsA}=0$.

Considering this procedure repeatedly, we find that a $k$-support operator $\bsA$ may have a nonzero coefficient only if $k$ is odd and it is an extended-doubling-product operator $\tb^1 \tb^2\tb^3\cdots \tb^{(k-1)/2}$ with $B^i\neq B^{i+1}$ for all $i$.
In addition, all the coefficients of the remaining $k$-support operators (extended-doubling-product operators) on odd sites are linearly connected, and this is also true for those on even sites.
Note that if $k$ is even, all $k$-support operators $\bsA$ are shown to have zero coefficients.

We first see why extended-doubling-product operators are special, with which we also confirm the latter fact; linear connection.
We here demonstrate how two extended-doubling-product operators are connected.
For example, $\tx\ty\tx\tz$ and $\ty\tx\tz\tx$ forms a pair by
\eq{
\ba{ccccccccccc}{
X&I&X&&&&&&&& \\
&&Y&I&Y&&&&&& \\
&&&&X&I&X&&&& \\
&&&&&&Z&I&Z&& \\ \hline \hline
&&&&&&&&X&I&X \\ \hline
}
\hspace{15pt}
\ba{ccccccccccc}{
&&Y&I&Y&&&&&& \\
&&&&X&I&X&&&& \\
&&&&&&Z&I&Z&& \\
&&&&&&&&X&I&X \\ \hline \hline
X&I&X&&&&&&&& \\ \hline
}.
}
This clearly shows that a pair of two operators are connected by removing an extended-doubling-operator from left/right and adding an extended-doubling-operator to right/left.
For example, $(\tx\ty\tx\tz)_1$ and $(\tz\tx\tz\ty)_1$ are connected as
\balign{
&(\tx\ty\tx\tz)_1\lr (\ty\tx\tz\tx)_3\lr (\tx\tz\tx\ty)_5\lr (\tz\tx\ty\tx)_7\lr (\tx\ty\tx\ty)_9 \nt \\
& \lr (\ty\tx\ty\tx)_7\lr (\tz\ty\tx\ty)_5\lr(\tx\tz\ty\tx)_3\lr (\tz\tx\tz\ty)_1,
}
where we explicitly write the position dependence to clarify the parity and displacement of operators.
In the first line, we append $\tx$ and $\ty$ alternately to the right, and in the second line we append $\ty, \tz, \tx, \tz$ to the left with removing the alternate $\tx$ and $\ty$ from the right in order to construct  $\tz\tx\tz\ty$.
Hence, any two extended-doubling-product operators on sites with the same parity are connected by properly removing and adding extended-doubling-operator.
In addition, as seen from how to connect two extended-doubling-product operators, the coefficient of $\tb^1 \tb^2\tb^3\cdots \tb^{(k-1)/2}$ is proportional to $\sigma (\tb^1, \tb^2,\tb^3,\ldots, \tb^{(k-1)/2})$, since any extended-doubling-product operator is constructed by appending $\tb^{(k-1)/2}, \ldots , \tb^2, \tb^1$ to the left.

We next see how a $k$-support operator which is not an extended-doubling-product operator vanishes.
Consider operator $XIZIZZYIXIY$ in the case of $k=11$ as an example, which can be expressed as
\eq{
XIZIZZYIXIY=
\ba{ccccccccccc}{
X&I&X&&&&&&&& \\
&&Y&I&Y&&&&&& \\
&&&&X&Z&X&&&& \\
&&&&&&Z&I&Z&& \\
&&&&&&&&Y&I&Y \\ \hline \hline
}.
}
Here, a ``defect" $XZX$ is inserted.
This defect lies in the following series of pairs
\eq{
\ba{ccccccccccc}{
X&I&X&&&&&&&& \\
&&Y&I&Y&&&&&& \\
&&&&X&Z&X&&&& \\
&&&&&&Z&I&Z&& \\
&&&&&&&&Y&I&Y \\ \hline \hline
}
\lr
\ba{ccccccccccc}{
Y&I&Y&&&&&&&& \\
&&X&Z&X&&&&&& \\
&&&&Z&I&Z&&&& \\
&&&&&&Y&I&Y&& \\
&&&&&&&&Z&I&Z \\ \hline \hline
}
\lr
\ba{ccccccccccc}{
X&Z&X&&&&&&&& \\
&&Z&I&Z&&&&&& \\
&&&&Y&I&Y&&&& \\
&&&&&&Z&I&Z&& \\
&&&&&&&&Y&I&Y \\ \hline \hline
}.
}
However, the last operator $XZYIXIXIXIY$ cannot form a pair because the two leftmost operators are $XZ\cdots$, not in the form of $XI\cdots$:
\eq{
\ba{ccccccccccccc}{
X&Z&Y&I&X&I&X&I&X&I&Y&& \\
&&&&&&&&&&Z&I&Z \\ \hline
X&Z&Y&I&X&I&X&I&X&I&X&I&X
},
\hspace{15pt}
\ba{ccccccccccccc}{
&&*&*&*&*&*&*&*&*&*&*&* \\
X&Z&*&&&&&&&&&& \\ \hline
X&Z&Y&I&X&I&X&I&X&I&X&I&X
}.
}
This fact implies
\eq{
q_{XZYIXIXIXIY}=0,
}
and hence the initial operator $XIZIZZYIXIY$ also has zero coefficient:
\eq{
q_{XIZIZZYIXIY}=0.
}
If a $k$-support operator is not an extended-doubling-product operator, then by removing extended-doubling-products from the left and adding proper extended-doubling-products to the right repeatedly, we arrive at an operator which cannot form a pair, resulting in a zero coefficient\fn{
\lb{f:rank3-twist}
Remark that additional care is required if the defect is in the form of $PIP'$ with $P,P'\in \{X,Y,Z\}$ and $P\neq P'$.
We explain by taking $\tx\ty(XIY)\tx\ty$ with the defect $XIY$ as an example.
In this case, we first remove extended doubling operators from right, not left, so that $(XIY)$ comes to the right end.
We then add $\tz$ to the right end.
After this, we remove extended doubling operators from the left by adding extended doubling operators to the right, whose details are not important.
The obtained sequence of pairs, for example, is
\balign{
\tx\ty(XIY)\tx\ty\lr \ty\tx\ty(XIY)\tx &\lr \tx\ty\tx\ty(XIY) \lr \ty\tx\ty(XIY)\tz  \nt \\
&\lr \tx\ty(XIY)\tz\ty \lr \ty(XIY)\tz\ty\tz \lr (XIY)\tz\ty\tz\ty,
}
where the last operator $(XIY)\tz\ty\tz\ty=XIXI\cdots$ does not form a pair when we add an extended doubling operator to the right.
In this argument, the choice of $\tz$ right next to $(XIY)$ is important, since owing to this choice the left end becomes $XI|Y\cdot Z|\cdots =XIX\cdots$, which has already been shown to have zero coefficient.
}.

Consider a Hamiltonian \eqref{rank3-standard} with $J_Z^2\neq 0$ and $J^1_{XX}\neq 0$.
In a candidate of a $k$-support conserved quantity $Q$, the coefficient of a $k$-support operator satisfies the following:

\blm{\lb{l:NNN-k}
Consider a Hamiltonian \eqref{rank3-standard} with $J^2_X, J^2_Y, J^2_Z\neq 0$ and $J^1\neq O$.
In a candidate of a $k$-support conserved quantity $Q$ expanded as \eqref{Qform}, a $k$-support operator $\bsA\in \calP^k$ which may have a nonzero coefficient is an extended-doubling-product operator written as 
\eqa{
\bsA=\prod_{i=1}^{(k-1)/2} \widetilde{B}^i.
}{rank3-k-form}
Otherwise, $\bsA$ has zero coefficient.

In addition, the coefficient of $\bsA$ in the form of \eref{rank3-k-form} is expressed as
\eqa{
q_{\bsA}=c^{3,k,a}\cdot \sigma(B^1,B^2,\ldots , B^{(k-1)/2}) \prod_{i=1}^{(k-1)/2} J^2_{B^i}
}{NNN-doubling-coefficient-1}
with two common constants $c^{3,k,a}$ corresponding to rank 3, $k$-support operator, and parity $a\in \{0,1\}$.
Here, $a=0$ (resp. $a=1$) means its nontrivial support on even (resp. odd) sites.
}

Clearly, a $k$-support conserved quantity with even $k$ vanishes.

\subsection{Restricting possible forms of $k-1$-support operators}\lb{s:NNN-step2}

We next restrict a possible form of $k-1$-support operator by considering commutators generates $k+1$-support operators.
If the nearest-neighbor interaction coefficient matrix $J^1$ has no off-diagonal term, then the required argument becomes completely the same as that presented in Ref.~\cite{Shi24}.
Therefore, the remaining case is that some off-diagonal elements have a nonzero coefficient.
Recalling the symmetry of $X$, $Y$, and $Z$, it suffices to treat only the case with $J^1_{YZ}=J^1_{ZY}\neq 0$.

We first notice that only the following two commutators generate a $k+1$-support operator $XIZI\cdots IXZ$:
\eqa{
\ba{ccccccccc}{
X&I&X&&&&&& \\
&&Y&I&Y&&&& \\
&&&&\ddots&&&& \\
&&&&&Z&I&Z& \\ \hline \hline
&&&&&&&Y&Z \\ \hline
}
, \ \ \ \
\ba{ccccccccc}{
&&Y&I&Y&&&& \\
&&&&\ddots&&&& \\
&&&&&Z&I&Z& \\
&&&&&&&Y&Z \\ \hline \hline
X&I&X&&&&&& \\ \hline
},
}{rank3-1-2-demo}
which implies the following relation of coefficients:
\eq{
-J^1_{YZ}q_{\tx\ty \cdots \tz}-J^2_Xq_{\ty\cdots \tz(YZ)}=0.
}
This relation connects the coefficient of a $k$-support operator and that of a $k-1$-support operator.

We can connect two $k-1$-support operators, e.g., ${\ty\tx\cdots \tz(YZ)}$ and ${\tx\cdots \tz(YZ)\ty}$, by considering the following two commutators:
\eq{
\ba{ccccccccccc}{
Y&I&Y&&&&&&&& \\
&&X&I&X&&&&&& \\
&&&&\ddots&&&&&& \\
&&&&&Z&I&Z&&& \\
&&&&&&&Y&Z&& \\ \hline \hline
&&&&&&&&Y&I&Y \\ \hline
}
,\ \ \
\ba{ccccccccccc}{
&&X&I&X&&&&&& \\
&&&&\ddots&&&&&& \\
&&&&&Z&I&Z&&& \\
&&&&&&&Y&Z&& \\
&&&&&&&&Y&I&Y \\ \hline  \hline
Y&I&Y&&&&&&&& \\ \hline
},
}
which implies
\eq{
J_Y^2q_{\ty\tx\cdots \tz(YZ)}+J_Y^2q_{\tx\cdots \tz(YZ)\ty}=0.
}
Using this procedure repeatedly, we find that any $k-1$-support operator written as the product of $(k-3)/2$ extended doubling operators and one nearest-neighbor interaction term are connected to $k$-support operators.
For example, in the case of $k=13$, $k-1$-support operator $\tx\ty\tx(YZ)\tx\tz=XIZIZIZYIYIZ$ is connected to a $k$-support operator as
\eq{
\tx\ty\tx(YZ)\tx\tz \lr \ty\tx\ty\tx(YZ)\tx \lr \tx\ty\tx\ty\tx(YZ) \lr \ty\tx\ty\tx\ty\tx ,
}
where the first three operators are $k-1$-support operators and the last one is a $k$-support operator.

We note that if a $k-1$-support operator is written as the above form, $(k-3)/2$ extended doubling operators and one nearest-neighbor interaction term are uniquely determined even when all matrix elements of $J^1$ are nonzero.
Extended doubling operators are determined automatically from left or right since next-nearest interactions are only $XIX$, $YIY$, or $ZIZ$.
Then, the remaining two operators, which constitute the nearest-neighbor interaction term, are determined.

Importantly, by a similar assertion, a $k-1$-support operator not in the above form is shown to have a zero coefficient.
To confirm this fact, we insert extended doubling operators from right repeatedly and move a non-extended-doubling-operator part to the left end.
If this operator is not written as $(P^1P^2)\tb^1\tb^2\cdots \tb^{(k-1)/2}$, then inserting an extended doubling operator from right and removing $(P^1P^2)$ as \eref{rank3-1-2-demo}, we find that the obtained $k$-support operator does not take the form of \eref{rank3-k-form} (in \lref{NNN-k}), which suggests that it has zero coefficient.
(See Sec.3.1 in Ref.~\cite{Shi24} for detailed discussion).

We remark that using the above trick two common factors, $c^{3,k,0}$ and $c^{3,k,1}$ in \lref{NNN-k}, are shown to take the same value.
This fact can be seen by connecting $k$-support operators on odd sites and those on even sites by forming pairs as
\eq{
(\tx\ty\tz)_1 \lr (\ty\tz(YZ))_3 \lr (\tz(YZ)\tx)_5 \lr ((YZ)\tx\ty)_7 \lr (\tx\ty\tz)_8,
}
where we employ the case of $J^1_{YZ}\neq 0$.
Other cases are treated in similar manners.

\blm{
Consider a Hamiltonian \eqref{rank3-standard} with $J^2_X, J^2_Y, J^2_Z\neq 0$ and $J^1\neq O$.
In a candidate of a $k$-support conserved quantity $Q$ expanded as \eqref{Qform}, two factors $c^{3,k,0}$ and $c^{3,k,1}$ in \lref{NNN-k} are the same:
\eq{
c^{3,k}:=c^{3,k,0}=c^{3,k,1}.
}
}

\blm{
Consider a Hamiltonian \eqref{rank3-standard} with $J^2_X, J^2_Y, J^2_Z\neq 0$ and $J^1\neq O$.
In a candidate of a $k$-support conserved quantity $Q$ expanded as \eqref{Qform}, a $k-1$-support operator $\bsA\in \calP^{k-1}$ which may have a nonzero coefficient is expressed as
\eqa{
\bsA=\tb^1\tb^2\cdots \tb^{m-1} \Psi \tb^{m}\cdots \tb^{(k-1)/2-1},
}{rank3-k-1-form}
where $\Psi=\Psi_1\Psi_2\in \{XX, YY, ZZ, XY, XZ, YX, YZ, ZX, ZY\}$ is a nearest-neighbor interaction term. 
Otherwise, the coefficient is zero: $q_{\bsA}=0$.

In addition, the coefficient of $\bsA$ in the form \eqref{rank3-k-1-form} is calculated as
\eqa{
q_{\bsA}=c^{3,k}\cdot \sigma(B^1, B^2,\ldots , B^{m-1}, \Psi^1)\sigma(\Psi^2, B^m, \ldots , B^{(k-1)/2-1}) J^1_{\Psi}\prod_{i=1}^{(k-1)/2-1} J^2_{B^i}
}{NNN-doubling-coefficient-2}
with a common constant $c^{3,k}$ to \eref{NNN-doubling-coefficient-1}.
}

\subsection{Demonstrating that the remaining $k$-support operators have zero coefficients}\lb{s:NNN-step3}

We finally demonstrate that the remaining $k$-support operator has zero coefficient by considering commutators generates a $k$-support operator.
We here take the case of $J^1_{YZ}\neq 0$ as an example.
Other cases can be treated similarly.

Following Ref.~\cite{Shi24}, we introduce the symbol $\hrzy$, representing commutation relations with $ZY$ at this position.
For example, a commutator
\eq{
\ba{ccccc}{
Y&I&Y&& \\
&&Z&I&Z \\ \hline \hline
&&Z&Y& \\ \hline
}
}
is expressed as
\eq{
\ty \hrzy \tz .
}
We also introduce the symbol $\brzy$, representing the multiplication of $ZY$ at this position.
For example, 
\eq{
YIYYZ=\ba{ccccc}{
Y&I&X&I&Z \\
&&Z&Y& \\ \hline \hline
}
=
\ba{ccccc}{
Y&I&Y&& \\
&&Z&I&Z \\
&&Z&Y& \\ \hline \hline
}
}
is expressed as
\eq{
YIYYZ=\ty \brzy \tz.
}
Here we regard the position of the vertical bar $|$ as the site with overlap of $\ty$ and $\tz$, which is also the position $Z$ in $ZY$ sits on.

We also introduce symbols ``$\rplus$" and ``$\lplus$", which mean that commutators act at the rightmost and leftmost sites, respectively.
For example, $\tx\tz \rplus \tx$ means a commutator where $\tx$ acts on the right end of $\tx\tz=XIYIZ$:
\eq{
\tx\tz \rplus \tx=\ba{ccccccc}{
X&I&X&&&& \\
&&Z&I&Z&& \\ \hline \hline 
&&&&X&I&X \\ \hline
}
}

We promise that if $\brzy$ sits at the right end and an extended-doubling-product is added from the right, then the left end of this extended-doubling-product acts on the second rightmost site where $Z$ in $\brzy$ acts.
For example, $\ty \brzy \rplus\tz$ means that $\tz$ acts on the right end of $\ty$ and generates 5-local operator $YIZYY$:
\eq{
\ty \brzy \rplus\tz=
\ba{ccccc}{
Y&I&Y&& \\
&&Z&Y& \\ \hline \hline
&&Z&I&Z \\ \hline
}.
}
Using these symbols, two commutators in \eref{rank3-1-2-demo} are expressed as
\eq{
\tx\ty \cdots \tz \rplus (YZ) , \ \ \ \tx \lplus\ty\cdots \tz(YZ) .
}

We further introduce symbols which represent alternating $\tx$ and $\ty$ defined as
\balign{
\tL^{2n}&:=\underbrace{\ty\tx\cdots \ty\tx}_{n \text{ copies of }\ty\tx}, \\
\tL^{2n+1}&:=\tx\underbrace{\ty\tx\cdots \ty\tx}_{n \text{ copies of }\ty\tx}, \\
\tR^{2n}&:=\underbrace{\tx\ty\cdots \tx\ty}_{n \text{ copies of }\tx\ty}, \\
\tR^{2n+1}&:=\underbrace{\tx\ty\cdots \tx\ty}_{n \text{ copies of }\tx\ty}\tx.
}
By definition, $\tL^{2n+1}=\tR^{2n+1}$ holds, and we use them interchangeably.

\bigskip

Now we construct a sequence of commutators, with which we can demonstrate that one of the remaining $k$-support operators has zero coefficient.
For the brevity of explanation, we only treat the case of $k\equiv 3\mod 4$ .
The extension to the case of $k\equiv 1\mod 4$ is straightforward.
We express $k=4r+3$ and consider the following sequence:
\eqa{
\ba{ccccccc}{
\tL^{2r} \hrzy \tz &\threeskip& \tL^{2r} \brzy \rplus\tz &&&\threeskip&  \ty \lplus \tL^{2r-1} \brzy \tz \\
\tL^{2r-1} \hrzy \tz\tR^1 &\threeskip&&& \tL^{2r-1} \brzy \tz \rplus\tx &\threeskip& \tx \lplus \tL^{2r-2} \brzy \tz\tR^1 \\
\tL^{2r-2} \hrzy \tz\tR^2 &\threeskip&&& \tL^{2r-2} \brzy \tz\tR^1 \rplus\ty &\threeskip& \ty \lplus \tL^{2r-3} \brzy \tz\tR^2 \\
\vdots &\threeskip&&& \vdots &\threeskip& \vdots \\
\tL^{2r-n} \hrzy \tz\tR^n &\threeskip&&& \tL^{2r-n} \brzy \tz \tR^{n-1} \rplus\ty &\threeskip& \ty \lplus \tL^{2r-n-1} \brzy \tz\tR^n \\
\tL^{2r-n-1} \hrzy \tz\tR^{n+1} &\threeskip&&& \tL^{2r-n-1} \brzy \tz\tR^n \rplus\tx &\threeskip& \tx \lplus \tL^{2r-n-2} \brzy \tz\tR^{n+1} \\
\vdots &\threeskip&&& \vdots &\threeskip& \vdots \\
\tL^2 \hrzy \tz\tR^{2r-2} &\threeskip&&& \tL^2 \brzy \tz\tR^{2r-3} \rplus\ty &\threeskip& \ty \lplus \tL^1 \brzy \tz\tR^{2r-2} \\
\tL^1 \hrzy \tz\tR^{2r-1} &\threeskip&&& \tL^1 \brzy \tz\tR^{2r-2}\rplus \tx &\threeskip& \\
}
}{sequence-NNN}
where $n$ is even.
Commutators in the same row generate the same operator.
The leftmost column has commutators between a $k$-body operator and 2-body operator $ZY$ (in the Hamiltonian), the second left column (in the first row) has a commutator between a $k-1$-body operator and 3-body operator $\tz$ in the Hamiltonian, and the two right columns show commutators between a $k-2$-body operator and a 3-body operator ($\tx$ or $\ty$) in the Hamiltonian.

We put two remarks:
First, in each row, commutators of a $k$-body operator and a 2-body operator appear only once.
To see this point, let us take $\tx  \hrzy \tz\tx$ as an example, which generates $\tL^1 \hrzy \tz \tR^1=XIXYYIX$.
With noting that remaining $k$-support operators take the form \eqref{rank3-k-form} stating $P^1IP^2I\cdots$ ($P^1,P^2,\ldots \in \{X,Y,Z\}$), another candidate of a commutator between a $k$-body operator and 2-body operator generating $XIXYYIX$ is
\eq{
\ba{ccccccc}{
X&I&X&I&*&I&X \\
&&&Y&*&& \\ \hline \hline
X&I&X&Y&Y&I&X \\ \hline
}.
}
However, this $k$-support operator does still not satisfy the form of \eref{rank3-k-form} (in \lref{NNN-k}), and thus this contribution does not exist.
Second, the last law generating $XIXYYI\cdots$ has only two elements, because we cannot obtain this operator by a commutator in the form of $\tx \lplus (k-2 \text{-support operator})$.

Hence, by employing the abbreviation $J_{\widetilde{\bsB}}:=\prod_i J^2_{B^i}$ with $\bsB=\tb^1\tb^2\cdots $, the relations obtained from the odd ($n+1$-th) row, namely
\eq{
\ba{ccccccc}{
\tL^{2r-n} \hrzy \tz\tR^n &\threeskip&&& \tL^{2r-n} \brzy \tz \tR^{n-1} \rplus\ty &\threeskip& \ty \lplus \tL^{2r-n-1} \brzy \tz\tR^n
},
\nt
}
except $n=0$ read
\balign{
J_{ZY}^1 J_{\tL^{2r-n}\tz\tR^n}\cdot c^{3,k}+&J_Y^2q_{ \tL^{2r-n}\brzy \tz \tR^{n-1}}+J_Y^2q_{\tL^{2r-n-1} \brzy \tz\tR^n}=0. \lb{NNN-step3-mid1}
}
In a similar manner to above, the relation on coefficients obtained from the $n+2$-th row 
\eq{
\ba{ccccccc}{
\tL^{2r-n-1} \hrzy \tz\tR^{n+1} &\threeskip&&& \tL^{2r-n-1} \brzy \tz\tR^n \rplus\tx &\threeskip& \tx \lplus \tL^{2r-n-2} \brzy \tz\tR^{n+1}
}, \nt
}
except $n=2r-2$ reads
\eqa{
J_{ZY}^1J_{\tL^{2r-n-1}\tz\tR^{n+1}} c^{3,k}-J_X^2q_{ \tL^{2r-n-1}\brzy \tz \tR^{n}} -J_X^2  q_{\tL^{2r-n-2} \brzy \tz\tR^{n+1}}=0.
}{NNN-step3-mid2}
In addition, the first row implies
\eq{
J_{ZY}^1J_{\tL^{2r}\tz}\cdot c^{3,k}+ J_Z^2 J_{\tL^{2r}\brzy}\cdot c^{3,k} + J_Y^2q_{ \tL^{2r-1}\brzy \tz}=0
}
and the last row implies
\eq{
J_{ZY}^1 J_{\tL^1 \tz\tR^{2r-1}}\cdot c^{3,k} - J_X^2 q_{\tL^1 \brzy \tz \tR^{2r-2}}=0.
}
Noticing that for even $n'$ and odd $n''$
\eq{
\frac{J_X^2}{J_Y^2}J_{\tL^{2r-n'}\tz\tR^{n'}}=J_{\tL^{2r-n''}\tz\tR^{n''}}
}
is satisfied, we find that the sum of these relations (Eqs.~\eqref{NNN-step3-mid1} and \eqref{NNN-step3-mid2}) from $n=0$ to $n=2r-1$ reads
\eq{
(2r+1) J_{ZY}^1c^{3,k}=0,
}
which implies that all the coefficients of $k$-support operators in $Q$ is zero.
This completes the proof for rank 3.

\bthm{
Consider a Hamiltonian \eqref{rank3-standard} with $J^2_X, J^2_Y, J^2_Z\neq 0$ and $J^1\neq O$.
This Hamiltonian has no $k$-local conserved quantity with $4\leq k\leq L/2$. 
}

\section{Rank 2}\lb{s:rank2}

The proof for the case of rank 2 is very similar to the case of rank 3 (See also Refs.~\cite{YCS24-1, YCS24-2} for the proof of rank 2 for nearest-neighbor interaction systems).
Without loss of generality, we assume that $J_X^2\neq 0$, $J_Y^2\neq 0$, and $J_Z^2=0$, with which the Hamiltonian is expressed as
\balign{
H=&\sum_i \mx{X_{i+2}&Y_{i+2}&Z_{i+2}}\mx{J_X^2 && \\ &J_Y^2& \\ &&0}\mx{X_i\\ Y_i\\ Z_i}+\sum_i \mx{X_{i+1}&Y_{i+1}&Z_{i+1}}\mx{J_{XX}^1 &J_{XY}^1&J_{XZ}^1 \\ J_{XY}^1&J_{YY}^1&J_{YZ}^1 \\ J_{XZ}^1&J_{YZ}^1&J_{ZZ}^1}\mx{X_i\\ Y_i\\ Z_i} \nt \\
&+\sum_i \mx{h_X&h_Y&h_Z}\mx{X_i\\ Y_i\\ Z_i} \lb{rank2-standard}
}
with $J^1\neq O$.

We shall show that this Hamiltonian has no  $k$-local conserved quantity with $4\leq k\leq L/2$.

\subsection{Restricting possible forms of $k$-support operators}

For a similar reason to the case of rank 3, the analysis of commutators generating $k+2$-support operators tells that an operator which may have a nonzero coefficient should take the form of $P^1IP^2IP^3I\cdots$ with $P^i\in \{X,Y,Z\}$.
In the case of rank 2, we can use only two types of extended doubling operators, $\tx$ and $\ty$, and thus we should insert these two alternatingly, which leads to four possible $k=2m+1$ operators which may have nonzero coefficients:
\eqa{
X(IZ)^{m-1}IX, \hspace{15pt} X(IZ)^{m-1}IY, \hspace{15pt} Y(IZ)^{m-1}IX, \hspace{15pt} Y(IZ)^{m-1}IY.
}{rank2-k-form}
Other $k$-support operators are shown to have zero coefficient by a similar argument to \sref{NNN-step1}.
Here, $(IZ)^{m-1}$ means $m-1$ copies of $IZ$.

Remark that, unlike the case of rank 3, two of the above four operators in \eref{rank2-k-form} are not extended-doubling-product operators, i.e., operators not expressed by an alternating product of $\tx$ and $\ty$.
More precisely, $X(IZ)^{m-1}IY$ and $Y(IZ)^{m-1}IX$ are not extended-doubling-product operators for odd $m$, and $X(IZ)^{m-1}IX$ and $Y(IZ)^{m-1}IY$ are not extended-doubling-product operators for even $m$.
Note that a similar situation is seen in nearest-neighbor interaction Hamiltonians~\cite{YCS24-1,YCS24-2}.

We here briefly explain why non-extended-doubling-product operators are not excluded at this stage.
As explained in footnote \ref{f:rank3-twist}, if the defect takes the form of $PIP'$ with $P\neq P'$, we need to put a proper extended doubling operator to its right so that when the defect $PIP'$ is at the left end, the operator is $PIPI\cdots$.
In the case of $PIP'=XIY$, the extended doubling operator to the right of $XIY$ should be $\tz$ in order to derive inconsistency.
The necessity of $\tz$ is also confirmed by observing that if $XIY$ sits in the alternating $\tx$ and $\ty$, there is ambiguity on the position of $XIY$.
For example, 9-support operator $XIZIZIZIX$ is expressed as
\eq{
\ba{ccccccccc}{
X&I&Y&&&&&& \\
&&X&I&X&&&& \\
&&&&Y&I&Y&& \\
&&&&&&X&I&X \\ \hline \hline
}
=
\ba{ccccccccc}{
X&I&X&&&&&& \\
&&Y&I&X&&&& \\
&&&&Y&I&Y&& \\
&&&&&&X&I&X \\ \hline \hline
}
=
\ba{ccccccccc}{
X&I&X&&&&&& \\
&&Y&I&Y&&&& \\
&&&&X&I&Y&& \\
&&&&&&X&I&X \\ \hline \hline
}
=
\ba{ccccccccc}{
X&I&X&&&&&& \\
&&Y&I&Y&&&& \\
&&&&X&I&X&& \\
&&&&&&Y&I&X \\ \hline \hline
}.
}
This ambiguity prohibits to move the defect to one end.
However, we only have $\tz$ and $\ty$ in the rank 2 Hamiltonian \eqref{rank2-standard}, which leads to the fact that we cannot derive inconsistency.

\bigskip

We note that we divide all remaining $k$-support operators into the following eight sets
\be{
\item A set $\calG^1_{ab}$ which consists of $(X(IZ)^{m-1}IY)_{4n+2a+b}$ and $(Y(IZ)^{m-1}IX)_{4n+2a+b+2}$ with $n\in \bbN$
\item A set $\calG^2_{ab}$ which consists of $(X(IZ)^{m-1}IX)_{4n+2a+b}$ and $(Y(IZ)^{m-1}IY)_{4n+2a+b+2}$ with $n\in \bbN$
}
with $a\in\{0, 1\}$ and $b\in \{0,1\}$.
Operators in the same set $\calG^n_{ab}$ are linearly connected.
More precisely, we find the following:

\blm{\lb{l:rank2-k+2}
Consider a Hamiltonian \eqref{rank2-standard} with $J^2_X, J^2_Y\neq 0$ and $J^1\neq O$.
In a candidate of a $k$-support conserved quantity $Q$, a $k$-support operator $\bsA\in \calP^k$ may have a nonzero coefficient only if $\bsA$ is one of \eref{rank2-k-form}.

In addition, their coefficients are linearly connected as
\balign{
q_{(X(IZ)^{m-1}IY)_{4n+2a+b}}&=-q_{(Y(IZ)^{m-1}IX)_{4n+2a+b+2}}, \\
q_{(X(IZ)^{m-1}IX)_{4n+2a+b}}&=\frac{J_X^2}{J_Y^2}q_{(Y(IZ)^{m-1}IY)_{4n+2a+b+2}}.
}
}

Under the present analysis, coefficients of operators in $\calG^1$ and those in $\calG^2$ are not linearly connected, while those in $\calG^n_{0b}$ and $\calG^n_{1b}$ with the same $n$ and $b$ are linearly connected if $L\equiv 2\mod 4$, and those in $\calG^n_{ab}$ with different $a$ and $b$ with the same $n$ are linearly connected if $L$ is odd.

\subsection{Restricting possible forms of $k-1$-support operators}\lb{s:rank2-k-1}

We next consider commutators generating $k+1$-support operators.
A similar argument to \sref{NNN-step2} suggests that a candidate of $k-1$-support operators with nonzero coefficient takes the form of 
\eqa{
\ba{cccccccccccc}{
W^1&I&Z&\cdots&I&W^2&&&&&& \\
&&&&&P^1&P^2&&&&& \\
&&&&&&W^3&I&Z&\cdots&I&W^4 \\ \hline
},
}{rank2-k-1-form}
where $W^1,\ldots , W^4 \in \{ X,Y\}$ and $P^1,P^2\in \{X,Y,Z\}$ such that $J_{P^1P^2}^1\neq 0$.
For any $P^1$ and $P^2$, we can choose proper $W^2$ and $W^3$ such that $P^1\neq W^2$ and $P^2\neq W^3$, which guarantees that operators in the above form indeed exist.

\blm{
Consider a Hamiltonian \eqref{rank2-standard} with $J^2_X, J^2_Y\neq 0$ and $J^1\neq O$.
In a candidate of a $k$-support conserved quantity $Q$, a $k-1$-support operator $\bsA\in \calP^{k-1}$ may have a nonzero coefficient only if $\bsA$ is expressed in the form of \eref{rank2-k-1-form}.
}

\subsection{Stronger linear relations on $k$-support and $k-1$-support operators in the case with $J^1_{ZP}\neq 0$ with some $P\in\{X,Y,Z\}$}

In the case with $J^1_{ZP}\neq 0$ ($P\in \{X,Y,Z\}$), we notice that non-extended-doubling-product operators (i.e., $X(IZ)^{m-1}IY$ and $Y(IZ)^{m-1}IX$ for odd $m$ and $X(IZ)^{m-1}IX$ and $Y(IZ)^{m-1}IY$ for even $m$) have zero coefficients.
Below, we demonstrate this fact by employing examples.
Its extension to general cases is redundant but straightforward.

We consider the case with $J^1_{ZX}\neq 0$ and show that the coefficient of non-extended-doubling-product operator $XIZIX$ is zero.
We first observe that $XIZIX$ forms a pair with $YIYX$ as\fn{
Here we need not consider the contribution of
\eq{
\ba{cccccc}{
X&I&Z&I&Z& \\
&&&&X&X \\ \hline
X&I&Z&I&Y&X
},
}
since the $k$-support operator $XIZIZ$ has already been shown to have zero coefficient in \lref{rank2-k+2}.
}
\eq{
\ba{cccccc}{
X&I&Z&I&X& \\
&&&&Z&X \\ \hline
X&I&Z&I&Y&X
}
\hspace{15pt}
\ba{cccccc}{
&&Y&I&Y&X \\
X&I&X&&& \\ \hline
X&I&Z&I&Y&X
}.
}
However, since $k+1$-support operator $YIYZIY$ is generated only by $YIYX$ as
\eq{
\ba{cccccc}{
Y&I&Y&X&& \\
&&&Y&I&Y \\ \hline
Y&I&Y&Z&I&Y
},
}
we conclude that $q_{YIYX}=q_{XIZIX}=0$.
In general, by adding $ZP$ to the right and moving the operator to the right, we can obtain a contradiction.

\bigskip

In addition, we demonstrate that all the remaining coefficients (i.e., $\calG^2_{ab}$ for odd $m$ and $\calG^1_{ab}$ for even $m$) are linearly connected regardless of $a$ and $b$.
To see this fact, we again take an example of $m=2$.
Its extension to general cases is straightforward.

We consider $(XIZIY)_i$.
This operator forms a sequence of pairs as
\eq{
(XIZIY)_i\lr (YIXX)_{i+2}\lr (ZZIY)_{i+4} \lr (YIZIX)_{i+5} \lr (XIZIY)_{i+3},
}
where we used $J_{ZX}\neq 0$.
Since 3 is a generator of the additive group with modulo 4, we conclude that all coefficients are linearly connected as $\calG^1_{00}\lr \calG^1_{11}\lr \calG^1_{10}\lr\calG^1_{01}\lr \calG^1_{00}$ regardless of their position.
Similar arguments hold for general $k$ (general $m$).
Hence, in the following we drop a subscript of operators representing its position.

\blm{\lb{l:rank2-Z-k+1}
Consider a Hamiltonian \eqref{rank2-standard} with $J^2_X, J^2_Y\neq 0$, $J^1\neq O$, and $J^1_{ZP}\neq 0$ with some $P$.
In a candidate of a $k$-support conserved quantity $Q$, a $k$-support operator $\bsA\in \calP^{k}$ may have a nonzero coefficient only if $\bsA\in \calG^2$ for odd $m$ and $\bsA\in \calG^1$ for even $m$.

In addition, coefficients are linearly connected as
\eq{
q_{X(IZ)^{m-1}IX}=\frac{J_X^2}{J_Y^2}q_{Y(IZ)^{m-1}IY}
}
for odd $m$, and
\eq{
q_{X(IZ)^{m-1}IY}=-q_{Y(IZ)^{m-1}IX}
}
for even $m$.
}

Moreover, under the condition that only a doubling-product operator may have a nonzero coefficient in the $k$-support operator as shown above, $P^1P^2$ in \eref{rank2-k-1-form} (the expression of $k-1$-support operator) is uniquely determined if two ends, $W^1$ and $W^4$, are known.
This implies that the coefficient of a $k-1$-support operator $\bsA\in \calP^{k-1}$ is given by \eref{NNN-doubling-coefficient-2}.

\blm{\lb{l:rank2-Z-k-1}
Consider a Hamiltonian \eqref{rank2-standard} with $J^2_X, J^2_Y\neq 0$, $J^1\neq O$, and $J^1_{ZP}\neq 0$ with some $P$.
In a candidate of a $k$-support conserved quantity $Q$, the coefficient of a $k-1$-support operator $\bsA\in \calP^{k-1}$ is given by \eref{NNN-doubling-coefficient-2}. 
}


\subsection{Demonstrating that the remaining $k$-support operators have zero coefficients}

We here divide possible Hamiltonians into three cases; (1) $J^1_{WZ}\neq 0$ with some $W\in \{X,Y\}$, (2) $J^1_{WZ}= 0$ for all $W\in \{X,Y\}$ and $J^1_{WW'}\neq 0$ with some $W,W'\in \{X,Y\}$, and (3) only $J^1_{ZZ}\neq 0$ and all other matrix elements of $J^1$ is zero.
In the following, we treat these three cases separately and show the absence of local conserved quantities in all cases.

\subsubsection{Case with $J^1_{WZ}\neq 0$ with $W\in \{X,Y\}$}

We first consider the case that $J^1_{WZ}\neq 0$ with some $W\in \{X,Y\}$.
Without loss of generality, we suppose $J^1_{XZ}\neq 0$.
In this case, \lref{rank2-Z-k+1} tells that the only remaining $k$-support operators which may have nonzero coefficients are doubling-product operators, i.e., $X(IZ)^mIY$ and $Y(IZ)^mIX$ with odd $m$ and $X(IZ)^mIX$ and $Y(IZ)^mIY$ with even $m$.
We shall show $X(IZ)^mIY$ with odd $m$ has a zero coefficient by examining commutators generating $k$-support operators.
A similar analysis holds for $Y(IZ)^mIY$ with even $m$.

We first consider commutators generating $k$-support operator $X(IZ)^{m-1}IYZY$ as
\balign{
\ba{cccccccc}{
X&I&Z&(IZ)^{m-2}&I&Z&I&Y \\
&&&&&X&Z& \\ \hline
X&I&Z&(IZ)^{m-2}&I&Y&Z&Y
}&
\hspace{15pt}
\ba{cccccccc}{
&&Y&(IZ)^{m-2}&I&Y&Z&Y \\
X&I&X&&&&& \\ \hline
X&I&Z&(IZ)^{m-2}&I&Y&Z&Y
}
\nt \\
\ba{cccccccc}{
X&I&Z&(IZ)^{m-2}&I&Y&X& \\
&&&&&&Y&Y \\ \hline
X&I&Z&(IZ)^{m-2}&I&Y&Z&Y
}&
\hspace{15pt}
\ba{cccccccc}{
X&I&Z&(IZ)^{m-2}&I&Y&Y& \\
&&&&&&X&Y \\ \hline
X&I&Z&(IZ)^{m-2}&I&Y&Z&Y
},
}
which leads to
\eqa{
J^1_{XZ}q_{X(IZ)^{m-1}IYZY}-J^2_Xq_{Y(IZ)^{m-2}IYZY}+J^1_{YY}q_{X(IZ)^{m-1}IYX}-J^1_{XY}q_{X(IZ)^{m-1}IYY}=0.
}{rank2-WZ-mid1}
We note that the coefficients of $X(IZ)^{m-1}IYX$ and $X(IZ)^{m-1}IYY$ have already been computed in \lref{rank2-Z-k-1}.

We next consider commutators generating $k$-support operators $Y(IZ)^{m-2n}IYZ(ZI)^{2n-1}X$ and $X(IZ)^{m-2n-1}IYZ(ZI)^{2n}Y$ as
\balign{
\ba{ccccccccccc}{
Y&I&Z&(IZ)^{m-2n-1}&I&Z&I&(ZI)^{2n-2}&Z&I&X \\
&&&&&X&Z&&&& \\ \hline
Y&I&Z&(IZ)^{m-2n-1}&I&Y&Z&(ZI)^{2n-2}&Z&I&X
}
&\hspace{15pt}
\ba{ccccccccccc}{
Y&I&Z&(IZ)^{m-2n-1}&I&Y&Z&(ZI)^{2n-2}&Y&& \\
&&&&&&&&X&I&X \\ \hline
Y&I&Z&(IZ)^{m-2n-1}&I&Y&Z&(ZI)^{2n-2}&Z&I&X
} \nt \\
\ba{ccccccccccc}{
&&X&(IZ)^{m-2n-1}&I&Y&Z&(ZI)^{2n-2}&Z&I&X \\
Y&I&Y&&&&&&&& \\ \hline
Y&I&Z&(IZ)^{m-2n-1}&I&Y&Z&(ZI)^{2n-2}&Z&I&X
},&
}
and
\balign{
\ba{ccccccccccc}{
X&I&Z&(IZ)^{m-2n-2}&I&Z&I&(ZI)^{2n-1}&Z&I&Y \\
&&&&&X&Z&&&& \\ \hline
X&I&Z&(IZ)^{m-2n-2}&I&Y&Z&(ZI)^{2n-1}&Z&I&Y
}
&\hspace{15pt}
\ba{ccccccccccc}{
X&I&Z&(IZ)^{m-2n-2}&I&Y&Z&(ZI)^{2n-1}&X&& \\
&&&&&&&&Y&I&Y \\ \hline
X&I&Z&(IZ)^{m-2n-2}&I&Y&Z&(ZI)^{2n-1}&Z&I&Y
} \nt \\
\ba{ccccccccccc}{
&&Y&(IZ)^{m-2n-2}&I&Y&Z&(ZI)^{2n-1}&Z&I&Y \\
X&I&X&&&&&&&& \\ \hline
X&I&Z&(IZ)^{m-2n-2}&I&Y&Z&(ZI)^{2n-1}&Z&I&Y
}& \lb{rank2-WZ-midcom1}
}
which lead to
\eqa{
J^1_{XZ}q_{Y(IZ)^mIX}-J^2_Xq_{Y(IZ)^{m-2n}IYZ(ZI)^{2n-2}Y}+J^2_Yq_{X(IZ)^{m-2n-1}IYZ(ZI)^{2n-1}X}=0
}{rank2-WZ-mid2}
and
\eqa{
J^1_{XZ}q_{X(IZ)^mIY}+J^2_Yq_{X(IZ)^{m-2n-1}IYZ(ZI)^{2n-1}X}-J^2_Xq_{Y(IZ)^{m-2n-2}IYZ(ZI)^{2n}Y}=0,
}{rank2-WZ-mid3}
respectively.

We notice that \eref{rank2-WZ-midcom1} with $n=(m-1)/2$ is replaced by
\balign{
\ba{cccccccc}{
X&I&Z&I&(ZI)^{m-2}&Z&I&Y \\
&&X&Z&&&& \\ \hline
X&I&Y&Z&(ZI)^{m-2}&Z&I&Y
}&
\hspace{15pt}
\ba{cccccccc}{
X&I&Y&Z&(ZI)^{m-2}&X&& \\
&&&&&Y&I&Y \\ \hline
X&I&Y&Z&(ZI)^{m-2}&Z&I&Y
} \hspace{15pt} 
\ba{cccccccc}{
&&Z&Z&(ZI)^{m-2}&Z&I&Y \\
X&I&X&&&&& \\ \hline
X&I&Y&Z&(ZI)^{m-2}&Z&I&Y
},
}
which leads to
\eqa{
J^1_{XZ}q_{X(IZ)^mIY}+J^2_Yq_{XIYZ(ZI)^{m-2}X}+J^2_Xq_{ZZ(ZI)^{m-1}Y}=0.
}{rank2-WZ-mid4}

To examine the coefficient of $k-2$-support operator $ZZ(ZI)^{m-1}Y$, we consider commutators generating $k$-support operator $ZZ(ZI)^{m}X$ as
\balign{
\ba{cccccccc}{
Y&I&(ZI)^{m-2}&Z&I&Z&I&X \\
X&Z&&&&&& \\ \hline
Z&Z&(ZI)^{m-2}&Z&I&Z&I&X
}
&\hspace{15pt}
\ba{cccccccc}{
Z&Z&(ZI)^{m-2}&Z&I&Y&& \\
&&&&&X&I&X \\ \hline
Z&Z&(ZI)^{m-2}&Z&I&Z&I&X
} \nt \\
\ba{cccccccc}{
&Y&(ZI)^{m-2}&Z&I&Z&I&X \\
Z&X&&&&&& \\ \hline
Z&Z&(ZI)^{m-2}&Z&I&Z&I&X
}&\hspace{15pt}
\ba{cccccccc}{
&X&(ZI)^{m-2}&Z&I&Z&I&X \\
Z&Y&&&&&& \\ \hline
Z&Z&(ZI)^{m-2}&Z&I&Z&I&X
},
}
which leads to
\eqa{
-J^1_{XZ}q_{Y(IZ)^mIX}-J^2_Xq_{ZZ(Z)^{m-1}Y}-J^1_{XZ}q_{Y(ZI)^mX}+J^1_{YZ}q_{X(ZI)^mX}=0.
}{rank2-WZ-mid5}
We note that the coefficients of $Y(ZI)^mX$ and $X(ZI)^mX$ have already been computed in \lref{rank2-Z-k-1}.

Summing \eref{rank2-WZ-mid1}, Eq.~\eqref{rank2-WZ-mid2} from $n=1$ to $n=(m-1)/2$ with multiplying $-1$, \eref{rank2-WZ-mid3} from $n=1$ to $n=(m-3)/2$, \eref{rank2-WZ-mid4}, and \eref{rank2-WZ-mid5}, and plugging \lref{rank2-Z-k+1} and \lref{rank2-Z-k-1}, we finally arrive at
\eq{
(m+1)J^1_{XZ}q_{X(IZ)^mIY}=0,
}
which clearly implies the absence of $k$-local conserved quantity.
Note that $J^1_{YY}J_{X(IZ)^{m-1}IYX}$ and $J^1_{XY}J_{X(IZ)^{m-1}IYY}$ in \eref{rank2-WZ-mid1} are canceled with $J^1_{XZ}q_{Y(ZI)^mX}$ and $J^1_{YZ}q_{X(ZI)^mX}$ in \eref{rank2-WZ-mid5}, respectively.

\bthm{
Consider a Hamiltonian \eqref{rank2-standard} with $J^2_X, J^2_Y\neq 0$, $J^1_{WZ}\neq0$ with some $W\in \{X,Y\}$.
This Hamiltonian has no $k$-local conserved quantity with $4\leq k\leq L/2$. 
}

\subsubsection{Case with $J^1_{WZ}=0$ with $W\in \{X,Y\}$ and $J^1_{WW'}\neq 0$ with some $W,W'\in \{X,Y\}$}

We here consider the case that $J^1_{WZ}=0$ for all $W\in \{X,Y\}$ and $J^1_{WW'}\neq 0$ with some $W,W'\in \{X,Y\}$.
Without loss of generality, we suppose $J^1_{XW}\neq 0$.
Unlike the case of $J^1_{WZ}\neq 0$ with $W\in \{X,Y\}$, we cannot exclude the possibility that a $k$-support non-extended-doubling operator in the form of $W(IZ)^mIW'$ ($W,W'\in \{X,Y\}$) may have a nonzero coefficient.

We shall show $X(IZ)^mIY$ has a zero coefficient by examining commutators generating $k$-support operators.
A similar analysis holds for $Y(IZ)^mIY$.

We first consider commutators generating $k$-support operator $X(IZ)^{m-1}IYWY$ as
\balign{
\ba{cccccccc}{
X&I&Z&(IZ)^{m-2}&I&Z&I&Y \\
&&&&&X&W& \\ \hline
X&I&Z&(IZ)^{m-2}&I&Y&W&Y
}&
\hspace{15pt}
\ba{cccccccc}{
&&Y&(IZ)^{m-2}&I&Y&W&Y \\
X&I&X&&&&& \\ \hline
X&I&Z&(IZ)^{m-2}&I&Y&W&Y
},
}
which leads to
\eqa{
J^1_{XW}q_{X(IZ)^{m-1}IYWY}-J^2_Xq_{Y(IZ)^{m-2}IYWY}=0.
}{rank2-WW-mid1}
Here, we need not consider the contributions of
\eq{
\ba{cccccccc}{
X&I&Z&(IZ)^{m-2}&I&Y&W^+& \\
&&&&&&W^-&Y \\ \hline
X&I&Z&(IZ)^{m-2}&I&Y&W&Y
}
\hspace{15pt}
\ba{cccccccc}{
X&I&Z&(IZ)^{m-2}&I&Y&W^-& \\
&&&&&&W^+&Y \\ \hline
X&I&Z&(IZ)^{m-2}&I&Y&W&Y
}
}
for the following reason, where $W^{\pm}$ are the two Pauli matrices not equal to $W$ defined as
\balign{
X^+:=Y&\hspace{15pt} X^-:=Z, \\
Y^+:=Z&\hspace{15pt} Y^-:=X.
}
One of $W^+$ or $W^-$ is $Z$, and thus one of $W^+Y$ or $W^-Y$ is $ZY$, which is assumed not to exist.
The other one is $X(IZ)^{m-1}IYZ$.
However, its coefficient is also shown to be zero for the following reason:
If $J^1_{ZZ}\neq 0$, then \lref{rank2-Z-k-1} claims that $q_{X(IZ)^{m-1}IYZ}$ is proportional to $J^1_{XZ}$, which is assumed to be zero.
If $J^1_{ZZ}=0$, then $k+1$-support operator $Y(IZ)^{m}IYZ$ is generated only by $X(IZ)^{m-1}IYZ$ (precisely, $\ty \lplus X(IZ)^{m-1}IYZ$), which again implies $q_{X(IZ)^{m-1}IYZ}=0$.

We next consider commutators generating $k$-support operators $X(IZ)^{m-2n-1}IYW(ZI)^{2n}Y$ and $Y(IZ)^{m-2n}IYW(ZI)^{2n-1}X$ as
\balign{
\ba{ccccccccccc}{
Y&I&Z&(IZ)^{m-2n-1}&I&Z&I&(ZI)^{2n-2}&Z&I&X \\
&&&&&X&W&&&& \\ \hline
Y&I&Z&(IZ)^{m-2n-1}&I&Y&W&(ZI)^{2n-2}&Z&I&X
}
&\hspace{15pt}
\ba{ccccccccccc}{
Y&I&Z&(IZ)^{m-2n-1}&I&Y&W&(ZI)^{2n-2}&Y&& \\
&&&&&&&&X&I&X \\ \hline
Y&I&Z&(IZ)^{m-2n-1}&I&Y&W&(ZI)^{2n-2}&Z&I&X
} \nt \\
\ba{ccccccccccc}{
&&X&(IZ)^{m-2n-1}&I&Y&W&(ZI)^{2n-2}&Z&I&X \\
Y&I&Y&&&&&&&& \\ \hline
Y&I&Z&(IZ)^{m-2n-1}&I&Y&W&(ZI)^{2n-2}&Z&I&X
}& \lb{rank2-WW-midcom1}
}
and
\balign{
\ba{ccccccccccc}{
X&I&Z&(IZ)^{m-2n-2}&I&Z&I&(ZI)^{2n-1}&Z&I&Y \\
&&&&&X&W&&&& \\ \hline
X&I&Z&(IZ)^{m-2n-2}&I&Y&W&(ZI)^{2n-1}&Z&I&Y
}
&\hspace{15pt}
\ba{ccccccccccc}{
X&I&Z&(IZ)^{m-2n-2}&I&Y&W&(ZI)^{2n-1}&X&& \\
&&&&&&&&Y&I&Y \\ \hline
X&I&Z&(IZ)^{m-2n-2}&I&Y&W&(ZI)^{2n-1}&Z&I&Y
} \nt \\
\ba{ccccccccccc}{
&&Y&(IZ)^{m-2n-2}&I&Y&W&(ZI)^{2n-1}&Z&I&Y \\
X&I&X&&&&&&&& \\ \hline
X&I&Z&(IZ)^{m-2n-2}&I&Y&W&(ZI)^{2n-1}&Z&I&Y
},& \lb{rank2-WW-midcom2}
}
which lead to
\eqa{
J^1_{XW}q_{Y(IZ)^mIX}-J^2_Xq_{Y(IZ)^{m-2n}IYW(ZI)^{2n-2}Y}+J^2_Yq_{X(IZ)^{m-2n-1}IYW(ZI)^{2n-1}X}=0
}{rank2-WW-mid2}
and
\eqa{
J^1_{XW}q_{X(IZ)^mIY}+J^2_Yq_{X(IZ)^{m-2n-1}IYW(ZI)^{2n-1}X}-J^2_Xq_{Y(IZ)^{m-2n-2}IYW(ZI)^{2n}Y}=0,
}{rank2-WW-mid3}
respectively.

Our next step depends on the parity of $m$.
If $m$ is even, \eref{rank2-WW-midcom1} with $n=m/2$ is replaced by
\eq{
\ba{cccccccc}{
Y&I&Z&I&(ZI)^{m-2}&Z&I&X \\
&&X&W&&&& \\ \hline
Y&I&Y&W&(ZI)^{m-2}&Z&I&X
}
\hspace{15pt}
\ba{cccccccc}{
Y&I&Y&W&(ZI)^{m-2}&Y&& \\
&&&&&X&I&X \\ \hline
Y&I&Y&W&(ZI)^{m-2}&Z&I&X
},
}
which leads to
\eqa{
J^1_{XW}q_{Y(IZ)^mIX}-J^2_Xq_{YIYW(ZI)^{m-2}Y}=0.
}{rank2-WW-mid4e}
Summing \eref{rank2-WW-mid1}, Eq.~\eqref{rank2-WW-mid2} from $n=1$ to $n=m/2-1$ with multiplying $-1$, \eref{rank2-WW-mid3} from $n=1$ to $n=m/2-1$, and \eref{rank2-WW-mid4e} with multiplying $-1$, and plugging \lref{rank2-Z-k+1} and \lref{rank2-Z-k-1}, we finally arrive at
\eq{
mJ^1_{XW}q_{Y(IZ)^mIX}=0,
}
which clearly implies the absence of $k$-local conserved quantity.

If $m$ is odd, \eref{rank2-WW-midcom2} with $n=(m-1)/2$ is replaced by
\balign{
\ba{cccccccc}{
X&I&Z&I&(ZI)^{m-2}&Z&I&Y \\
&&X&W&&&& \\ \hline
X&I&Y&W&(ZI)^{m-2}&Z&I&Y
}&
\hspace{15pt}
\ba{cccccccc}{
X&I&Y&W&(ZI)^{m-2}&X&& \\
&&&&&Y&I&Y \\ \hline
X&I&Y&W&(ZI)^{m-2}&Z&I&Y
},
}
which leads to
\eqa{
J^1_{XW}q_{X(IZ)^mIY}+J^2_Yq_{XIYW(ZI)^{m-2}X}=0.
}{rank2-WW-mid4o}
Here we need not consider the contribution of
\eq{
\ba{cccccccc}{
&&Z&W&(ZI)^{m-2}&Z&I&Y \\
X&I&X&&&&& \\ \hline
X&I&Y&W&(ZI)^{m-2}&Z&I&Y
},
}
since the coefficient of $ZW(ZI)^{m-1}Y$, as computed in \lref{rank2-Z-k-1}, contains $J^1_{YZ}$ and thus is zero.
Summing \eref{rank2-WW-mid1}, Eq.~\eqref{rank2-WW-mid2} from $n=1$ to $n=(m-1)/2$ with multiplying $-1$, \eref{rank2-WW-mid3} from $n=1$ to $n=(m-3)/2$, and \eref{rank2-WW-mid4o}, and plugging \lref{rank2-Z-k+1} and \lref{rank2-Z-k-1}, we finally arrive at
\eq{
mJ^1_{XW}q_{Y(IZ)^mIX}=0,
}
which clearly implies the absence of $k$-local conserved quantity.

\bthm{
Consider a Hamiltonian \eqref{rank2-standard} with $J^2_X, J^2_Y\neq 0$, $J^1_{WZ}=0$ for all $W\in \{X,Y\}$ and $J^1_{WW'}\neq 0$ with some $W,W'\in \{X,Y\}$.
This Hamiltonian has no $k$-local conserved quantity with $4\leq k\leq L/2$. 
}

\subsubsection{Case with $J^1_{ZZ}\neq0$ and $J^1_{PP'}=0$ for all $(P,P')\neq (Z,Z)$}\lb{s:rank2-ZZ}

We finally consider the case that $J^1_{ZZ}\neq0$ and all other matrix elements of $J^1$ is zero (i.e., $J^1_{PP'}=0$ for all $(P,P')\neq (Z,Z)$).
In this case, we can again apply \lref{rank2-Z-k+1} and obtain that the only remaining $k$-support operators which may have nonzero coefficients are doubling-product operators, i.e., $X(IZ)^mIY$ and $Y(IZ)^mIX$ with odd $m$ and $X(IZ)^mIX$ and $Y(IZ)^mIY$ with even $m$.
We shall show $X(IZ)^mIY$ with odd $m$ has a zero coefficient by examining commutators generating $k$-support operators.
A similar analysis holds for $Y(IZ)^mIY$ with even $m$.

Thanks to this uniqueness, $k-1$-support operator $Y(IZ)^{m-1}IXZ$ is connected to $k$-support operator $X(IZ)^mIY$ as
\eq{
\ba{ccccccc}{
X&I&Z&(IZ)^{m-1}&I&Y& \\
&&&&&Z&Z \\ \hline
X&I&Z&(IZ)^{m-1}&I&X&Z
}
\hspace{15pt}
\ba{ccccccc}{
&&Y&(IZ)^{m-1}&I&X&Z \\
X&I&X&&&& \\ \hline
X&I&Z&(IZ)^{m-1}&I&X&Z
},
}
which leads to
\eq{
J^1_{ZZ}q_{X(IZ)^mIY}-J^2_Xq_{Y(IZ)^{m-1}IXZ}=0.
}
Hence, to prove the absence of $k$-local conserved quantity it suffices to demonstrate $q_{Y(IZ)^{m-1}IXZ}=0$.
Note that following arguments in \sref{rank2-k-1} and \lref{rank2-Z-k-1}, the coefficients of $k-1$-support operators $Y(IZ)^{m-2n-1}IXX(IZ)^{2n-1}IX$ and $X(IZ)^{m-2n-2}IXX(IZ)^{2n}IY$ are also shown to be connected to that of $Y(IZ)^{m-1}IXZ$.

To this end, we first consider commutators generating $k-1$-support operator $Y(IZ)^{m-1}ZYZ$ as
\fn{
Here we need not consider the contribution of
\eq{
\ba{cccccccc}{
Y&I&Z&(IZ)^{m-3}I&Z&Z&Y&Z  \\
&&&&Y&I&Y& \\ \hline
Y&I&Z&(IZ)^{m-3}I&Z&Z&Y&Z 
},
}
since this $k-1$-support operator $Y(IZ)^{m-1}ZYZ$ does not satisfy the form of \eref{rank2-k-1-form} and thus has a zero coefficient.
We also need not consider the shrinking case that a commutator with $k$-support operator and 3-support or 2-support operator generates a $k-1$-support operator, since no $k$-support operator with nonzero coefficient generate $k-1$-support operator $Y(IZ)^{m-1}ZYZ$ by this shrinking commutation relation.
A similar observation is used in the following examinations to restrict possible commutators.
}
\balign{
\ba{ccccccc}{
Y&I&Z&(IZ)^{m-2}&I&X&Z \\
&&&&Z&Z& \\ \hline
Y&I&Z&(IZ)^{m-2}&Z&Y&Z 
}
\hspace{15pt}
\ba{ccccccc}{
Y&I&Z&(IZ)^{m-2}&Z&X& \\
&&&&&Z&Z \\ \hline
Y&I&Z&(IZ)^{m-2}&Z&Y&Z 
}
\hspace{15pt}
\ba{ccccccc}{
&&X&(IZ)^{m-2}&Z&Y&Z \\
Y&I&Y&&&& \\ \hline
Y&I&Z&(IZ)^{m-2}&Z&Y&Z 
},
}
which leads to
\eqa{
-J^1_{ZZ}q_{Y(IZ)^{m-1}IXZ}-J^1_{ZZ}q_{Y(IZ)^{m-1}ZX}+J^2_Yq_{X(IZ)^{m-2}ZYZ}=0.
}{rank2-ZZ-mid1}

The middle $k-2$-support operator $Y(IZ)^{m-1}ZX$ in \eref{rank2-ZZ-mid1} is evaluated by considering commutators generating $k$-support operator $X(IZ)^{m}ZX$ as
\eq{
\ba{ccccccc}{
&&Y&(IZ)^{m-2}I&Z&Z&X \\
X&I&X&&&& \\ \hline
X&I&Z&(IZ)^{m-2}I&Z&Z&X
}
\hspace{15pt}
\ba{ccccccc}{
X&I&Z&(IZ)^{m-2}I&Z&I&Y \\
&&&&&Z&Z \\ \hline
X&I&Z&(IZ)^{m-2}I&Z&Z&X
}
\hspace{15pt}
\ba{ccccccc}{
X&I&Z&(IZ)^{m-2}I&Y&Z& \\
&&&&X&I&X \\ \hline
X&I&Z&(IZ)^{m-2}I&Z&Z&X
},
}
which leads to
\eqa{
-J^2_Xq_{Y(IZ)^{m-1}ZX}+J^1_{ZZ}q_{X(IZ)^mIY}-J^2_Xq_{X(IZ)^{m-1}IYZ}=0.
}{rank2-ZZ-mid2}
Notice that both $q_{X(IZ)^mIY}$ and $q_{X(IZ)^{m-1}IYZ}$ have already been computed in \lref{rank2-Z-k+1} and \lref{rank2-Z-k-1}, respectively.

To evaluate the last $k-3$-support operator $X(IZ)^{m-2}ZYZ$ in \eref{rank2-ZZ-mid1}, we consider the following sequence of commutators\fn{
In the following arguments, we used the fact that we need not consider the shrinking case that a commutator with $k$-support operator and 3-support or 2-support operator generates a $k-1$-support operator, since no $k$-support operator with nonzero coefficient generate $k-1$-support operator $Y(IZ)^{m-1}ZYZ$ by this shrinking commutation relation.
}:
We first consider commutators generating $k-1$-support operator $X(IZ)^{m-2}ZYXIY$ as
\eq{
\ba{ccccccccc}{
X&I&Z&(IZ)^{m-3}&I&X&X&I&Y \\
&&&&Z&Z&&& \\ \hline
X&I&Z&(IZ)^{m-3}&Z&Y&X&I&Y
}
\hspace{15pt}
\ba{ccccccccc}{
X&I&Z&(IZ)^{m-3}&Z&Y&Z&& \\
&&&&&&Y&I&Y \\ \hline
X&I&Z&(IZ)^{m-3}&Z&Y&X&I&Y
}
\hspace{15pt}
\ba{ccccccccc}{
&&Y&(IZ)^{m-3}&Z&Y&X&I&Y \\
X&I&X&&&&&& \\ \hline
X&I&Z&(IZ)^{m-3}&Z&Y&X&I&Y
},
}
which leads to
\eqa{
-J^1_{ZZ}q_{X(IZ)^{m-2}IXXIY}-J^2_Yq_{X(IZ)^{m-2}ZYZ}-J^2_Xq_{Y(IZ)^{m-3}ZYXIY}=0.
}{rank2-ZZ-mid3}

We next consider commutators generating $k-1$-support operators $Y(IZ)^{m-2n-3}ZYX(IZ)^{2n+1}IX$ and $X(IZ)^{m-2n-4}ZYX(IZ)^{2n+2}IY$ as
\balign{
\ba{cccccccccccc}{
Y&I&Z&(IZ)^{m-2n-4}&I&X&X&(IZ)^{2n}&I&Z&I&X \\
&&&&Z&Z&&&&&& \\ \hline
Y&I&Z&(IZ)^{m-2n-4}&Z&Y&X&(IZ)^{2n}&I&Z&I&X
}
&\hspace{15pt}
\ba{cccccccccccc}{
Y&I&Z&(IZ)^{m-2n-4}&Z&Y&X&(IZ)^{2n}&I&Y&& \\
&&&&&&&&&X&I&X \\ \hline
Y&I&Z&(IZ)^{m-2n-4}&Z&Y&X&(IZ)^{2n}&I&Z&I&X
}
\nt \\
\ba{cccccccccccc}{
&&X&(IZ)^{m-2n-4}&Z&Y&X&(IZ)^{2n}&I&Z&I&X \\
Y&I&Y&&&&&&&&& \\ \hline
Y&I&Z&(IZ)^{m-2n-4}&Z&Y&X&(IZ)^{2n}&I&Z&I&X
}& \lb{rank2-ZZ-midcom2}
}
and
\balign{
\ba{cccccccccccc}{
X&I&Z&(IZ)^{m-2n-5}&I&X&X&(IZ)^{2n+1}&I&Z&I&Y \\
&&&&Z&Z&&&&&& \\ \hline
X&I&Z&(IZ)^{m-2n-5}&Z&Y&X&(IZ)^{2n+1}&I&Z&I&Y
}
&\hspace{15pt}
\ba{cccccccccccc}{
X&I&Z&(IZ)^{m-2n-5}&Z&Y&X&(IZ)^{2n+1}&I&X&& \\
&&&&&&&&&Y&I&Y \\ \hline
X&I&Z&(IZ)^{m-2n-5}&Z&Y&X&(IZ)^{2n+1}&I&Z&I&Y
}
\nt \\
\ba{cccccccccccc}{
&&Y&(IZ)^{m-2n-5}&Z&Y&X&(IZ)^{2n+1}&I&Z&I&Y \\
X&I&X&&&&&&&&& \\ \hline
X&I&Z&(IZ)^{m-2n-5}&Z&Y&X&(IZ)^{2n+1}&I&Z&I&Y
},&
}
which leads to
\eqa{
-J^1_{ZZ}q_{Y(IZ)^{m-2n-3}IXX(IZ)^{2n+1}IX}-J^2_Xq_{Y(IZ)^{m-2n-3}IXX(IZ)^{2n}IY}+J^2_Yq_{X(IZ)^{m-2n-4}IXX(IZ)^{2n+1}IX}=0
}{rank2-ZZ-mid4}
and
\eqa{
-J^1_{ZZ}q_{X(IZ)^{m-2n-4}IXX(IZ)^{2n+2}IY}+J^2_Yq_{X(IZ)^{m-2n-4}IXX(IZ)^{2n+1}IX}-J^2_Xq_{Y(IZ)^{m-2n-5}IXX(IZ)^{2n+2}IY}=0,
}{rank2-ZZ-mid5}
respectively.

For $n=(m-3)/2$, \eref{rank2-ZZ-midcom2} is replaced by
\balign{
\ba{ccccccccc}{
Y&I&X&X&(IZ)^{m-1}&I&Z&I&X \\
&Z&Z&&&&&& \\ \hline
Y&Z&Y&X&(IZ)^{m-1}&I&Z&I&X
}
\hspace{15pt}
\ba{ccccccccc}{
X&I&Y&X&(IZ)^{m-1}&I&Z&I&X \\
Z&Z&&&&&&& \\ \hline
Y&Z&Y&X&(IZ)^{m-1}&I&Z&I&X
}
\hspace{15pt}
\ba{ccccccccc}{
Y&Z&Y&X&(IZ)^{m-1}&I&Y&& \\
&&&&&&X&I&X \\ \hline
Y&Z&Y&X&(IZ)^{m-1}&I&Z&I&X
},
}
which leads to
\eqa{
-J^1_{ZZ}q_{YIXX(IZ)^mIX}-J^1_{ZZ}q_{XIYX(IZ)^mIX}-J^2_Xq_{YZYX(IZ)^{m-1}IY}=0.
}{rank2-ZZ-mid6}

Summing \eref{rank2-ZZ-mid1}, \eref{rank2-ZZ-mid2} with multiplying $-J^1_{ZZ}/J^2_{X}$, \eref{rank2-ZZ-mid3}, \eref{rank2-ZZ-mid4} from $n=0$ to $n=(m-5)/2$ with multiplying $-1$, \eref{rank2-ZZ-mid5} from $n=0$ to $n=(m-5)/2$, and \eref{rank2-ZZ-mid6} with ultiplying $-1$,  and plugging \lref{rank2-Z-k+1} and \lref{rank2-Z-k-1}, we finally arrive at
\eq{
-(m-1)J^1_{ZZ}q_{Y(IZ)^{m-1}IXZ}=0,
}
which directly implies the absence of $k$-local conserved quantity.
Here, we notice that the term $J^2_Xq_{X(IZ)^{m-1}IYZ}$ in \eref{rank2-ZZ-mid2} is canceled with the term $J^1_{ZZ}q_{XIYX(IZ)^mIX}$ in \eref{rank2-ZZ-mid6}.

\bigskip

In summary, we establish that all the rank 2 Hamiltonian are non-integrable as long as $J^1\neq O$.

\bthm{
Consider a Hamiltonian \eqref{rank2-standard} with $J^2_X, J^2_Y\neq 0$ and $J^1\neq O$.
This Hamiltonian has no $k$-local conserved quantity with $4\leq k\leq L/2$. 
}

\section{Rank 1: general preliminary analysis}\lb{s:rank1-gen}

We shall treat the case of rank 1, where only $J^2_Z$ is nonzero and $J^2_X=J^2_Y=0$.
In this case, we can further diagonalize the $X$ and $Y$ submatrix of $J^1$ (i.e., $\mx{J^1_{XX}& J^1_{XY}\\ J^1_{XY} & J^1_{YY}}$), with which the Hamiltonian is expressed as
\balign{
H=&\sum_i \mx{X_{i+2}&Y_{i+2}&Z_{i+2}}\mx{0 && \\ &0& \\ &&J_Z^2}\mx{X_i\\ Y_i\\ Z_i}+\sum_i \mx{X_{i+1}&Y_{i+1}&Z_{i+1}}\mx{J_{XX}^1 &0&J_{XZ}^1 \\ 0&J_{YY}^1&J_{YZ}^1 \\ J_{XZ}^1&J_{YZ}^1&J_{ZZ}^1}\mx{X_i\\ Y_i\\ Z_i} \nt \\
&+\sum_i \mx{h_X&h_Y&h_Z}\mx{X_i\\ Y_i\\ Z_i} \lb{rank1-standard}
}
with nonzero $J_Z^2$ and $J^1\neq O$.

We divide the rank 1 case by whether one of $J^1_{XX}$ or $J^1_{YY}$ is nonzero or both of them are zero.
In the latter case, by applying a proper orthogonal matrix in the $XY$ space we can set $J^1_{YZ}=0$.
Then we further divide the cases by whether $J^1_{XZ}$ is nonzero or zero.
In summary, by using the symmetry of $X$ and $Y$, all cases are reduced to one of the following three cases:
\be{
\renewcommand{\labelenumi}{\Alph{enumi}.}
\item $J^1_{XX}\neq 0$ ($J^1_{YY}$, $J^1_{XZ}$, $J^1_{YZ}$, $J^1_{ZZ}$ can be both zero or nonzero).
\item $J^1_{XX}=J^1_{YY}=0$.
\be{
\renewcommand{\labelenumii}{\arabic{enumii}.}
\item $J^1_{XZ}\neq 0$ and $J^1_{YZ}=0$ ($J^1_{ZZ}$ can be both zero or nonzero).
\item $J^1_{XZ}= J^1_{YZ}=0$ and $J^1_{ZZ}\neq 0$.
}
}
Here we omit two trivial cases: a classical case (only $J^1_{ZZ}$ and $h_Z$ may take nonzero values and all other coefficients are zero) and no nearest-neighbor interaction case ($J^1=O$).
The latter case is reduced to the classification theorem with nearest-neighbor interaction, which has already been studied in Refs.~\cite{YCS24-1, YCS24-2}.

We treat these three cases one by one with different proof ideas.
In the following, we first present a general statement which holds for all of these three cases and then treat them separately.
Our conclusion is that all models in case A and case B2 are non-integrable, while some models in case B1 are integrable, and the remaining models in case B1 are non-integrable.

\subsection{Some restrictions on $k$-support and $k-1$-support operators}

We first notice that a $k$-support operator where both ends are $X$ or $Y$, has zero coefficient.
This fact is confirmed by observing the fact that 
\eq{
\ba{ccccc}{
W&\cdots &W'&& \\
&&Z&I&Z \\ \hline
W&\cdots&{W'}^{\rm c}&I&Z
}
}
is the only commutator generating a $k+2$-support operator $W\cdots{W'}^{\rm c}IZ$.
Here, $W,W'\in \{X,Y\}$, and ${W'}^{\rm c}$ is an observable in $\{X,Y\}$ which is not $W'$.

In a similar manner, a $k$-support operator where one end is $X$ or $Y$, and the other end is $ZP$ ($P\in \{X,Y,Z\}$ also has zero coefficient.
This fact is confirmed by observing the fact that
\eq{
\ba{cccccc}{
Z&P&\cdots &W&& \\
&&&Z&I&Z \\ \hline
Z&P&\cdots&{\Wc}&I&Z
}
}
is the only commutator generating a $k+2$-support operator $ZP\cdots{\Wc}IZ$.

\blm{\lb{l:rank1-k+2}
Consider a Hamiltonian \eqref{rank1-standard} with nonzero $J_Z^2$.
In a candidate of a $k$-support conserved quantity $Q$, a $k$-support operator which takes the form of $W\cdots W'$ ($W,W'\in \{X,Y\}$), $ZP\cdots W$ ($P\in \{X,Y,Z\}$) or $W\cdots PZ$, has zero coefficient.
}

The remaining operators which may have nonzero coefficients take the form of $ZI\cdots W$ (and its reflection) or $Z\cdots Z$.

Using a similar assertion, we find that some $k-1$-support operators also have zero coefficients.
Consider a $k-1$-support operator in the form of $W\cdots IW'$ with $W,W'\in \{X,Y\}$.
Then, $k+1$-support operator $ZI\Wc \cdots IW'$ is generated only by the following commutator:
\eq{
\ba{cccccc}{
&&W&\cdots&I&W' \\
Z&I&Z&&& \\ \hline
Z&I&\Wc&\cdots &I&W'
},
}
which directly implies $q_{W\cdots IW'}=0$.

\blm{\lb{l:rank1-k-1-WIW}
Consider a Hamiltonian \eqref{rank1-standard} with nonzero $J_Z^2$.
In a candidate of a $k$-support conserved quantity $Q$, a $k-1$-support operator in the form of $W\cdots IW'$ or $WI\cdots W'$ ($W,W'\in \{X,Y\}$) has zero coefficient.
}

\section{Rank 1: Case with $J^1_{XX}\neq 0$ (Case A)}\lb{s:rank1-A}

To treat case A, we divide $k$-support operators in $Q$ into those in the form of $Z\cdots W$ (and its reflection) and those in the form of $Z\cdots Z$.
We treat $Z\cdots W$ (and its reflection) in \sref{rank1-A-ZW} and $Z\cdots Z$ in \sref{rank1-A-ZZ}.

\subsection{Analysis of $Z\cdots W$}\lb{s:rank1-A-ZW}
\subsubsection{Restricting possible forms of some $k$-support operators}

First, we further investigate a $k$-support operator $ZI\cdots W$ by analyzing commutators generating $k+1$-support operators.
We consider commutators generating $k+1$-support operator $XYI\cdots W$ as\fn{
Here we need not take into account the following commutator
\eqa{
\ba{cccccc}{
X&Y&I&\cdots &*''& \\
&&&&Z&W \\ \hline
X&Y&I&\cdots &*&W
},
}{rank1-A-ZW-k+1-mid1}
since $*''$ should take $X$ or $Y$, and a $k$-support operator in this form has already been shown to have zero coefficient in \lref{rank1-k+2}.
}
\eq{
\ba{cccccc}{
&Z&I&\cdots &*&W \\
X&X&&&& \\ \hline
X&Y&I&\cdots &*&W
}
\hspace{15pt}
\ba{cccccc}{
X&Y&I&\cdots &*'& \\
&&&&W&W \\ \hline
X&Y&I&\cdots &*&W
}.
}

In order to make the $k$-support operator $XYI\cdots *'$ in \eref{rank1-A-ZW-k+1-mid1} with a nonzero coefficient, $*'$ should be $Z$, and \lref{rank1-k+2} suggests that the operator should take the form of $XYI\cdots IZ$.
Thus, we find that $ZI\cdots W$ may have a nonzero coefficient only if it takes the form of $ZI\cdots W^{\rm c}W$.
In this case, $ZI\cdots W^{\rm c}W$ forms a pair as 
\eq{
ZI\cdots W^{\rm c}W\lr XYI\cdots IZ.
}
(If $J^1_{YY}=0$, then we have a further constraint that $W$ should be $X$.)

Now we consider commutators generating $k+2$-support operator $ZIYYI\cdots IZ$ as
\eq{
\ba{ccccccccc}{
&&X&Y&I&\cdots&*&I&Z \\
Z&I&Z&&&&&& \\ \hline
Z&I&Y&Y&I&\cdots &*&I&Z
}
\hspace{15pt}
\ba{ccccccccc}{
Z&I&Y&Y&I&\cdots&*'&& \\
&&&&&&Z&I&Z \\ \hline
Z&I&Y&Y&I&\cdots &*&I&Z
}.
}
The second commutator exists only if $*'$ is $X$ or $Y$.
This relation implies that our initial operator $ZI\cdots W$  may have a nonzero coefficient only if it takes the form of $ZI\cdots W'IW^{\rm c}W$.
In this case, we have a sequence of pairs:
\eq{
ZI\cdots W'IW^{\rm c}W \lr XYI\cdots W'IZ \lr ZIYYI\cdots W'^{\rm c}.
}

Considering this procedure repeatedly, we arrive at the fact that a $k$-support operator in the form of $ZI\cdots W$ with nonzero coefficient should be expressed as $\tz \w^1 \tz \w^2 \cdots \w^{m-1}\tz \w^m$, where $W^1,W^2\ldots , W^m\in \{X,Y\}$ if $J^1_{YY}\neq 0$, and $W^1,W^2\ldots , W^m=X$ if $J^1_{YY}=0$.
As a direct consequence, a $k$-support operator in the form of $ZI\cdots W$ may have a nonzero coefficient only if $k\equiv 1\mod 3$.

\blm{\lb{l:rank1-ZW-k+1}
Consider a Hamiltonian \eqref{rank1-standard} with $J_Z^2\neq 0$ and $J^1_{XX}\neq 0$.
In a candidate of a $k$-support conserved quantity $Q$, a $k$-support operator in the form of $\bsA=Z\cdots W$ may have a nonzero coefficient only if $\bsA$ is expressed as 
\eqa{
\bsA=\tz \w^1 \tz \w^2 \cdots \w^{m-1}\tz \w^m,
}{rank1-ZW-DP-1}
with $k=3m+1$ and $W^1,W^2\ldots , W^m\in \{X,Y\}$.
If $J^1_{YY}=0$, then $W^1,W^2\ldots , W^m=X$ holds.
In addition, their coefficients are linearly connected as
\eqa{
q_{\bsA}=c^{1,k,a}_{XX-ZW}\cdot \sigma(Z, W^m) (-1)^{m-1}(J^2_Z)^m \prod_{j=1}^m J^1_{W^jW^j}
}{rank1-ZW-expand-1}
with a common factor $c^{1,k,a}$, where $a\in \{0,1,2\}$ represents the sector of the spatial position of this operator in terms of modulo 3.

Similarly, in a candidate of a $k$-support conserved quantity $Q$, a $k$-support operator in the form of $\bsA=W\cdots Z$ may have a nonzero coefficient only if $\bsA$ is expressed as 
\eqa{
\bsA=\w^1 \tz \w^2 \cdots \w^{m-1}\tz \w^m \tz,
}{rank1-ZW-DP-2}
with $k=3m+1$ and $W^1,W^2\ldots , W^m\in \{X,Y\}$.
If $J^1_{YY}=0$, then $W^1,W^2\ldots , W^m=X$ holds.
In addition, their coefficients are linearly connected as
\eqa{
q_{\bsA}=c^{1,k,a}_{XX-ZW}\cdot \sigma(W^1, Z) (-1)^{m-1}(J^2_Z)^m \prod_{j=1}^m J^1_{W^jW^j}
}{rank1-ZW-expand-2}
with the same common factor $c^{1,k,a}$.
}

We here clarify the meaning of the sector $a\in \{0,1,2\}$.
Consider $(\tz \x \tz \x)_1$ as an example.
This operator is connected as
\eq{
(\tz \x \tz \x)_1\lr (\x\tz\x\tz)_3 \lr (\tz \x \tz \x)_4\lr (\x\tz\x\tz)_6 \lr(\tz \x \tz \x)_7\cdots ,
}
where only $(\tz \x \tz \x)_i$ with $i\equiv 1\mod 3$ and $(\x\tz\x\tz)_i$ with $i\equiv 0\mod 3$ appears.
Thus, $(\tz \w \tz \w)_i$ with $i\equiv 1\mod 3$ and $(\w\tz\w\tz)_i$ with $i\equiv 0\mod 3$ share the same coefficient $c^{1,k,a}_{XX-ZW}$, while others do not share this coefficient.
Similar arguments hold for the other two sectors.
In this subsection, we assign the coefficient $c^{1,k,a}_{XX-ZW}$ to $(\tz \w^1 \tz \w^2 \cdots \w^{m-1}\tz \w^m)_{3n+a}$ and $(\w^1 \tz \w^2 \cdots \w^{m-1}\tz \w^m \tz)_{3n+a+2}$.

If the system size $L$ is not a multiple of 3, then these three coefficients automatically coincide by using the periodic boundary condition.
We shall show in \sref{rank1-ZW-common} that these three coefficients coincide even if $k$ is a multiple of 3, while it requires some additional elaborated arguments.

\subsubsection{Restricting possible forms of some specific types of $k$-support and $k-1$-support operators}

In a similar manner to \lref{rank1-ZW-k+1}, coefficients of some $k$-support operators where both ends are $Z$, $Z\cdots Z$, are also determined.
Take a $k$-support operator expressed as $\bsA=\tz \w^1 \tz \w^2 \cdots \w^{m-1}\tz (XZ)$ as an example.
Considering commutators generating $\x \tz \w^1 \tz \w^2 \cdots \w^{m-1}\tz (XZ)$, we find
\eq{
J^1_{XX}q_{\tz \w^1 \cdots \w^{m-1}\tz (XZ)}+J^1_{XZ}q_{\x \tz \w^1 \cdots \w^{m-1}\tz }=0.
}
Since $q_{\x \tz \w^1 \cdots \w^{m-1}\tz }$ has already been computed in \eref{rank1-ZW-expand-2}, we conclude
\eq{
q_{\tz \w^1 \cdots \w^{m-1}\tz (XZ)}=c^{1,k,a}_{XX-ZW}\cdot \sigma(Z,X) (-1)^{m-1}(J^2_Z)^m J^1_{XZ}\prod_{j=1}^{m-1} J^1_{W^jW^j}.
}
In this line,  $\bsA=\tz \w^1 \tz \w^2 \cdots \w^{m-1}\tz (YZ)$ and its reflection can also be computed.

We can also show that a $k$-support operator in the form of $ZI\cdots IWZ$ but not expressed as $\tz \w^1 \tz \w^2 \cdots \w^{m-1}\tz (\Wc Z)$ has zero coefficient.
This fact can be shown by considering commutators generating $k+1$-support operator $XYI\cdots IWZ$ as\fn{
Here we need not consider
\eq{
\ba{ccccccc}{
X&Y&I&\cdots&I&\Wc& \\
&&&&&Z&Z \\ \hline
X&Y&I&\cdots&I&W&Z
},
}
since $k$-support operator $XYI\cdots I\Wc$ has already been shown to have zero coefficient in \lref{rank1-k+2}.
}
\eq{
\ba{ccccccc}{
&Z&I&\cdots&I&W&Z \\
X&X&&&&& \\ \hline
X&Y&I&\cdots&I&W&Z
}
\hspace{15pt}
\ba{ccccccc}{
X&Y&I&\cdots&I&Z& \\
&&&&&\Wc&Z \\ \hline
X&Y&I&\cdots&I&W&Z
}.
}
By assumption, the latter $k$-support operator $XYI\cdots IZ$ is not in the form of Eqs.~\eqref{rank1-ZW-DP-1} or \eqref{rank1-ZW-DP-2}, implying $q_{XYI\cdots IZ}=0$.
Since $q_{ZI\cdots IWZ}$ is linearly connected to $q_{XYI\cdots IZ}$, we conclude that $q_{ZI\cdots IWZ}$ is also zero.

\blm{\lb{l:rank1-ZW-ZZ-k+1}
Consider a Hamiltonian \eqref{rank1-standard} with $J_Z^2\neq 0$ and $J^1_{XX}\neq 0$.
In a candidate of a $k$-support conserved quantity $Q$, the coefficient of a $k$-support operator in the form of $\bsA=\tz \w^1 \tz \w^2 \cdots \w^{m-1}\tz (W^mZ)$ ($W^i\in \{X,Y\}$) is computed as
\eqa{
q_{\tz \w^1 \tz \w^2 \cdots \w^{m-1}\tz (W^mZ)}=c^{1,k}_{XX-ZW}\cdot \sigma(Z,W^m)(-1)^m (J^2_Z)^m J^1_{W^mZ}\prod_{j=1}^{m-1} J^1_{W^jW^j}.
}{rank1-ZW-ZZ-k-expand}
The coefficient of its reflection is computed similarly (with multiplying $-1$).

In addition, if a $k$-support operator is in the form of $ZI\cdots IWZ$ ($W\in \{X,Y\}$) but not expressed as $\tz \w^1 \tz \w^2 \cdots \w^{m-1}\tz (\Wc Z)$ (or its reflection), then this operator has zero coefficient in $Q$.
}

We further specify some of the coefficients of $k-1$-support operators.
Let 
\eq{
\Psi=\Psi^1\Psi^2\in \calN:= \{ XX, YY, ZZ, XZ, YZ, ZX, ZY, ZZ\}
}
be nearest-neighbor interaction terms in the Hamiltonian \eqref{rank1-standard} in consideration.
Here, for $\Psi=ZY$ as an example we express $\Psi^1=Z$ and $\Psi^2=Y$.
Using this symbol, we consider a $k-1$-support operator given by
\eqa{
\bsA=\w^1\tz\w^2\tz\cdots \w^l \tz \Psi^1\Psi^2\tz  \w^{l+1}\tz\cdots \tz\w^{m-1},
}{rank1-XX-k-1-PsiPsi}
where $\Psi^1, \Psi^2\in \calN$ follows the same rule as the doubling-product operators, i.e., the support of $\Psi^2$ is one-site shift of that of $\Psi^1$, and $\Psi^1_1\neq Z$, $\Psi^1_2\neq \Psi^2_1$, and $\Psi^2_2\neq Z$.
An example is
\eq{
\x\tz(XZ)\y\tz\x=
\ba{ccccccccc}{
X&X&&&&&&& \\
&Z&I&Z&&&&& \\
&&&X&Z&&&& \\
&&&&Y&Y&&& \\
&&&&&Z&I&Z& \\
&&&&&&&X&X \\ \hline \hline
X&Y&I&Y&X&X&I&Y&X
}.
}

Remark that different sequences $\Psi^1\Psi^2$ and ${\Psi'}^1{\Psi'}^2$ may produce the same operator.
This happens in the cases of\fn{
If we remove the condition $\Psi^1_1\neq Z$ and $\Psi^2_2\neq Z$, we further have examples as
\eq{
\ba{ccc}{
X&X& \\
&Z&Z \\ \hline
X&Y&Z
}
\hspace{15pt}
\ba{ccc}{
X&Z& \\
&X&Z \\ \hline
X&Y&Z
}.
}
}
\eq{
\ba{ccc}{
X&X& \\
&Z&X \\ \hline
X&Y&X
}
\hspace{15pt}
\ba{ccc}{
X&Z& \\
&X&X \\ \hline
X&Y&X
}
}
and
\eq{
\ba{ccc}{
Y&Y& \\
&Z&Y \\ \hline
Y&X&Y
}
\hspace{15pt}
\ba{ccc}{
Y&Z& \\
&Y&Y \\ \hline
Y&X&Y
}.
}
We confirm by an exhaustive search that no other 3-support operator has this type of multiple expression.

We claim that $k-1$-support operators in the form of \eref{rank1-XX-k-1-PsiPsi} except for $(\Psi^1, \Psi^2)=(XX,XZ)$ and $(YY, YZ)$ is connected to $k$-support operators in the form of \eref{rank1-ZW-DP-1}, which implies that the coefficients of these $k-1$-support operators are also given by equations similar to \eref{rank1-ZW-expand-1}.
For example, $\x\tz(XZ)\y\tz\x$ with $k=10$ has a sequence of pairs as
\eq{
\x\tz(XZ)\y\tz\x \lr \tz\x\tz(XZ)\y\tz \lr \x\tz\x\tz(XZ)\y \lr  \tz\x\tz\x\tz(XZ), 
}
where the last $k$-support operator has already been examined in \lref{rank1-ZW-ZZ-k+1}.

We examine pairs in this sequence.
The first pair comes from the analysis of commutators generating $k+1$-support operator $\tz\x\tz(XZ)\y\tz\x=ZIYYIYXXIYX$ as
\eq{
\ba{ccccccccccc}{
&&X&Y&I&Y&X&X&I&Y&X \\
Z&I&Z&&&&&&&& \\ \hline
Z&I&Y&Y&I&Y&X&X&I&Y&X
}
\hspace{15pt}
\ba{ccccccccccc}{
Z&I&Y&Y&I&Y&X&X&I&Z& \\
&&&&&&&&&X&X \\ \hline
Z&I&Y&Y&I&Y&X&X&I&Y&X
}.
}
Here we do not consider the contribution of
\eq{
\ba{ccccccccccc}{
Z&I&Y&Y&I&Y&X&X&I&X& \\
&&&&&&&&&Z&X \\ \hline
Z&I&Y&Y&I&Y&X&X&I&Y&X
},
}
since the $k$-support operator is in the form of $Z\cdots W$ while it does not satisfy \eref{rank1-ZW-DP-1}, implying zero coefficient.
Similar arguments hold when removing/adding $\w$ and adding/removing $\tz$, including the second pair in this sequence.

The last pair comes from the analysis of commutators generating $k+1$-support operator $\tz\x\tz\x\tz(XZ)\y=ZIYYIYYIYXY$ as
\eq{
\ba{ccccccccccc}{
&&X&Y&I&Y&Y&I&Y&X&Y \\
Z&I&Z&&&&&&&& \\ \hline
Z&I&X&Y&I&Y&Y&I&Y&X&Y
}
\hspace{15pt}
\ba{ccccccccccc}{
Z&I&X&Y&I&Y&Y&I&Y&Z&\\
&&&&&&&&&Y&Y \\ \hline
Z&I&X&Y&I&Y&Y&I&Y&X&Y
}.
}
Here we do not consider the contribution of
\eq{
\ba{ccccccccccc}{
Z&I&X&Y&I&Y&Y&I&Y&Y&\\
&&&&&&&&&Z&Y \\ \hline
Z&I&X&Y&I&Y&Y&I&Y&X&Y
}
}
since the $k$-support operator is in the form of $Z\cdots W$ while it does not satisfy \eref{rank1-ZW-DP-1}, implying zero coefficient.
Similar arguments hold for general $k-1$-support operators in the form of \eref{rank1-XX-k-1-PsiPsi} except for $(\Psi^1, \Psi^2)=(XX,ZX)$ and $(YY, ZY)$.

We next treat the remaining case $(\Psi^1, \Psi^2)=(XX,ZX)$ and $(YY, ZY)$.
We shall show that an operator in the form of \eref{rank1-XX-k-1-PsiPsi} with $(\Psi^1, \Psi^2)=(XX,ZX)$ or $(YY, ZY)$ has zero coefficient.
We demonstrate this fact by the minimum example; $\tx\tz\x(ZX)=\tx\tz(XZ)\x=XYIYYX$ with $k=7$.
To analyze it, we consider commutators generating $ZIYYIYYX$ as
\eq{
\ba{cccccccc}{
&&X&Y&I&Y&Y&X \\
Z&I&Z&&&&& \\ \hline
Z&I&Y&Y&I&Y&Y&X
}
\hspace{15pt}
\ba{cccccccc}{
Z&I&Y&Y&I&Y&Z& \\
&&&&&&X&X \\ \hline
Z&I&Y&Y&I&Y&Y&X
}
\hspace{15pt}
\ba{cccccccc}{
Z&I&Y&Y&I&Y&X& \\
&&&&&&Z&X \\ \hline
Z&I&Y&Y&I&Y&Y&X
},
}
which leads to
\eqa{
-J^2_Zq_{XYIYYX}+J^1_{XX}q_{ZIYYIYZ}-J^1_{XZ}q_{ZIYYIYX}=0.
}{rank1-ZW-Psi1Psi2-zero-mid}
Both $q_{ZIYYIYZ}$ and $q_{ZIYYIYX}$ have already been computed in \eref{rank1-ZW-expand-1} and \eref{rank1-ZW-ZZ-k-expand} as
\balign{
q_{ZIYYIYZ}&=c_{XX-ZW}^{1,7,a}(J^2_Z)^2J^1_{XX}J^1_{XZ}, \\
q_{ZIYYIYX}&=c_{XX-ZW}^{1,7,a}(J^2_Z)^2(J^1_{XX})^2.
}
By plugging these two expressions into \eref{rank1-ZW-Psi1Psi2-zero-mid}, the second term of \eref{rank1-ZW-Psi1Psi2-zero-mid}, $J^1_{XX}q_{ZIYYIYZ}$, and the third term of  \eref{rank1-ZW-Psi1Psi2-zero-mid}, $-J^1_{XZ}q_{ZIYYIYX}$ cancel, and \eref{rank1-ZW-Psi1Psi2-zero-mid} turns out to be
\eq{
-J^2_Zq_{XYIYYX}=0.
}
Similar arguments hold for other cases with $(\Psi^1, \Psi^2)=(XX,XZ)$ and $(YY, YZ)$.
Thus we have the following theorem.

\blm{\lb{l:rank1-ZW-PsiPsi-k-1}
Consider a Hamiltonian \eqref{rank1-standard} with $J_Z^2\neq 0$ and $J^1_{XX}\neq 0$.
In a candidate of a $k$-support conserved quantity $Q$, the coefficient of a $k-1$-support operator in the form of 
\eq{
\bsA=\w^1\tz\w^2\tz\cdots \w^l \tz \Psi^1\Psi^2\tz  \w^{l+1}\tz\cdots \tz\w^{m-1}
}
(which is the same as \eref{rank1-XX-k-1-PsiPsi}) except for $(\Psi^1, \Psi^2)=(XX,XZ)$ and $(YY, YZ)$ is computed as
\eq{
q_{\bsA}=c^{1,k,a}_{XX-ZW}\cdot \sigma(W^1, Z)\sigma(Z, \Psi^1_1)\sigma(\Psi^1_2, \Psi^2_1)\sigma(\Psi^2_2,Z) \sigma(Z,W^{m-1}) (-1)^{m-2}  (J^2_Z)^{m-1} J^1_{\Psi^1}J^1_{\Psi^2}\prod_{j=1}^{m-1} J^1_{W^jW^j}.
}

In addition, a $k-1$-support operator in the form of \eref{rank1-XX-k-1-PsiPsi} with $(\Psi^1, \Psi^2)=(XX,ZX)$ or $(YY, ZY)$ has zero coefficient.
}

We further show that a $k-1$-support operator in the form of $WI\cdots Z$ but not expressed as $(WIZ)\w^1 \tz \w^2 \cdots \w^{m-1}\tz$ have zero coefficient.
To demonstrate this, we first consider commutators generating $k$-support operator $WI\cdots YX$ as\fn{
Here we used the fact that we need not consider the contribution of
\eq{
\ba{ccccc}{
Z&I&\cdots&Y&X \\
\Wc&&&& \\ \hline
W&I&\cdots&Y&X
}
}
for the following reason.
This $k$-support operator $ZI\cdots YX$ may have nonzero coefficient only when it is expressed as $\tz\w^1 \tz \w^2 \cdots \w^{m-1}\tz\x$.
However, this means that the $k-1$-support operator $WI\cdots Z$ in consideration is expressed as $(WIZ)\w^1 \tz \w^2 \cdots \w^{m-1}\tz$, which contradicts the assumption.
}
\eq{
\ba{ccccc}{
W&I&\cdots&Z& \\
&&&X&X \\ \hline
W&I&\cdots&Y&X
}
\hspace{15pt}
\ba{ccccc}{
W&I&\cdots&X& \\
&&&Z&X \\ \hline
W&I&\cdots&Y&X
}.
}
However, $k+1$-support operator $WI\cdots YIZ$ is generated only by the following commutator
\eq{
\ba{cccccc}{
W&I&\cdots&X&& \\
&&&Z&I&Z \\ \hline
W&I&\cdots&Y&I&Z
},
}
which implies $q_{WI\cdots X}=0$ and hence $q_{WI\cdots Z}$ is also zero.

\blm{\lb{l:rank1-XX-k-1-zero}
Consider a Hamiltonian \eqref{rank1-standard} with $J_Z^2\neq 0$ and $J^1_{XX}\neq 0$.
In a candidate of $k$-support conserved quantity $Q$, a $k-1$-support operator in the form of $WI\cdots Z$ but not expressed as $(WIZ)\w^1 \tz \w^2 \cdots \w^{m-1}\tz$ have zero coefficient.
}

\subsubsection{Commonness of coefficients}\lb{s:rank1-ZW-common}

We next show that the three coefficients, $c^{1,k,0}_{XX-ZW}$, $c^{1,k,1}_{XX-ZW}$, and $c^{1,k,2}_{XX-ZW}$ are the same.
We demonstrate it by dividing Hamiltonians into two cases: (i) one of $J^1_{YY}$ or $J^1_{YZ}$ is nonzero, and (ii) both of $J^1_{YY}$ and $J^1_{YZ}$ are zero.

We first treat the former case (i): one of $J^1_{YY}$ or $J^1_{YZ}$ is nonzero.
In this case, using \lref{rank1-ZW-PsiPsi-k-1}, we can connect operators in different sectors.
For example, in the case with $J^1_{YY}\neq 0$, $(\tz\x\tz\x)_1$ is connected to other sectors as
\eq{
(\tz\x\tz\x)_1\lr (\x\tz\x\y)_3\lr (\tz\x\y\tz)_4 \lr (\x\y\tz\x)_6\lr (\y\tz\x\tz)_7 \lr (\tz\x\tz\x)_8,
}
where the sector with $1\mod 3$ and that with $8\equiv 2\mod 3$ is connected, implying $c^{1,k,1}_{XX-ZW}=c^{1,k,2}_{XX-ZW}$.
Performing this connecting process again, we have $c^{1,k,1}_{XX-ZW}=c^{1,k,2}_{XX-ZW}=c^{1,k,0}_{XX-ZW}$.
Obviously, this connecting process works for general $k$ and the case of $J^1_{YZ}\neq 0$.

We next treat the latter case (ii): both $J^1_{YY}$ and $J^1_{YZ}$ are zero.
In this case, we connect operators in different sectors using a different idea from above.
We take $(\x\tz)_1=(XYIZ)_1$ with $k=4$ as an example.
We first consider commutators generating $(XZXZ)_1$ as\fn{
Here we need not consider the contribution of
\eq{
\ba{cccc}{
X&Z&Y& \\
&&Z&Z \\ \hline
X&Z&X&Z
},
}
since $XYZ$ is shown to have a zero coefficient as follows:
We first notice that $XZY$ satisfies the form \eqref{rank1-XX-k-1-PsiPsi} by setting $\Psi^1=XX$ and $\Psi^2=YY$.
As shown in \lref{rank1-ZW-PsiPsi-k-1}, the coefficient of $q_{XZY}$ is specified as $q_{XZY}=c^{1,4,a}_{XX-ZW}J^1_{XX}J^1_{YY}$, while we set $J^1_{YY}=0$, implying $q_{XZY}=0$.
}
\eqa{
\ba{cccc}{
X_1&Y&I&Z \\
&X&X& \\ \hline
X&Z&X&Z
}
\hspace{15pt}
\ba{cccc}{
&Y_2&X&Z \\
X&X&& \\ \hline
X&Z&X&Z
},
}{rank1-A-common-com1}
where we put subscripts in the first Pauli operators to clarify the position of operators.

To treat the latter operator $(YXZ)_2$, we consider commutators generating $(YYZZ)_2$ as\fn{
Here we need not consider contributions from 
\eq{
\ba{cccc}{
Y&X&I&Z \\
&Z&Z& \\ \hline
Y&Y&Z&Z
}
\hspace{15pt}
\ba{cccc}{
Y&Z&I&Z \\
&X&Z& \\ \hline
Y&Y&Z&Z
}
}
since these 4-support operators have already been shown to have zero coefficients in \lref{rank1-ZW-k+1}.
}
\eqa{
\ba{cccc}{
Y_2&X&Z& \\
&Z&I&Z \\ \hline
Y&Y&Z&Z
}
\hspace{15pt}
\ba{cccc}{
X_2&Y&I&Z \\
Z&I&Z& \\ \hline
Y&Y&Z&Z
}.
}{rank1-A-common-com2}
Here we need not consider contributions from 
\eq{
\ba{cccc}{
Y&Y&Y& \\
&&X&Z \\ \hline
Y&Y&Z&Z
},
}
since $YYY$ is shown to have a zero coefficient, which is shown by the fact that $ZIXYY$ is generated only by
\eq{
\ba{ccccc}{
&&Y&Y&Y \\
Z&I&Z&& \\ \hline
Z&I&X&Y&Y
}.
}

Summarizing Eqs.~\eqref{rank1-A-common-com1} and \eqref{rank1-A-common-com2}, we find a sequence of pairs as
\eq{
(XYIZ)_1\lr (YXZ)_2\lr (XYIZ)_2,
}
which clearly shows the connection between different sectors.
Similar arguments hold for general $k$, which leads to the commonness of coefficients.

\blm{\lb{l:rank1-XX-common-ZW}
Consider a Hamiltonian \eqref{rank1-standard} with $J_Z^2\neq 0$ and $J^1_{XX}\neq 0$.
Three coefficients, $c^{1,k,0}_{XX-ZW}$, $c^{1,k,1}_{XX-ZW}$, and $c^{1,k,2}_{XX-ZW}$, appearing in \lref{rank1-ZW-k+1} are equal, which we simply denote by $c^{1,k}_{XX-ZW}$:
\eq{
c^{1,k}_{XX-ZW}:=c^{1,k,0}_{XX-ZW}=c^{1,k,1}_{XX-ZW}=c^{1,k,2}_{XX-ZW}.
}
}

In the following, we express the results in \lref{rank1-ZW-k+1}, \lref{rank1-ZW-ZZ-k+1}, and \lref{rank1-ZW-PsiPsi-k-1} simply with $c^{1,k}_{XX-ZW}$.

\subsubsection{Demonstration with the case of $k=4$}

Before going to the general analysis, we demonstrate our proof in the simplest case, $k=4$.
Thanks to \lref{rank1-ZW-k+1} (and \lref{rank1-XX-common-ZW}), it suffices to show $q_{ZIYX}=0$.

We first observe that $ZXZX$ is generated by the following five commutators\fn{
Here we do not consider following commutators
\eq{
\ba{cccc}{
Z&Z&I&X \\
&Y&Z& \\ \hline
Z&X&Z&X
}
\hspace{15pt}
\ba{cccc}{
Z&Y&I&X \\
&Z&Z& \\ \hline
Z&X&Z&X
}
\hspace{15pt}
\ba{cccc}{
Z&X&I&Y \\
&&Z&Z \\ \hline
Z&X&Z&X
},
}
since these 4-support operators do not satisfy the condition in \lref{rank1-ZW-k+1} and thus have zero coefficients.
}:
\eqa{
\ba{cccc}{
Z&I&Y&X \\
&X&X& \\ \hline
Z&X&Z&X
}
\hspace{15pt}
\ba{cccc}{
Z&X&I&Z \\
&&Z&Y \\ \hline
Z&X&Z&X
}
\hspace{15pt}
\ba{cccc}{
Z&X&Y& \\
&&X&X \\ \hline
Z&X&Z&X
}
\hspace{15pt}
\ba{cccc}{
&Z&Z&X \\
Z&Y&& \\ \hline
Z&X&Z&X
}
\hspace{15pt}
\ba{cccc}{
&Y&Z&X \\
Z&Z&& \\ \hline
Z&X&Z&X
}.
}{rank1-ZW-midcom1}
Here we need not take into account the following commutator
\eq{
\ba{cccc}{
Z&X&Z&Z \\
&&&Y \\ \hline
Z&X&Z&X
}
}
because $ZXZYX$ is generated by
\eq{
\ba{ccccc}{
Z&X&Z&Z& \\
&&&X&X \\ \hline\hline
Z&X&Z&Y&X
}
\hspace{15pt}
\ba{ccccc}{
Z&X&Z&X& \\
&&&Z&X \\ \hline\hline
Z&X&Z&Y&X
}
\hspace{15pt}
\ba{ccccc}{
&Z&Z&Y&X \\
Z&Y&&& \\ \hline\hline
Z&X&Z&Y&X
},
}
while both $ZXZX$ and $ZZYX$ are not in the form of Eqs.~\eqref{rank1-ZW-DP-1} or \eqref{rank1-ZW-DP-2} and thus have zero coefficients.

Then, \eref{rank1-ZW-midcom1} leads to the relation
\eqa{
-J^1_{XX}q_{ZIYX}-J^1_{YZ}q_{ZXIZ}-J^1_{XX}q_{ZXY}-J^1_{YZ}q_{ZZX}+J^1_{ZZ}q_{YZX}=0.
}{rank1-ZW-midmain}
The 4-support operator $ZXIZ$ satisfies the condition of \lref{rank1-ZW-ZZ-k+1}, and thus we already have the expression of $q_{ZXIZ}$ as
\eqa{
q_{ZXIZ}=c^{1,4}_{ZW-XX}J^1_{YZ}J^2_Z.
}{rank1-ZW-mid4}
Thus, below we connect the latter three coefficients of 3-support operators, $ZXY$, $ZZX$, and $YZX$, to those of 4-support operators in the form of Eqs.~\eqref{rank1-ZW-DP-1} or \eqref{rank1-ZW-DP-2}.

The latter three 3-support operators, $ZXY$, $ZZX$, and $YZX$, are connected to 4-support operators by considering generation of $ZXXIZ$, $ZZYIZ$, and $YZYIZ$ as\fn{
Here we need not consider the contribution of
\eq{
\ba{ccccc}{
&Y&Y&I&Z \\
Z&X&&& \\ \hline
Z&Z&Y&I&Z
},
}
since the 4-support operator $YYIZ$ does not satisfy the form of Eqs.~\eqref{rank1-ZW-DP-1} nor \eqref{rank1-ZW-DP-2}.
}
\eq{
\ba{ccccc}{
Z&X&Y&& \\
&&Z&I&Z \\ \hline
Z&X&X&I&Z
}
\hspace{15pt}
\ba{ccccc}{
&X&Y&I&Z \\
Z&I&Z&& \\ \hline
Z&X&X&I&Z
}
\hspace{15pt}
\ba{ccccc}{
&Z&X&I&Z \\
Z&Y&&& \\ \hline
Z&X&X&I&Z
}
\hspace{15pt}
\ba{ccccc}{
&Y&X&I&Z \\
Z&Z&&& \\ \hline
Z&X&X&I&Z
},
}
\eq{
\ba{ccccc}{
Z&Z&X&& \\
&&Z&I&Z \\ \hline
Z&Z&Y&I&Z
}
\hspace{15pt}
\ba{ccccc}{
&Z&X&I&Z \\
Z&I&Z&& \\ \hline
Z&Z&Y&I&Z
}
\hspace{15pt}
\ba{ccccc}{
&X&Y&I&Z \\
Z&Y&&& \\ \hline
Z&Z&Y&I&Z
},
}
\eq{
\ba{ccccc}{
Y&Z&X&& \\
&&Z&I&Z \\ \hline
Y&Z&Y&I&Z
}
\hspace{15pt}
\ba{ccccc}{
&X&Y&I&Z \\
Y&Y&&& \\ \hline
Y&Z&Y&I&Z
}.
}
They lead to the following relations of coefficients:
\balign{
J^2_Zq_{ZXY}+J^2_Zq_{XYIZ}-J^1_{YZ}q_{ZXIZ}+J^1_{ZZ}q_{YXIZ}&=0, \\
-J^2_Z q_{ZZX}-J^2_Zq_{ZXIZ}+J^1_{YZ}q_{XYIZ}&=0, \\
-J^2_Zq_{YZX}+J^1_{YY}q_{XYIZ}&=0,
}
or equivalently
\balign{
q_{ZXY}=&-q_{XYIZ}+\frac{J^1_{YZ}}{J^2_Z}q_{ZXIZ}-\frac{J^1_{ZZ}}{J^2_Z}q_{YXIZ}, \lb{rank1-ZW-mid1} \\
q_{ZZX}=&q_{ZXIZ}-\frac{J^1_{YZ}}{J^2_Z}q_{XYIZ},\lb{rank1-ZW-mid2} \\
q_{YZX}=&\frac{J^1_{YY}}{J^2_Z}q_{XYIZ}.\lb{rank1-ZW-mid3}
}
Thus, using these three relations we can erase all the coefficients of 3-support operators.

Plugging Eqs.~\eqref{rank1-ZW-mid1}, \eqref{rank1-ZW-mid2}, \eqref{rank1-ZW-mid3}, and \eqref{rank1-ZW-mid4} into \eref{rank1-ZW-midmain} and employing Eqs.~\eqref{rank1-ZW-expand-1} and \eqref{rank1-ZW-expand-2}, we arrive at
\eq{
-2(J^1_{XX})^2 J^2_Z c^{1,4}_{ZW-XX}=0.
}
Since we assumed $J^2_Z\neq 0$ and $J^1_{XX}\neq 0$, we arrive at the desired result: $c^{1,4}_{ZW-XX}=0$.

\subsubsection{Demonstrating that the remaining $k$-support operators have zero coefficients}

Now we treat $k$-support operators in the form of $Z\cdots W$ with general $k=3m+1$.
Owing to \lref{rank1-ZW-k+1} (and \lref{rank1-XX-common-ZW}), for the proof of the absence of $k$-local conserved quantity in the form of $Z\cdots W$ it suffices to demonstrate that the coefficient of $(\tz\x)^m=Z(IYY)^{m-1}IYX$ is zero.

We first observe
\balign{
\ba{cccccccc}{
Z&I&Y&(YIY)^{m-2}&Y&I&Y&X \\
&&&&&X&X& \\ \hline
Z&I&Y&(YIY)^{m-2}&Y&X&Z&X
}&
\hspace{15pt}
\ba{cccccccc}{
Z&I&Y&(YIY)^{m-2}&Y&X&Y& \\
&&&&&&X&X \\ \hline
Z&I&Y&(YIY)^{m-2}&Y&X&Z&X
} \nt \\
\hspace{15pt}
\ba{cccccccc}{
&&X&(YIY)^{m-2}&Y&X&Z&X \\
Z&I&Z&&&&& \\ \hline
Z&I&Y&(YIY)^{m-2}&Y&X&Z&X
}&
\hspace{15pt}
\ba{cccccccc}{
Z&I&Y&(YIY)^{m-2}&Y&X&I&Z \\
&&&&&&Z&Y \\ \hline
Z&I&Y&(YIY)^{m-2}&Y&X&Z&X
},
}
which leads to
\eqa{
-J^1_{XX}q_{Z(IYY)^{m-1}IYX}-J^1_{XX}q_{Z(IYY)^{m-1}XY}-J^2_Z q_{X(YIY)^{m-2}YXZX}-J^1_{YZ}q_{Z(IYY)^{m-1}XIZ}=0.
}{rank1-XX-k-mid1}
Below we compute the latter three coefficients, $Z(IYY)^{m-1}XY$, $X(YIY)^{m-2}YXZX$, and $Z(IYY)^{m-1}XIZ$, one by one.

\bigskip

\underline{Coefficient of $Z(IYY)^{m-1}XIZ$}

We first examine the last $k$-support operator $Z(IYY)^{m-1}XIZ$.
This operator forms a pair as\fn{
Here we need not consider the contribution of
\eq{
\ba{ccccccccc}{
&X&I&Y&(YIY)^{m-2}&Y&X&I&Z \\
X&Z&&&&&&& \\ \hline
X&Y&I&Y&(YIY)^{m-2}&Y&X&I&Z
},
}
since this $k$-support operator has already been shown to have zero coefficient in \lref{rank1-ZW-k+1}.
In the following, we use similar assertions.
}
\eq{
\ba{ccccccccc}{
&Z&I&Y&(YIY)^{m-2}&Y&X&I&Z \\
X&X&&&&&&& \\ \hline
X&Y&I&Y&(YIY)^{m-2}&Y&X&I&Z
}
\hspace{15pt}
\ba{ccccccccc}{
X&Y&I&Y&(YIY)^{m-2}&Y&Y&& \\
&&&&&&Z&I&Z \\ \hline
X&Y&I&Y&(YIY)^{m-2}&Y&X&I&Z
}.
}
The latter $k-1$-support operator also forms a pair as
\eq{
\ba{ccccccccc}{
&&X&Y&I&Y&(YIY)^{m-2}&Y&Y \\
Z&I&Z&&&&&& \\ \hline
Z&I&Y&Y&I&Y&(YIY)^{m-2}&Y&Y
}
\hspace{15pt}
\ba{ccccccccc}{
Z&I&Y&Y&I&Y&(YIY)^{m-2}&X& \\
&&&&&&&Z&Y \\ \hline
Z&I&Y&Y&I&Y&(YIY)^{m-2}&Y&Y
}.
}
Notice that the last operator $ZIY(YIY)^{m-1}X$ is in the form of \eref{rank1-ZW-DP-1}, and thus its coefficient has already been computed in \eref{rank1-ZW-expand-1}.
These two pairs of commutators lead to
\eqa{
J^1_{XX}q_{Z(IYY)^{m-1}XIZ}-J^1_{YZ}q_{ZIY(YIY)^{m-1}X}=0.
}{rank1-XX-k-mid1-4th}

\bigskip

\underline{Coefficient of $X(YIY)^{m-2}YXZX$}

To compute the coefficient of $k-2$-support operator $X(YIY)^{m-2}YXZX$, we shall move an irregular term $YXZ$ in it from right to left.
We here provide general relations connecting $k-2$-support operators $XY(IYY)^{m-n-2}XZ(YIY)^nX$ and $XY(IYY)^{m-n-3}XZ(YIY)^{n+1}X$ by considering commutators generating $k$-support operator $XY(IYY)^{m-n-2}XZ(YIY)^nYIZ$ and $k$-support operator $Z(IYY)^{m-n-2}XZ(YIY)^{n+1}X$ as\fn{
Here we need not consider the contribution of
\eq{
\ba{ccccccccc}{
&X&(IYY)^{m-n-2}&X&Z&(YIY)^n&Y&I&Z \\
X&Z&&&&&&& \\ \hline
X&Y&(IYY)^{m-n-2}&X&Z&(YIY)^n&Y&I&Z
}
\hspace{15pt}
\ba{ccccccccc}{
Z&Y&(IYY)^{m-n-2}&X&Z&(YIY)^n&Y&I&Z \\
Y&&&&&&&& \\ \hline
X&Y&(IYY)^{m-n-2}&X&Z&(YIY)^n&Y&I&Z
}
}
for the following reason.
The latter $k$-support operator $ZY(IYY)^{m-n-2}XZ(YIY)^nYIZ$ has already been shown to have zero coefficient in \lref{rank1-ZW-ZZ-k+1}.
The former $k-1$-support operator $X(IYY)^{m-n-2}XZ(YIY)^nYIZ$ has also already been shown to have zero coefficient in \lref{rank1-XX-k-1-zero}.
}
\balign{
\ba{ccccccccc}{
X&Y&(IYY)^{m-n-2}&X&Z&(YIY)^n&X&& \\
&&&&&&Z&I&Z \\ \hline
X&Y&(IYY)^{m-n-2}&X&Z&(YIY)^n&Y&I&Z
}&
\hspace{15pt}
\ba{ccccccccc}{
&Z&(IYY)^{m-n-2}&X&Z&(YIY)^n&Y&I&Z \\
X&X&&&&&&& \\ \hline
X&Y&(IYY)^{m-n-2}&X&Z&(YIY)^n&Y&I&Z
} 
\nt \\
\ba{ccccccccc}{
X&Y&(IYY)^{m-n-2}&I&Y&(YIY)^n&Y&I&Z \\
&&&X&X&&&& \\ \hline
X&Y&(IYY)^{m-n-2}&X&Z&(YIY)^n&Y&I&Z
} 
\lb{rank1-XX-midcom1}
}
and
\balign{
\ba{cccccccccccc}{
Z&I&Y&Y&(IYY)^{m-n-3}&X&Z&(YIY)^n&Y&I&Z& \\
&&&&&&&&&&X&X \\ \hline
Z&I&Y&Y&(IYY)^{m-n-3}&X&Z&(YIY)^n&Y&I&Y&X
}&
\hspace{15pt}
\ba{cccccccccccc}{
&&X&Y&(IYY)^{m-n-3}&X&Z&(YIY)^n&Y&I&Y&X \\
Z&I&Z&&&&&&&&& \\ \hline
Z&I&Y&Y&(IYY)^{m-n-3}&X&Z&(YIY)^n&Y&I&Y&X
}, \nt \\
\ba{cccccccccccc}{
Z&I&Y&Y&(IYY)^{m-n-3}&I&Y&(YIY)^n&Y&I&Y&X \\
&&&&&X&X&&&&& \\ \hline
Z&I&Y&Y&(IYY)^{m-n-3}&X&Z&(YIY)^n&Y&I&Y&X
}&
\lb{rank1-XX-midcom2}
}
where $X(YIY)^{m-2}YXZX$ meets the case of $n=0$.
These two sets of commutators lead to
\balign{
-J^2_Zq_{XY(IYY)^{m-n-2}XZ(YIY)^nX}+J^1_{XX}q_{Z(IYY)^{m-n-2}XZ(YIY)^nYIZ}-J^1_{XX}q_{X(YIY)^{m-1}YIZ}&=0, \lb{rank1-XX-midcom1-eq} \\
J^1_{XX}q_{Z(IYY)^{m-n-2}XZ(YIY)^nYIZ}-J^2_Zq_{XY(IYY)^{m-n-3}XZ(YIY)^{n+1}X}-J^1_{XX}q_{Z(IYY)^{m-1}IYX}&=0.
}
Subtracting the upper line from the lower line, we obtain
\eqa{
J^2_Zq_{XY(IYY)^{m-n-2}XZ(YIY)^nX}-J^2_Zq_{XY(IYY)^{m-n-3}XZ(YIY)^{n+1}X}-2J^1_{XX}q_{Z(IYY)^{m-1}IYX}=0.
}{rank1-XX-k-nth}
Here we used $q_{Z(IYY)^{m-1}IYX}=-q_{X(YIY)^{m-1}YIZ}$, which follows from \lref{rank1-ZW-k+1}.

\bigskip

Observe that \eref{rank1-XX-k-nth} holds from $n=0$ to $n=m-3$, but the case with $n=m-2$ requires some modification.
In this case, \eref{rank1-XX-midcom1} (and thus \eref{rank1-XX-midcom1-eq}) holds as it is.
On the other hand, \eref{rank1-XX-midcom2} is modified as\fn{
Here we need not consider the contribution of
\eq{
\ba{cccccccc}{
Z&X&Z&(YIY)^{m-2}&Y&I&X& \\
&&&&&&Z&X \\ \hline
Z&X&Z&(YIY)^{m-2}&Y&I&Y&X
},
}
since this $k-1$-support operator has already been shown to have a zero coefficient in \lref{rank1-XX-k-1-zero}.
}
\balign{
\ba{cccccccc}{
Z&X&Z&(YIY)^{m-2}&Y&I&Z& \\
&&&&&&X&X \\ \hline
Z&X&Z&(YIY)^{m-2}&Y&I&Y&X
}&
\hspace{15pt}
\ba{cccccccc}{
&Z&Z&(YIY)^{m-2}&Y&I&Y&X \\
Z&Y&&&&&& \\ \hline
Z&X&Z&(YIY)^{m-2}&Y&I&Y&X
}, \nt \\
\ba{cccccccc}{
&Y&Z&(YIY)^{m-2}&Y&I&Y&X \\
Z&Z&&&&&& \\ \hline
Z&X&Z&(YIY)^{m-2}&Y&I&Y&X
}&
\hspace{15pt}
\ba{cccccccc}{
Z&I&Y&(YIY)^{m-2}&Y&I&Y&X \\
&X&X&&&&& \\ \hline
Z&X&Z&(YIY)^{m-2}&Y&I&Y&X
}, 
\lb{rank1-XX-midcom3}
}
which leads to
\eqa{
J^1_{XX}q_{ZXZ(YIY)^{m-2}YIZ}-J^1_{YZ}q_{ZZ(YIY)^{m-1}X}+J^1_{ZZ}q_{YZ(YIY)^{m-1}X}-J^1_{XX}q_{ZIY(YIY)^{m-1}X}=0.
}{rank1-XX-ZW-gen3-last-main}
The first $k-1$ operator $ZXZ(YIY)^{m-2}YIZ$ is what we want to connect.
The last $k$-support operator is in the form of \eref{rank1-ZW-DP-1}.
Thus, we shall treat the remaining two $k-1$-support operators, $ZZ(YIY)^{m-1}X$ and $YZ(YIY)^{m-1}X$.

The former $k-1$-support operator in \eref{rank1-XX-ZW-gen3-last-main}, $YZ(YIY)^{m-1}X$, is connected to a $k$-support operator as
\eq{
\ba{ccccccccc}{
Y&Z&(YIY)^{m-2}&Y&I&Y&X&& \\
&&&&&&Z&I&Z \\ \hline
Y&Z&(YIY)^{m-2}&Y&I&Y&Y&I&Z
}
\hspace{15pt}
\ba{ccccccccc}{
&X&(YIY)^{m-2}&Y&I&Y&Y&I&Z \\
Y&Y&&&&&&& \\ \hline
Y&Z&(YIY)^{m-2}&Y&I&Y&Y&I&Z
},
}
which leads to
\eqa{
-J^2_Z q_{YZ(YIY)^{m-1}X}+J^1_{YY}q_{X(YIY)^{m-1}YIZ}=0.
}{rank1-XX-ZW-gen3-last1}
This $k$-support operator $X(YIY)^{m-1}YIZ$ is in the form of \eref{rank1-ZW-DP-2}, and thus its coefficient has already been computed in \eref{rank1-ZW-expand-2}.

We next treat the latter $k-1$-support operator in \eref{rank1-XX-ZW-gen3-last-main}, $ZZ(YIY)^{m-2}YIYX$.
We consider a $k+1$-support operator $ZZ(YIY)^{m-1}YIZ$, which is generated by\fn{
Here we need not consider the contribution of
\eq{
\ba{ccccccccc}{
&Y&Y&I&Y&(YIY)^{m-2}&Y&I&Z \\
Z&X&&&&&&& \\ \hline
Z&Z&Y&I&Y&(YIY)^{m-2}&Y&I&Z
}
}
because this $k$-support operator does not take the form of Eqs.~\eqref{rank1-ZW-DP-1} or \eqref{rank1-ZW-DP-2}.
}
\eq{
\ba{ccccccccc}{
Z&Z&Y&I&Y&(YIY)^{m-2}&X&& \\
&&&&&&Z&I&Z \\ \hline
Z&Z&Y&I&Y&(YIY)^{m-2}&Y&I&Z
}
\hspace{15pt}
\ba{ccccccccc}{
&X&Y&I&Y&(YIY)^{m-2}&Y&I&Z \\
Z&Y&&&&&&& \\ \hline
Z&Z&Y&I&Y&(YIY)^{m-2}&Y&I&Z
}
\hspace{15pt}
\ba{ccccccccc}{
&Z&X&I&Y&(YIY)^{m-2}&Y&I&Z \\
Z&I&Z&&&&&& \\ \hline
Z&Z&Y&I&Y&(YIY)^{m-2}&Y&I&Z
},
}
which leads to
\eqa{
-J^2_Z q_{ZZ(YIY)^{m-1}X} +J^1_{YZ}q_{X(YIY)^{m-1}YIZ}-J^2_Z q_{ZXIY(YIY)^{m-2}YIZ}=0.
}{rank1-XX-ZW-gen3-last2}
In these two $k$-support operators, the first one, $X(YIY)^{m-2}YIZ$ takes the form of \eref{rank1-ZW-DP-2}, and thus its coefficient is explicitly written down in \eref{rank1-ZW-expand-2}.
The second one, $ZXIY(YIY)^{m-2}YIZ$, is treated in \lref{rank1-ZW-ZZ-k+1}, and its coefficient has already been computed.

Plugging Eqs.~\eqref{rank1-XX-ZW-gen3-last1} and \eqref{rank1-XX-ZW-gen3-last2} into \eref{rank1-XX-ZW-gen3-last-main} and employing \lref{rank1-ZW-k+1} and \lref{rank1-ZW-ZZ-k+1}, we have
\eqa{
J^1_{XX}q_{ZXZ(YIY)^{m-2}YIZ}+\( \frac{-J^1_{YY}J^1_{ZZ}+(J^1_{YZ})^2}{J^2_Z}+\frac{(J^1_{YZ})^2}{J^1_{XX}}-J^1_{XX}\) q_{ZIY(YIY)^{m-1}X}=0.
}{rank1-XX-ZW-gen3-last-fin}

Now we combine our findings.
Summing \eref{rank1-XX-k-nth} from $n=0$ to $n=m-3$ and \eref{rank1-XX-ZW-gen3-last-fin}, and subtracting \eref{rank1-XX-midcom1-eq} with $n=m-2$, we finally have
\eqa{
J^2_Zq_{XY(IYY)^{m-2}XZX}+\( \frac{-J^1_{YY}J^1_{ZZ}+(J^1_{YZ})^2}{J^2_Z}+\frac{(J^1_{YZ})^2}{J^1_{XX}}-(2m-2)J^1_{XX}\) q_{ZIY(YIY)^{m-1}X}=0.
}{rank1-XX-k-mid1-3rd}

\bigskip

\underline{Coefficient of $Z(IYY)^{m-1}XY$}

We finally treat the remaining $k-1$-support operator in \eref{rank1-XX-k-mid1}; $Z(IYY)^{m-1}XY$.
First, considering commutators generating $k+1$-support operator $Z(IYY)^{m-1}XXIZ$, we have
\eqa{
\ba{cccccccc}{
Z&I&Y&Y&(IYY)^{m-2}X&Y&& \\
&&&&&Z&I&Z \\ \hline
Z&I&Y&Y&(IYY)^{m-2}X&X&I&Z
}
\hspace{15pt}
\ba{cccccccc}{
&&X&Y&(IYY)^{m-2}X&X&I&Z \\
Z&I&Z&&&&& \\ \hline
Z&I&Y&Y&(IYY)^{m-2}X&X&I&Z
},
}{rank1-XX-ZW-midcom3}
which leads to
\eqa{
J^2_Z q_{Z(IYY)^{m-1}XY}-J^2_Zq_{XY(IYY)^{m-2}XXIZ}=0.
}{rank1-XX-ZW-gen2-mid1}
In addition, considering commutators generating $k$-support operator $XY(IYY)^{m-n-2}XX(IYY)^nIYX$, we have\fn{
Here, we need not consider contributions from commutators of a $k$-support operator and a term in the Hamiltonian, since $XY(IYY)^{m-n-2}XX(IYY)^nIYX$ cannot be obtained by applying at most 3-support operators to a $k$-support operator in the form of Eqs.~\eqref{rank1-ZW-DP-1} or \eqref{rank1-ZW-DP-2}.
We also need not consider the contribution of
\eq{
\ba{cccccccc}{
X&Y&(IYY)^{m-n-2}X&X&(IYY)^n&I&X& \\
&&&&&&Z&X \\ \hline
X&Y&(IYY)^{m-n-2}X&X&(IYY)^n&I&Y&X
}
\hspace{15pt}
\ba{cccccccc}{
&X&(IYY)^{m-n-2}X&X&(IYY)^n&I&Y&X \\
X&Z&&&&&& \\ \hline
X&Y&(IYY)^{m-n-2}X&X&(IYY)^n&I&Y&X
},
}
because a $k-1$-support operator in the form of $W\cdots IW'$ with $W,W'\in \{X,Y\}$ has zero coefficient as shown in \lref{rank1-k-1-WIW}.
}
\eq{
\ba{cccccccc}{
X&Y&(IYY)^{m-n-2}X&X&(IYY)^n&I&Z& \\
&&&&&&X&X \\ \hline
X&Y&(IYY)^{m-n-2}X&X&(IYY)^n&I&Y&X
}
\hspace{15pt}
\ba{cccccccc}{
&Z&(IYY)^{m-n-2}X&X&(IYY)^n&I&Y&X \\
X&X&&&&&& \\ \hline
X&Y&(IYY)^{m-n-2}X&X&(IYY)^n&I&Y&X
},
}
which leads to
\eqa{
J^1_{XX}q_{XY(IYY)^{m-n-2}XX(IYY)^nIZ}+J^1_{XX}q_{Z(IYY)^{m-n-2}XX(IYY)^nIYX}=0.
}{rank1-XX-ZW-gen2-mid2}

Moreover, we consider commutators generating $k+1$-support operator $Z(IYY)^{m-n-2}XX(IYY)^{n+1}IZ$, which is a counterpart of \eref{rank1-XX-ZW-midcom3}, and obtain
\eq{
\ba{cccccccccccc}{
Z&I&Y&Y&(IYY)^{m-n-3}X&X&(IYY)^n&I&Y&X&& \\
&&&&&&&&&Z&I&Z \\ \hline
Z&I&Y&Y&(IYY)^{m-n-3}X&X&(IYY)^n&I&Y&Y&I&Z
}
\hspace{15pt}
\ba{cccccccccccc}{
&&X&Y&(IYY)^{m-n-3}X&X&(IYY)^n&I&Y&Y&I&Z \\
Z&I&Z&&&&&&&&& \\ \hline
Z&I&Y&Y&(IYY)^{m-n-3}X&X&(IYY)^n&I&Y&Y&I&Z
},
}
which leads to
\eqa{
-J^2_Zq_{Z(IYY)^{m-n-2}XX(IYY)^{n}IYX}-J^2_Zq_{XY(IYY)^{m-n-3}XX(IYY)^{n+1}IZ}=0.
}{rank1-XX-ZW-gen2-mid3}
Combining Eqs.~\eqref{rank1-XX-ZW-gen2-mid2} and \eqref{rank1-XX-ZW-gen2-mid3}, we find
\eqa{
q_{XY(IYY)^{m-n-2}XX(IYY)^{n}IZ}=-q_{Z(IYY)^{m-n-2}XX(IYY)^{n}IYX}=q_{XY(IYY)^{m-n-3}XX(IYY)^{n+1}IZ}.
}{rank1-XX-ZW-gen2-mid4}
Applying this relation repeatedly, we can connect $XY(IYY)^{m-2}XXIZ$ to $XYXX(IYY)^{m-2}IZ$.

We finally consider commutators generating $k+1$-support operator $ZXX(IYY)^{m-1}IYX$ as
\balign{
\ba{cccccccccccc}{
Z&X&X&(IYY)^{m-2}&I&Y&X&& \\
&&&&&&Z&I&Z \\ \hline
Z&X&X&(IYY)^{m-2}&I&Y&Y&I&Z
}&
\hspace{15pt}
\ba{cccccccccccc}{
&X&Y&(IYY)^{m-2}&I&Y&Y&I&Z \\
Z&I&Z&&&&&& \\ \hline
Z&X&X&(IYY)^{m-2}&I&Y&Y&I&Z
}
\nt \\
\ba{cccccccccccc}{
&Z&X&(IYY)^{m-2}&I&Y&Y&I&Z \\
Z&Y&&&&&&& \\ \hline
Z&X&X&(IYY)^{m-2}&I&Y&Y&I&Z
}&
\hspace{15pt}
\ba{cccccccccccc}{
&Y&X&(IYY)^{m-2}&I&Y&Y&I&Z \\
Z&Z&&&&&&& \\ \hline
Z&X&X&(IYY)^{m-2}&I&Y&Y&I&Z
},
}
which leads to
\eqa{
-J^2_Zq_{ZXX(IYY)^{m-2}IYX}+J^2_Zq_{XY(IYY)^{m-1}IZ}-J^1_{YZ}q_{ZX(IYY)^{m-1}IZ}+J^1_{ZZ}q_{YX(IYY)^{m-1}IZ}=0.
}{rank1-XX-ZW-gen2-mid5}
This relation is a counterpart of \eref{rank1-XX-ZW-gen2-mid3} in the case of $n=m-2$.
The coefficients of these three $k$-support operators, $q_{XY(IYY)^{m-1}IZ}$, $q_{ZX(IYY)^{m-1}IZ}$, and $q_{YX(IYY)^{m-1}IZ}$, have already been computed in \lref{rank1-ZW-k+1} and \lref{rank1-ZW-ZZ-k+1}.
Plugging these expressions into \eref{rank1-XX-ZW-gen2-mid5}, we find
\eqa{
-J^2_Zq_{ZXX(IYY)^{m-2}IYX}+\( J^2_Z+\frac{(J^1_{YZ})^2}{J^1_{XX}}-\frac{J^1_{ZZ}J^1_{YY}}{J^1_{XX}} \)q_{XY(IYY)^{m-1}IZ}=0.
}{rank1-XX-ZW-gen2-mid6}

We shall summarize our observations.
Using \eref{rank1-XX-ZW-gen2-mid4} repeatedly and plugging it into \eref{rank1-XX-ZW-gen2-mid6}, we arrive at
\eqa{
-J^2_Zq_{Z(IYY)^{m-2}XY}+\( J^2_Z+\frac{(J^1_{YZ})^2}{J^1_{XX}}-\frac{J^1_{ZZ}J^1_{YY}}{J^1_{XX}} \)q_{XY(IYY)^{m-1}IZ}=0.
}{rank1-XX-k-mid1-2nd}

\bigskip

Now we combine our findings.
Plugging Eqs.~\eqref{rank1-XX-k-mid1-2nd}, \eqref{rank1-XX-k-mid1-3rd}, and \eqref{rank1-XX-k-mid1-4th} into \eref{rank1-XX-k-mid1}, we arrive at
\balign{
&-J^1_{XX}q_{Z(IYY)^{m-1}IYX}-\frac{J^1_{XX}}{J^2_Z}\( J^2_Z+\frac{(J^1_{YZ})^2}{J^1_{XX}}-\frac{J^1_{ZZ}J^1_{YY}}{J^1_{XX}} \)q_{XY(IYY)^{m-1}IZ} \nt \\
&+\( \frac{-J^1_{YY}J^1_{ZZ}+(J^1_{YZ})^2}{J^2_Z}+\frac{(J^1_{YZ})^2}{J^1_{XX}}-(2m-2)J^1_{XX}\) q_{ZIY(YIY)^{m-1}X}-\frac{(J^1_{YZ})^2}{J^1_{XX}}q_{ZIY(YIY)^{m-1}X} \nt \\
=&-(2m-2)J^1_{XX} q_{ZIY(YIY)^{m-1}X}=0.
}
Since we have $m\geq 2$ in our treatment and $J^1_{XX}\neq 0$ is satisfied by assumption, we arrive at the desired result $q_{ZIY(YIY)^{m-1}X}=0$, or equivalently, $c^{1,k}_{XX-ZW}=0$.
This implies that all $k$-support operators in the form of $Z\cdots W$ with $W\in \{X,Y\}$ have zero coefficients.

\blm{\lb{l:rank1-XX-ZW}
Consider a Hamiltonian \eqref{rank1-standard} with $J_Z^2\neq 0$ and $J^1_{XX}\neq 0$.
In a candidate of a $k$-support conserved quantity $Q$, the coefficient of a $k$-support operator one of whose end is not $Z$ has zero coefficient.
}

\subsection{Analysis of $Z\cdots Z$}\lb{s:rank1-A-ZZ}

\subsubsection{Restricting possible forms of $k$-support operators}

In a similar manner to the case of $Z\cdots W$ ($W\in \{X,Y\}$), we first specify the form of a $k$-support operator $Z\cdots Z$ with nonzero coefficient.

We first examine the case that a $k$-support operator is in the form of $Z\cdots PZ$ with $P\in \{X,Y,Z\}$.
In this case, we consider commutators generating $W\Wc \cdots PZ$ as
\eq{
\ba{cccccc}{
&Z&\cdots &*&P&Z \\
W&W&&&& \\ \hline
W&W^{\rm c}&\cdots &*&P&Z
}
\hspace{15pt}
\ba{cccccc}{
&W&\cdots &*&P&Z \\
W&Z&&&& \\ \hline
W&W^{\rm c}&\cdots &*&P&Z
}
\hspace{15pt}
\ba{cccccc}{
W&W^{\rm c}&\cdots &*&*''& \\
&&&&*'''&Z \\ \hline
W&W^{\rm c}&\cdots &*&P&Z
}
\hspace{15pt}
\ba{cccccc}{
W&W^{\rm c}&\cdots &*&P& \\
&&&Z&I&Z \\ \hline
W&W^{\rm c}&\cdots &*&P&Z
}.
}
However, \lref{rank1-XX-ZW} tells that the latter three $k$-support operators have zero coefficients, and hence $Z\cdots PZ$ also has a zero coefficient.

\blm{
Consider a Hamiltonian \eqref{rank1-standard} with $J_Z^2\neq 0$ and $J^1_{XX}\neq 0$.
In a candidate of a $k$-support conserved quantity $Q$, a $k$-support operator $\bsA$ may have a nonzero coefficient only if $\bsA$ takes the form of $ZI\cdots IZ$.
}

Next, similarly to the case of $Z\cdots W$, we consider commutators generating $W\Wc \cdots IZ$ as\fn{
Here we need not consider the commutator
\eq{
\ba{cccccc}{
&W&\cdots &*&I&Z \\
W&Z&&&& \\ \hline
W&W^{\rm c}&\cdots &*&I&Z
}
}
because \lref{rank1-XX-ZW} tells that this $k$-support operator has a zero coefficient.
}
\eq{
\ba{cccccc}{
&Z&\cdots &*&I&Z \\
W&W&&&& \\ \hline
W&W^{\rm c}&\cdots &*&I&Z
}
\hspace{15pt}
\ba{cccccc}{
W&W^{\rm c}&\cdots &W'&& \\
&&&Z&I&Z \\ \hline
W&W^{\rm c}&\cdots &*&I&Z
}.
}

To examine the $k-1$-support operator $W\Wc \cdots W'$, we consider commutators generating $k+1$-support operator $ZI\Wc\Wc \cdots W'$ as
\fn{
Here we need not consider the commutator
\eq{
\ba{ccccccc}{
Z&I&\Wc&\Wc&\cdots &W'& \\
&&&&&Z&W' \\ \hline
Z&I&\Wc&\Wc&\cdots &{W'}^{\rm c}&W'
}
}
because \lref{rank1-XX-ZW} tells that this $k$-support operators have zero coefficient.
}
\eq{
\ba{ccccccc}{
&&W&\Wc&\cdots &*&W' \\
Z&I&Z&&&& \\ \hline
Z&I&\Wc&\Wc&\cdots &*&W'
}
\hspace{15pt}
\ba{ccccccc}{
Z&I&\Wc&\Wc&\cdots &Z& \\
&&&&&W'&W' \\ \hline
Z&I&\Wc&\Wc&\cdots &{W'}^{\rm c}&W'
}.
}

Through these two pairs, we obtain a connection between two $k$-support operators in the form of $ZI\cdots IZ$.
Using this connection repeatedly, we arrive at the following result, which is a counterpart of \lref{rank1-ZW-k+1}:

\blm{\lb{l:rank1-ZZ-k+1}
Consider a Hamiltonian \eqref{rank1-standard} with $J_Z^2\neq 0$ and $J^1_{XX}\neq 0$.
In a candidate of a $k$-support conserved quantity $Q$, a $k$-support operator may have nonzero coefficient only if $\bsA$ is expressed as 
\eqa{
\bsA=\tz \w^1 \tz \w^2 \cdots \w^{m-1}\tz \w^m \tz,
}{rank1-ZZ-DP}
with $k=3m+3$ and $W^1,W^2\ldots , W^m\in \{X,Y\}$.
In addition, if $J^1_{YY}=0$, then $W^1,W^2\ldots , W^m=X$ holds.

The coefficients are linearly connected as
\eqa{
q_{\bsA}=c^{1,k,a}_{XX-ZZ}\cdot (-1)^m (J^2_Z)^{m-1} \prod_{j=1}^m J^1_{W^jW^j}
}{rank1-ZZ-expand}
with a constant $c^{1,k,a}_{XX-ZZ}$, where $a\in \{0,1,2\}$ is the label of sectors.
}

\subsubsection{Restricting possible forms of some $k-1$-support operators}

A similar assertion to the above restricts a possible form of $k-1$-support operators where both ends are not $Z$ (i.e., $W\cdots W'$ with $W,W'\in \{X,Y\}$).
Let us consider commutators generating $k+1$-support operator $ZI\Wc \cdots W'$ as
\eq{
\ba{cccccc}{
&&W&\cdots &*&W' \\
Z&I&Z&&& \\ \hline
Z&I&\Wc&\cdots &{W'}^{\rm c}&W'
}
\hspace{15pt}
\ba{cccccc}{
Z&I&\Wc&\cdots &Z& \\
&&&&W'&W' \\ \hline
Z&I&\Wc&\cdots &{W'}^{\rm c}&W'
}.
}
The latter $k$-support operator should satisfy \eref{rank1-ZZ-DP} (in \lref{rank1-ZZ-k+1}), and thus its coefficient is expressed as \eref{rank1-ZZ-expand}.
From this fact, we conclude the following lemma:

\blm{\lb{l:rank1-ZZ-k+1-k-1}
Consider a Hamiltonian \eqref{rank1-standard} with $J_Z^2\neq 0$ and $J^1_{XX}\neq 0$.
In a candidate of a $k$-support conserved quantity $Q$, a $k-1$-support operator $\bsA$ both of whose ends are not $Z$ may have nonzero coefficient only if $\bsA$ is expressed as 
\eqa{
\bsA=\w^1 \tz \w^2 \cdots \w^{m-1}\tz \w^m \tz \w^{m+1}
}{rank1-ZZ-DP-k-1}
with $k=3m+3$ and $W^1,W^2\ldots , W^{m+1}\in \{X,Y\}$.
Otherwise, a $k-1$-support operator has a zero coefficient.

The coefficient of a $k-1$-support operator in the form of \eref{rank1-ZZ-DP-k-1} is calculated as
\eq{
q_{\bsA}=c^{1,k,a}_{XX-ZZ}\cdot \sigma(W^1,Z)\sigma(Z, W^{m+1}) (-1)^{m-1}(J^2_Z)^{m} \prod_{j=1}^{m+1} J^1_{W^jW^j}.
}
}

A similar argument suggests the following slight generalization:

\blm{\lb{l:rank1-ZZ-k+1-k-1-PsiPsi}
Consider a Hamiltonian \eqref{rank1-standard} with $J_Z^2\neq 0$ and $J^1_{XX}\neq 0$.
In a candidate of a $k$-support conserved quantity $Q$, a $k-1$-support operator $\bsA$ expressed as 
\eq{
\bsA=\Psi^1 \tz \w^2 \cdots \w^{m-1}\tz \w^m \tz \Psi^2
}
with $k=3m+3$, $W^2\ldots , W^{m}\in \{X,Y\}$ and $\Psi^1,\Psi^2\in \{ XX, YY, ZZ, XZ, YZ, ZX, ZY\}$ has its coefficient calculated as
\eq{
q_{\bsA}=c^{1,k,a}_{XX-ZZ}\cdot \sigma(\Psi^1_2,Z)\sigma(Z, \Psi^2_1) (-1)^{m-1}(J^2_Z)^{m} J^1_{\Psi^1}J^1_{\Psi^2}\prod_{j=2}^{m} J^1_{W^jW^j}.
}
}

If one or two ends of a $k-1$-support operator $\bsA$ is $Z$, the treatment becomes slightly complicated.
We here consider $\bsA$ in the form of $ZI\cdots W$ ($W\in \{X,Y\}$) and $ZI\cdots P'Z$ ($P'\in \{X,Y,Z\}$).
In the following, we alternatingly add $\x$ and $\tz$ to the left end so that a $k-1$-support operator $\bsA$ is reduced to the known $k$-support operator $\tz\x\tz\x\cdots$.
The following arguments confirm that this procedure indeed works.

In both cases of $ZI\cdots W$ and $ZI\cdots P'Z$, we consider the following commutators\fn{
Here we need not consider the contribution of
\eq{
\ba{cccccc}{
&X&I&\cdots&P'&P \\
X&Z&&&& \\ \hline
X&Y&I&\cdots&P'&P
}
}
for the following reason.
If $P=X$ or $Y$, $k-1$-support operator $XI\cdots P'P$ has already been shown to have zero coefficient in \lref{rank1-k-1-WIW}.
If $P=Z$, $XI\cdots P'Z$ forms a pair with $k$-support operator $ZIYI\cdots $ as
\eq{
\ba{ccccccc}{
&&X&I&\cdots&P'&Z \\
Z&I&Z&&&& \\ \hline
Z&I&Y&I&\cdots&P'&Z
}
\hspace{15pt}
\ba{ccccccc}{
Z&I&Y&I&\cdots&*& \\
&&&&&*'&Z \\ \hline
Z&I&Y&I&\cdots&P'&Z
},
}
while this $k$-support operator $ZIYI\cdots $ does not satisfy the form in \lref{rank1-ZZ-k+1}, implying zero coefficient.
}:
\eqa{
\ba{cccccc}{
&Z&I&\cdots&P'&P \\
X&X&&&& \\ \hline
X&Y&I&\cdots&P'&P
}
\hspace{15pt}
\ba{cccccc}{
X&Y&I&\cdots&*& \\
&&&&*&P \\ \hline
X&Y&I&\cdots&P'&P
}
\hspace{15pt}
\ba{ccccccc}{
X&Y&I&\cdots&*&P'& \\
&&&&P&I&P \\ \hline
X&Y&I&\cdots&*&P'&P
}.
}{rank1-XX-ZZ-nonunique}
We remark that the latter two commutators might not be unique.
In this case
We shall treat this ambiguity later.

The important fact in this relation is that the operator $XYI\cdots$ in the latter two commutators is always $k-1$-support, not $k-2$-support.
Therefore, by considering commutators generating $k+1$-support operator $ZIYYI\cdots *$ as
\eq{
\ba{ccccccc}{
&&X&Y&I&\cdots &* \\
Z&I&Z&&&& \\ \hline
Z&I&Y&Y&I&\cdots &*
},
}
we find that $k-1$-support operator $XYI\cdots *$ takes two possible forms of the right end.

One possible form of the right end is $\cdots IZ$.
In this case, $k-1$-support operator $XYI\cdots IZ$ forms a pair as
\eq{
\ba{ccccccccc}{
&&X&Y&I&\cdots &\Wc&I&Z \\
Z&I&Z&&&&&& \\ \hline
Z&I&Y&Y&I&\cdots &\Wc&I&Z
}
\hspace{15pt}
\ba{ccccccccc}{
Z&I&Y&Y&I&\cdots &W&& \\
&&&&&&Z&I&Z \\ \hline
Z&I&Y&Y&I&\cdots &\Wc&I&Z
}.
}
The newly obtained $k-1$-support operator $ZIYYI\cdots W$ again satisfies the first form of our supposition $ZI\cdots W$ ($W\in \{X,Y\}$), and thus we can repeatedly apply this treatment.
However, since the length of this operator is not a multiple of 3, we cannot apply this treatment eternally, and at some point, we move to the next, second case.

Another possible form of the right end is $\cdots IWP$ ($W\in \{X,Y\}$, $P\in \{X,Y,Z\}$).
In this case, $k-1$-support operator $XYI\cdots IWP$ forms a pair as
\eq{
\ba{ccccccccc}{
&&X&Y&I&\cdots &I&W&P \\
Z&I&Z&&&&&& \\ \hline
Z&I&Y&Y&I&\cdots &I&W&P
}
\hspace{15pt}
\ba{ccccccccc}{
Z&I&Y&Y&I&\cdots &I&Z& \\
&&&&&&&\Wc&P \\ \hline
Z&I&Y&Y&I&\cdots &I&W&P
}.
}
Since the operator $ZIYYI\cdots IZ$ in the latter commutator is a $k$-support operator, this should take the form of \eref{rank1-ZZ-DP}, whose coefficient has already been calculated.
Using this expression, the coefficient of the original $k-1$-support operator $\bsA$ is also determined.

Now we examine the remaining task; the non-uniqueness of \eref{rank1-XX-ZZ-nonunique}.
The non-uniqueness appears in the second case ($XYI\cdots IWP$) in the above treatment\fn{
In the former case, the end is always $\cdots IZ$, and thus the second commutator is uniquely determined.
}.
We first present some examples for $k=6$.
The ``pair" of 5-support operator $ZIYYZ$ {\it bifurcates} into the following three operators as
\eq{
\ba{cccccc}{
&Z&I&Y&Y&Z \\
X&X&&&& \\ \hline
X&Y&I&Y&Y&Z 
}
\hspace{15pt}
\ba{cccccc}{
X&Y&I&Y&Z& \\
&&&&X&Z \\ \hline
X&Y&I&Y&Y&Z 
}
\hspace{15pt}
\ba{cccccc}{
X&Y&I&Y&X& \\
&&&&Z&Z \\ \hline
X&Y&I&Y&Y&Z 
}
\hspace{15pt}
\ba{cccccc}{
X&Y&I&X&Y& \\
&&&Z&I&Z \\ \hline
X&Y&I&Y&Y&Z 
}.
}
We note that all these three 5-support operators may have nonzero coefficients at present, and these coefficients have already been computed in \lref{rank1-ZZ-k+1-k-1} and \lref{rank1-ZZ-k+1-k-1-PsiPsi}.
This non-uniqueness comes from the structure that $ZIYYZ$ is expressed as
\eq{
ZIYYZ=
\ba{ccccc}{
Z&I&Z&& \\
&&X&Y&Z \\ \hline \hline
},
}
and $XYZ$ is obtained by the following three products
\eqa{
\ba{ccc}{
X&Z& \\
&X&Z \\ \hline \hline
X&Y&Z
} \hspace{15pt}
\ba{ccc}{
X&X& \\
&Z&Z \\ \hline  \hline
X&Y&Z
} \hspace{15pt}
\ba{ccc}{
Y&Y& \\
Z&I&Z \\ \hline  \hline
X&Y&Z
}.
}{XYZ-product}
For this reason, $ZIYYZ$ is expressed by products of operators in the Hamiltonian not uniquely but in three different ways.
Correspondingly, the coefficient $q_{ZIYYZ}$ is calculated as
\eq{
q_{ZIYYZ}=c_{XX-ZZ}^{1,6}\( J^2_Z J^1_{XZ}J^1_{XZ}-J^2_ZJ^1_{XX}J^1_{ZZ}+J^2_ZJ^1_{YY}J^2_Z\) ,
}
where we used \lref{rank1-ZZ-k+1-k-1} and \lref{rank1-ZZ-k+1-k-1-PsiPsi} to the coefficients of compute 5-support operators.
In general, if an operator has a bifurcation of pairs, then its coefficient is given by the sum of products of interaction coefficients.

In a similar manner to above, the coefficient of $k-1$-support operator $\bsA$ in the form of $ZI\cdots W$ ($W\in \{X,Y\}$) and $ZI\cdots PZ$ ($P\in \{X,Y,Z\}$) can be computed.
To compute the coefficient efficiently, we introduce a symbol representing three Pauli matrices $T:=P^1P^2P^3$ ($P^i\in \{X,Y,Z\}$).
A $k-1$-support operator $\bsA$ in the form of $ZI\cdots W$ ($W\in \{X,Y\}$) or $ZI\cdots P'Z$ ($P'\in \{X,Y,Z\}$) may have a nonzero coefficient only if it is expressed as
\eq{
\bsA=\tz \w^1\tz \w^2\cdots \w^l \tz T \tz \w^{l+1} \cdots \w^{m-2} \tz \Psi
}
with $k=3m+3$, where $W^i\in \{X,Y\}$ and $\Psi\in \{XX,YY,ZZ,XZ,YZ,ZX,ZY,ZZ\}$, or
\eqa{
\bsA=\tz \w^1\tz \w^2\cdots \tz \w^{m-1} \tz T.
}{rank1-A-ZZ-k-1-endT-pre}

Whether one of the ends of $T$ is $Z$ or not determines whether a bifurcation occurs or not.
We first consider the case that both ends of $T$ are $X$ or $Y$.
This case has already been analyzed in \lref{rank1-ZW-PsiPsi-k-1} in a slightly different appearance.
What we have shown in this lemma is as follows:
(1) If $T$ is neither $XYX$ nor $YXY$, then $T$ does not have a bifurcation.
(2) If $T$ is $XYX$ or $YXY$, then this $k-1$-support operator has zero coefficient.

We next consider the case that one of the ends of $T$ is $Z$.
We first notice that one of the ends of $T$ can be $Z$ only in \eref{rank1-A-ZZ-k-1-endT-pre}, and the left end of $T$ is $X$ or $Y$.
This is because otherwise $T$ is sandwiched by $\tz$, and thus both ends of $T$ cannot be $Z$.

Thus, possible form of $T$ in this case is $XXZ$, $XYZ$, $XZZ$, $YXZ$, $YYZ$, and $YZZ$.
In these 6 operators, $XYZ$ and $YXZ$ are obtained by three products.
The case of $XYZ$ is seen in \eref{XYZ-product}, and $YXZ$ is obtained as
\eq{
\ba{ccc}{
Y&Z& \\
&Y&Z \\ \hline \hline
Y&X&Z
} \hspace{15pt}
\ba{ccc}{
Y&Y& \\
&Z&Z \\ \hline  \hline
Y&X&Z
} \hspace{15pt}
\ba{ccc}{
X&X& \\
Z&I&Z \\ \hline  \hline
Y&X&Z
}.
}
In the remaining 4 operators, $XZZ$ and $YZZ$ are obtained by two products as
\eq{
\ba{ccc}{
X&X& \\
&Y&Z \\ \hline  \hline
X&Z&Z
} \hspace{15pt}
\ba{ccc}{
Y&Z& \\
Z&I&Z \\ \hline  \hline
X&Z&Z
}
}
and
\eq{
\ba{ccc}{
Y&Y& \\
&X&Z \\ \hline  \hline
Y&Z&Z
} \hspace{15pt}
\ba{ccc}{
X&Z& \\
Z&I&Z \\ \hline  \hline
Y&Z&Z
},
}
and the remaining operators, $XXZ$ and $YYZ$, are obtained by a single product as
\eq{
\ba{ccc}{
X&Z& \\
&Y&Z \\ \hline \hline
X&X&Z
}
}
and
\eq{
\ba{ccc}{
Y&Z& \\
&X&Z \\ \hline \hline
Y&Y&Z
},
}
respectively.

Here, we denote by $(\Psi, \Phi)\in \calS(T)$ the set of two operators yielding $T$.
For example, $\calS(YXZ)$ consists of $(\Psi, \Phi)=(YZ, YZ), \ (YY, ZZ), \ (XX, ZIZ)$.
In addition, we introduce the sign $\sigma'_Z(\Psi, \Phi)$ defined as follows:
If $\Phi \neq ZIZ$, then $\Phi$ is a nearest-neighbor interaction term $\Phi_1\Phi_2$, and we simply define $\sigma'_Z(\Psi, \Phi)$ as
\eqa{
\sigma'_Z(\Psi, \Phi):=\sigma(\Psi_2, \Phi_1).
}{sigma'Z-def1}
On the other hand, if $\Phi=ZIZ$, we define $\sigma'_Z(\Psi, \Phi)$ as
\eqa{
\sigma'_Z(\Psi, \Phi):=\sigma( \abs{Z\cdot \Psi_1}, Z).
}{sigma'Z-def2}
Using this symbol, the coefficient of a $k-1$-support operator in the form of $ZI\cdots W$ ($W\in \{X,Y\}$) and $ZI\cdots PZ$ ($P\in \{X,Y,Z\}$) can be expressed as follows:

\blm{\lb{l:rank1-ZZ-k+1-k-1-2nd}
Consider a Hamiltonian \eqref{rank1-standard} with $J_Z^2\neq 0$ and $J^1_{XX}\neq 0$.
In a candidate of a $k$-support conserved quantity $Q$, a $k-1$-support operator $\bsA$ in the form of $ZI\cdots W$ ($W\in \{X,Y\}$) and $ZI\cdots PZ$ ($P\in \{X,Y,Z\}$) may have a nonzero coefficient only if $\bsA$ is expressed as
\eqa{
\bsA=\tz \w^1\tz \w^2\cdots \w^l \tz T \tz \w^{l+1} \cdots \w^{m-2} \tz \Psi
}{rank1-A-ZZ-k-1-normal}
with $k=3m+3$, where $W^i\in \{X,Y\}$, $\Psi\in \{XX,YY,ZZ,XZ,YZ,ZX,ZY,ZZ\}$, and $T=T_1T_2T_3=W^1P^2P^3$ with $W^1\in \{X,Y\}$ and $P^2, P^3\in \{X,Y,Z\}$, or
\eqa{
\bsA=\tz \w^1\tz \w^2\cdots \tz \w^{m-1} \tz T.
}{rank1-A-ZZ-k-1-endT}
In both cases, $T\neq XYX, YXY$ holds.
In addition, $P^3$ is also $X$ or $Y$ in the case of \eref{rank1-A-ZZ-k-1-normal}.

In the case of \eref{rank1-A-ZZ-k-1-normal}, $\calS(T)$ has a single element $(\Psi^1, \Psi^2)$, and the coefficient of $\bsA$ is computed as
\eq{
q_{\bsA}=c^{1,k}_{XX-ZZ}\cdot (-1)^{m-2} \sigma(Z, \Psi_1) \sigma(T) (J^2_Z)^m J^1_{\Psi^1}J^1_{\Psi^2} J^1_{\Psi} \prod_{i=1}^{m-2} J^1_{\w^l}
}
with
\eq{
\sigma(T)=\sigma(Z, \Psi^1_1)\sigma(\Psi^1_2,\Psi^2_1)\sigma(\Psi^2_2, Z).
}
In the case of \eref{rank1-A-ZZ-k-1-endT}, the coefficient of $\bsA$ is computed as
\eq{
q_{\bsA}=c^{1,k}_{XX-ZZ}\cdot (-1)^{m-2} (J^2_Z)^m \prod_{i=1}^{m-1} J^1_{\w^l} \( \sum_{(\Psi, \Phi)\in \calS(T)}\sigma(Z, \Psi^1) \sigma'(\Psi, \Phi)J^1_{\Psi}J_{\Phi}  \) ,
} 
where $\sigma'(\Psi, \Phi)$ is defined as Eqs.~\eqref{sigma'Z-def1} and \eqref{sigma'Z-def2}.

The case of a $k-1$-support operator $\bsA$ in the form of $W\cdots IZ$ ($W\in \{X,Y\}$) and $ZP\cdots IZ$ ($P\in \{X,Y,Z\}$) is treated in  a similar manner.

}

\subsubsection{Commonness of coefficients}

We shall show that the three coefficients in \lref{rank1-ZZ-k+1}, $c^{1,k,0}_{XX-ZZ}$, $c^{1,k,1}_{XX-ZZ}$, and $c^{1,k,2}_{XX-ZZ}$, are equal.
The essence of our proof idea is the same as \sref{rank1-ZW-common}.
We again divide Hamiltonians into two cases: (i) one of $J^1_{YY}$ or $J^1_{YZ}$ is nonzero, and (ii) both of $J^1_{YY}$ and $J^1_{YZ}$  are zero.

We first treat the case (i): one of $J^1_{YY}$ or $J^1_{YZ}$ is nonzero.
In this case, \lref{rank1-ZZ-k+1-k-1-2nd} plays a similar role to \lref{rank1-ZW-PsiPsi-k-1} and we can connect operators in different sectors.
For example, in the case with $J^1_{YY}\neq 0$, $(\tz\x\tz)_1$ is connected to other sectors as
\eq{
(\tz\x\tz)_1\lr (\x\tz\x)_3\lr (\tz\x\y)_4 \lr (\x\y\tz)_6\lr (\y\tz\x)_7\lr (\tz\x\tz)_8,
}
where the sector with $1\mod 3$ and that with $2\mod 3$ is connected, implying $c^{1,k,1}_{XX-ZZ}=c^{1,k,2}_{XX-ZZ}$.
Performing this connecting process again, we have $c^{1,k,1}_{XX-ZZ}=c^{1,k,2}_{XX-ZZ}=c^{1,k,0}_{XX-ZZ}$.
Obviously, this connecting process works for general $k$ and the case of $J^1_{YZ}\neq 0$.

We next treat the case (ii): both of $J^1_{YY}$ and $J^1_{YZ}$ are zero.
We explain our idea by taking $(\tz\x\tz)_1=(ZIYYIZ)_1$ with $k=6$ as an example.
We first consider commutators generating $(ZIYYXY)_1$ as
\eq{
\ba{cccccc}{
Z_1&I&Y&Y&I&Z \\
&&&&X&X \\ \hline
Z&I&Y&Y&X&Y
}
\hspace{15pt}
\ba{cccccc}{
&&X_3&Y&X&Y \\
Z&I&Z&&& \\ \hline
Z&I&Y&Y&X&Y
}.
}

We next consider commutators generating $k$-support operator $(XYXXIZ)_3$ as\fn{
Here we need not consider the contribution from 
\eq{
\ba{cccccc}{
&X&X&X&I&Z \\
X&Z&&&& \\ \hline
X&Y&X&X&I&Z
},
}
since this $k-1$-support operator $XXXIZ$ satisfies the form of \lref{rank1-ZZ-k+1-k-1-2nd}, and this lemma implies that the coefficient of this operator is proportional to $J^1_{XZ}J^1_{YY}J^2_Z$, which is zero in our setup.
}
\eq{
\ba{cccccc}{
X_3&Y&X&Y&& \\
&&&Z&I&Z \\ \hline
X&Y&X&X&I&Z
}
\hspace{15pt}
\ba{cccccc}{
&Z_4&X&X&I&Z \\
X&X&&&& \\ \hline
X&Y&X&X&I&Z
}
}

We finally consider commutators generating $(ZXXIYX)_4$ as\fn{
Here we need not consider the contribution from
\eq{
\ba{cccccc}{
&Y&X&I&Y&X \\
Z&Z&&&& \\ \hline
Z&X&X&I&Y&X
},
}
since this $k-1$-support operator $XXXIZ$ satisfies the form of \lref{rank1-ZZ-k+1-k-1-2nd}, and this lemma implies that the coefficient of this operator is proportional to $J^1_{XX}J^1_{YY}J^2_Z$, which is zero in our setup.
}
\eq{
\ba{cccccc}{
Z_4&X&X&I&Z& \\
&&&&X&X \\ \hline
Z&X&X&I&Y&X
}
\hspace{15pt}
\ba{cccccc}{
&X_5&Y&I&Y&X \\
Z&I&Z&&& \\ \hline
Z&X&X&I&Y&X
}.
}

Noting that $(XYIYX)_5$ is connected to $(ZIYYIZ)_6$, we find a sequence of pairs as
\eq{
(ZIYYXY)_1\lr (XYXXIZ)_3\lr (ZXXIYX)_4\lr (XYIYX)_5\lr (ZIYYIZ)_6,
}
which clearly shows the connection between different sectors.
Similar arguments hold for general $k$, which leads to the commonness of coefficients.

\blm{\lb{l:rank1-XX-common-ZZ}
Consider a Hamiltonian \eqref{rank1-standard} with $J_Z^2\neq 0$ and $J^1_{XX}\neq 0$.
Three coefficients, $c^{1,k,0}_{XX-ZZ}$, $c^{1,k,1}_{XX-ZZ}$, and $c^{1,k,2}_{XX-ZZ}$, appearing in \lref{rank1-ZZ-k+1} are equal, which we simply denote by $c^{1,k}_{XX-ZZ}$:
\eq{
c^{1,k}_{XX-ZZ}:=c^{1,k,0}_{XX-ZZ}=c^{1,k,1}_{XX-ZZ}=c^{1,k,2}_{XX-ZZ}.
}
}

In the following, we express the results in \lref{rank1-ZZ-k+1}, \lref{rank1-ZZ-k+1-k-1}, \lref{rank1-ZZ-k+1-k-1-PsiPsi}, and \lref{rank1-ZZ-k+1-k-1-2nd} simply with $c^{1,k}_{XX-ZZ}$.

\subsubsection{Demonstration with the case of $k=6$}

Before going to the general analysis on commutators generating $k$-support operators, we demonstrate our proof by employing the simplest case, $k=6$, as an example.
Our goal is to show $c^{1,6}_{XX-ZZ}=0$.

To this end, we first consider commutators generating $ZIYYXY$ as
\eq{
\ba{cccccc}{
Z&I&Y&Y&I&Z \\
&&&&X&X \\ \hline
Z&I&Y&Y&X&Y
}
\hspace{15pt}
\ba{cccccc}{
&&X&Y&X&Y \\
Z&I&Z&&& \\ \hline
Z&I&Y&Y&X&Y
}
\hspace{15pt}
\ba{cccccc}{
Z&I&Y&Y&Z& \\
&&&&Y&Y \\ \hline
Z&I&Y&Y&X&Y
}
\hspace{15pt}
\ba{cccccc}{
Z&I&Y&Y&Y& \\
&&&&Z&Y \\ \hline
Z&I&Y&Y&X&Y
},
}
which leads to
\eqa{
J^1_{XX}q_{ZIYYIZ}-J^2_Zq_{XYXY}-J^1_{YY}q_{ZIYYZ}+J^1_{YZ}q_{ZIYYY}=0.
}{rank1-XX-ZZ-ex-mid1}
The coefficients of $ZIYYZ$ and $ZIYYY$ have already been computed in \lref{rank1-ZZ-k+1-k-1-2nd}.
To compute the remaining coefficient $q_{XYXY}$, we consider commutators generating $XYXZX$ as\fn{
Here, we need not consider contributions from commutators with a 6-support operator and a 2-support operator as
\eq{
\ba{cccccc}{
X&Y&X&Z&Y&Z \\
&&&&Z&Z \\ \hline
X&Y&X&Z&X&
},
}
since we know that $XYXZYZ$ has a zero coefficient and therefore we need not consider this contribution.
In general, all 6-support operators which can generate $XYXZX$ by these commutators have already been shown to have zero coefficients.

We also need not consider contributions from commutators with a 5-support operator and a 1-support operator as
\eq{
\ba{ccccc}{
X&Y&X&Z&Z \\
&&&&Y \\ \hline
X&Y&X&Z&X
}
}
for a similar reason as above.
In general, all 5-support operators which can generate $XYXZX$ by these commutators have already been shown to have zero coefficients.
}
\eq{
\ba{ccccc}{
X&Y&X&Y& \\
&&&X&X \\ \hline
X&Y&X&Z&X
}
\hspace{15pt}
\ba{ccccc}{
&Z&X&Z&X \\
X&X&&& \\ \hline
X&Y&X&Z&X
}
\hspace{15pt}
\ba{ccccc}{
&X&X&Z&X \\
X&Z&&& \\ \hline
X&Y&X&Z&X
}
\hspace{15pt}
\ba{ccccc}{
Z&I&X&Z&X \\
Y&Y&&& \\ \hline
X&Y&X&Z&X
}
\hspace{15pt}
\ba{ccccc}{
X&Y&X&I&Z \\
&&&Z&Y \\ \hline
X&Y&X&Z&X
}
\hspace{15pt}
\ba{ccccc}{
X&Y&I&Y&X \\
&&X&X& \\ \hline
X&Y&X&Z&X
},
}
which leads to
\eqa{
-J^1_{XX}q_{XYXY}+J^1_{XX}q_{ZXZX}-J^1_{XZ}q_{XYZX}-J^1_{YY}q_{ZIXZX}-J^1_{YZ}q_{XYXIZ}-J^1_{XX}q_{XYIYX}=0.
}{rank1-XX-ZZ-ex-mid2}
The coefficients of $ZIXZX$ and $XYXIZ$ (the fourth and fifth terms in \eref{rank1-XX-ZZ-ex-mid2}) are computed in  \lref{rank1-ZZ-k+1-k-1-2nd}, and that of $XYIYX$ is computed in  \lref{rank1-ZZ-k+1-k-1}.

To evaluate the remaining coefficients in \eref{rank1-XX-ZZ-ex-mid2}, those of $ZXZX$ and $XXZX$, we further consider the commutators generating $ZXZYIZ$ and $XXZYIZ$ as
\eq{
\ba{cccccc}{
Z&X&Z&X&& \\
&&&Z&I&Z \\ \hline
Z&X&Z&Y&I&Z
}
\hspace{15pt}
\ba{cccccc}{
Z&I&Y&Y&I&Z \\
&X&X&&& \\ \hline
Z&X&Z&Y&I&Z
}
\hspace{15pt}
\ba{cccccc}{
&Z&Z&Y&I&Z \\
Z&Y&&&& \\ \hline
Z&X&Z&Y&I&Z
}
\hspace{15pt}
\ba{cccccc}{
&Y&Z&Y&I&Z \\
Z&Z&&&& \\ \hline
Z&X&Z&Y&I&Z
}
}
and
\eq{
\ba{cccccc}{
X&X&Z&X&& \\
&&&Z&I&Z \\ \hline
X&X&Z&Y&I&Z
}
\hspace{15pt}
\ba{cccccc}{
&Y&Z&Y&I&Z \\
X&Z&&&& \\ \hline
X&X&Z&Y&I&Z
},
}
which lead to
\eqa{
-J^2_Zq_{ZXZX}-J^1_{XX}q_{ZIYYIZ}-J^1_{YZ}q_{ZZYIZ}+J^1_{ZZ}q_{YZYIZ}=0
}{rank1-XX-ZZ-ex-mid3}
and
\eqa{
-J^2_Zq_{XXZX}+J^1_{XZ}q_{YZYIZ}=0,
}{rank1-XX-ZZ-ex-mid4}
respectively.
Note that the coefficients of $ZIYYIZ$, $ZZYIZ$, $YZYIZ$, and $YZYIZ$ have already been computed in \lref{rank1-ZZ-k+1} and \lref{rank1-ZZ-k+1-k-1-2nd}.

Combining Eqs.~\eqref{rank1-XX-ZZ-ex-mid1}, \eqref{rank1-XX-ZZ-ex-mid2}, \eqref{rank1-XX-ZZ-ex-mid3}, and \eqref{rank1-XX-ZZ-ex-mid4}, and using \lref{rank1-ZZ-k+1}, \lref{rank1-ZZ-k+1-k-1}, and \lref{rank1-ZZ-k+1-k-1-2nd}, we obtain
\eqa{
c_{XX-ZZ}^{1,6}\[ -3(J^1_{XX})^2+(J^1_{YY})^2-(J^1_{YZ})^2\] (J^2_Z)^2=0.
}{rank1-XX-ZZ-ex-fin1}
Here, if $J^1_{YY}=0$, then the sum inside the square bracket is strictly negative (since $J^1_{XX}\neq 0$ by assumption), implying that it is not zero.
Recalling $J^2_Z\neq 0$ by assumption, we arrive at the desired result $c_{XX-ZZ}^{1,6}=0$.
Thus, it suffices to treat the case with $J^1_{YY}\neq 0$.

In the case with $J^1_{YY}\neq 0$, we can follow the same argument as above by exchanging $X$ and $Y$.
The obtained relation is
\eqa{
c_{XX-ZZ}^{1,6}\[ -3(J^1_{YY})^2+(J^1_{XX})^2-(J^1_{XZ})^2\] (J^2_Z)^2=0.
}{rank1-XX-ZZ-ex-fin2}
Summing Eqs.~\eqref{rank1-XX-ZZ-ex-fin1} and \eqref{rank1-XX-ZZ-ex-fin2}, we finally have
\eq{
c_{XX-ZZ}^{1,6}\[ -2(J^1_{YY})^2-2(J^1_{XX})^2-(J^1_{YZ})^2-(J^1_{XZ})^2\] (J^2_Z)^2=0.
}
Now the sum inside the square bracket is strictly negative, and thus is nonzero, which directly implies the desired result $c_{XX-ZZ}^{1,6}=0$.

\subsubsection{Demonstrating that the remaining $k$-support operators have zero coefficients}

We now treat the case of general $k=3m$.
Our goal is to show the coefficient of $Z(IYY)^{m-1}IZ$ zero.

We first consider commutators generating $k$-support operator $Z(IYY)^{m-1}XY$ as
\balign{
\ba{ccccccc}{
Z&I&Y&Y&(IYY)^{m-2}&I&Z \\
&&&&&X&X \\ \hline
Z&I&Y&Y&(IYY)^{m-2}&X&Y
}
\hspace{15pt}&
\ba{ccccccc}{
&&X&Y&(IYY)^{m-2}&X&Y \\
Z&I&Z&&&& \\ \hline
Z&I&Y&Y&(IYY)^{m-2}&X&Y
} \nt \\
\ba{ccccccc}{
Z&I&Y&Y&(IYY)^{m-2}&Z& \\
&&&&&Y&Y \\ \hline
Z&I&Y&Y&(IYY)^{m-2}&X&Y
}
\hspace{15pt}&
\ba{ccccccc}{
Z&I&Y&Y&(IYY)^{m-2}&Y& \\
&&&&&Z&Y \\ \hline
Z&I&Y&Y&(IYY)^{m-2}&X&Y
},
}
which leads to
\eqa{
J^1_{XX}q_{Z(IYY)^{m-1}IZ}-J^2_Zq_{XY(IYY)^{m-2}XY}-J^1_{YY}q_{Z(IYY)^{m-1}Z}+J^1_{YZ}q_{Z(IYY)^{m-1}Y}=0.
}{rank1-XX-ZZ-gen-mid1}
Here, the coefficients of $k-1$-support operators $Z(IYY)^{m-1}Z$ and $Z(IYY)^{m-1}Y$ have already been computed in \lref{rank1-ZZ-k+1-k-1-2nd}.

To treat the remaining $k-2$-support operator $XY(IYY)^{m-2}XY$ in \eref{rank1-XX-ZZ-gen-mid1}, we consider commutators generating $k-1$-support operator $XY(IYY)^{m-2}XZX$ as\fn{
\lb{f:rank1-k-2-zero}
Here we need not consider the contribution of
\eq{
\ba{cccccc}{
&X&(IYY)^{m-2}&X&Z&X \\
X&Z&&&& \\ \hline
X&Y&(IYY)^{m-2}&X&Z&X
},
}
for the following reason.
This $k-2$-support operator $X(IYY)^{m-2}XZX$ forms a pair with $k$-support operator $Z(IYY)^{m-2}XZYIZ$ as
\eq{
\ba{ccccccc}{
X&(IYY)^{m-2}&X&Z&X&& \\
&&&&Z&I&Z \\ \hline
X&(IYY)^{m-2}&X&Z&Y&I&Z
}
\hspace{15pt}
\ba{ccccccc}{
Z&(IYY)^{m-2}&X&Z&Y&I&Z \\
Y&&&&&& \\ \hline
X&(IYY)^{m-2}&X&Z&Y&I&Z
},
}
while the latter $k$-support operator $Z(IYY)^{m-2}XZYIZ$ does not satisfy the form of \lref{rank1-ZZ-k+1} and thus has zero coefficient.
}
\balign{
\ba{cccccc}{
X&Y&(IYY)^{m-2}&X&Y& \\
&&&&X&X \\ \hline
X&Y&(IYY)^{m-2}&X&Z&X
}
\hspace{15pt}&
\ba{cccccc}{
&Z&(IYY)^{m-2}&X&Z&X \\
X&X&&&& \\ \hline
X&Y&(IYY)^{m-2}&X&Z&X
} \nt \\
\ba{cccccc}{
X&Y&(IYY)^{m-2}&X&I&Z \\
&&&&Z&Y \\ \hline
X&Y&(IYY)^{m-2}&X&Z&X
}
\hspace{15pt}&
\ba{cccccc}{
X&Y&(IYY)^{m-2}&I&Y&Z \\
&&&X&X& \\ \hline
X&Y&(IYY)^{m-2}&X&Z&X
},
}
which leads to
\eqa{
-J^1_{XX}q_{XY(IYY)^{m-2}XY}+J^1_{XX}q_{Z(IYY)^{m-2}XZX}-J^1_{YZ}q_{XY(IYY)^{m-2}XIZ}-J^1_{XX}q_{XY(IYY)^{m-2}IYX}=0.
}{rank1-XX-ZZ-gen-mid2}
Here, the coefficients of $k-1$-support operators $XY(IYY)^{m-2}XIZ$ and $XY(IYY)^{m-2}IYX$ have already been computed in \lref{rank1-ZZ-k+1-k-1-2nd}.

To treat the remaining $k-2$-support operator $Z(IYY)^{m-2}XZX$ in \eref{rank1-XX-ZZ-gen-mid2}, we consider a more generalized setup that commutators generate $k$-support operator $Z(IYY)^{m-2-n}XZ(YIY)^n YIZ$ expressed as
\balign{
\ba{ccccccccccc}{
Z&I&Y&Y&(IYY)^{m-3-n}&X&Z&(YIY)^n&X&& \\
&&&&&&&&Z&I&Z \\ \hline
Z&I&Y&Y&(IYY)^{m-3-n}&X&Z&(YIY)^n&Y&I&Z
}&
\hspace{15pt}
\ba{ccccccccccc}{
&&X&Y&(IYY)^{m-3-n}&X&Z&(YIY)^n&Y&I&Z\\
Z&I&Z&&&&&&&& \\ \hline
Z&I&Y&Y&(IYY)^{m-3-n}&X&Z&(YIY)^n&Y&I&Z
} \nt \\
\ba{ccccccccccc}{
Z&I&Y&Y&(IYY)^{m-3-n}&I&Y&(YIY)^n&Y&I&Z \\
&&&&&X&X&&&&\\ \hline
Z&I&Y&Y&(IYY)^{m-3-n}&X&Z&(YIY)^n&Y&I&Z
},&
}
which leads to 
\eqa{
-J^2_Z q_{Z(IYY)^{m-2-n}XZ(YIY)^nX}-J^2_Z q_{XY(IYY)^{m-3-n}XZ(YIY)^nYIZ}-J^1_{XX}q_{Z(IYY)^{m-1}IZ}=0.
}{rank1-XX-ZZ-gen-mid3}
Recall that $q_{Z(IYY)^{m-1}IZ}$ is what we want to show to be zero.

In addition to this, to treat the coefficient of $XY(IYY)^{m-3-n}XZ(YIY)^nYIZ$, we further consider commutators generating $k-1$-support operator $XY(IYY)^{m-3-n}XZY(IYY)^nIYX$ as\fn{
Here we need not consider the contribution of commutators of a $k$-support operator and a 2-support operator, since the remaining $k$-support operator takes the form of $ZI\cdots IZ$ (shown in \lref{rank1-ZZ-k+1}, and thus commutation with a 2-support operator cannot generate $k-1$-support operator.

We also need not consider the contribution of commutators of a $k-1$-support operator and a 2-support operator or 1-support operator except for the presented one for the following reason.
The remaining $k-1$-support operator which may have a nonzero coefficient takes the form of \lref{rank1-ZZ-k+1-k-1-PsiPsi} or \lref{rank1-ZZ-k+1-k-1-2nd} if one of the ends is not $Z$.
However, no further $k-1$-support operator in the above forms does not generate $k-1$-support operator $XY(IYY)^{m-3-n}XZY(IYY)^nIYX$.

Moreover, we need not consider the contribution of commutators as
\eq{
\ba{cccccccc}{
X&Y&(IYY)^{m-3-n}&X&Z&Y(IYY)^nI&X& \\
&&&&&&Z&X \\ \hline
X&Y&(IYY)^{m-3-n}&X&Z&Y(IYY)^nI&Y&X
}
\hspace{15pt}
\ba{cccccccc}{
&X&(IYY)^{m-3-n}&X&Z&Y(IYY)^nI&Y&X \\
X&Z&&&&&& \\ \hline
X&Y&(IYY)^{m-3-n}&X&Z&Y(IYY)^nI&Y&X
},
}
since these $k-2$-support operators are shown to have zero coefficients in a similar manner to footnote~\ref{f:rank1-k-2-zero}.
}
\balign{
\ba{cccccccc}{
X&Y&(IYY)^{m-3-n}&X&Z&Y(IYY)^nI&Z& \\
&&&&&&X&X \\ \hline
X&Y&(IYY)^{m-3-n}&X&Z&Y(IYY)^nI&Y&X
}
&\hspace{15pt}
\ba{cccccccc}{
&Z&(IYY)^{m-3-n}&X&Z&Y(IYY)^nI&Y&X \\
X&X&&&&&& \\ \hline
X&Y&(IYY)^{m-3-n}&X&Z&Y(IYY)^nI&Y&X
} \nt \\
\ba{cccccccc}{
X&Y&(IYY)^{m-3-n}&I&Y&Y(IYY)^nI&Y&X \\
&&&X&X&&& \\ \hline
X&Y&(IYY)^{m-3-n}&X&Z&Y(IYY)^nI&Y&X
},
}
which leads to
\eqa{
J^1_{XX}q_{XY(IYY)^{m-3-n}XZ(YIY)^nYIZ}+J^1_{XX}q_{Z(IYY)^{m-3-n}XZ(YIY)^{n+1}X}-J^1_{XX}q_{X(YIY)^{m-1}X}=0.
}{rank1-XX-ZZ-gen-mid4}
Here, the coefficient of $k-1$-operator $X(YIY)^{m-1}X$ has already been computed in \lref{rank1-ZZ-k+1-k-1}.

Combining Eqs.~\eqref{rank1-XX-ZZ-gen-mid3} and \eqref{rank1-XX-ZZ-gen-mid4}, we find
\eqa{
q_{Z(IYY)^{m-2-n}XZ(YIY)^nX}-q_{Z(IYY)^{m-3-n}XZ(YIY)^{n+1}X}+\frac{J^1_{XX}}{J^2_Z}q_{Z(IYY)^{m-1}IZ}+q_{X(YIY)^{m-1}X}=0.
}{rank1-XX-ZZ-gen-mid5}
This relation clearly shows the connection between the coefficients of 
\eq{
Z(IYY)^{m-2-n}XZ(YIY)^nX \lr Z(IYY)^{m-3-n}XZ(YIY)^{n+1}X,
}
i.e., $n\lr n+1$.
Notice that \eref{rank1-XX-ZZ-gen-mid5} holds from $n=0$ to $n=m-4$, which connects the coefficients of $Z(IYY)^{m-2}XZX$ and $Z(IYY)XZ(YIY)^{m-3}X$.
At $n=m-3$, \eref{rank1-XX-ZZ-gen-mid3} holds as it is, while \eref{rank1-XX-ZZ-gen-mid4} requires the following modification:
Commutators generating $k-1$-support operator $XYXZ(YIY)^{m-2}X$ are\fn{
Here we need not consider the contribution of
\eq{
\ba{ccccccc}{
X&Y&X&Z&Y(IYY)^{m-3}I&X& \\
&&&&&Z&X \\ \hline
X&Y&X&Z&Y(IYY)^{m-3}I&Y&X 
},
}
since this $k-2$-support operator $XYXZY(IYY)^{m-3}IX$ is shown to have zero coefficient in a similar manner to footnote~\ref{f:rank1-k-2-zero}.

We also need not consider the contributions of
\eq{
\ba{ccccccc}{
Y&I&X&Z&Y(IYY)^{m-3}I&Y&X \\
Z&Y&&&&&\\ \hline
X&Y&X&Z&Y(IYY)^{m-3}I&Y&X 
}
\hspace{15pt}
\ba{ccccccc}{
X&Y&I&Z&Y(IYY)^{m-3}I&Y&X \\
&Z&X&&&&\\ \hline
X&Y&X&Z&Y(IYY)^{m-3}I&Y&X 
},
}
since these $k-1$-support operators have already been shown to have zero coefficients in \lref{rank1-ZZ-k+1-k-1}.
}
\balign{
\ba{ccccccc}{
X&Y&X&Z&Y(IYY)^{m-3}I&Z& \\
&&&&&X&X \\ \hline
X&Y&X&Z&Y(IYY)^{m-3}I&Y&X 
}
&\hspace{15pt}
\ba{ccccccc}{
&Z&X&Z&Y(IYY)^{m-3}I&Y&X  \\
X&X&&&&& \\ \hline
X&Y&X&Z&Y(IYY)^{m-3}I&Y&X 
} \nt \\
\ba{ccccccc}{
X&Y&I&Y&Y(IYY)^{m-3}I&Y&X \\
&&X&X&&&\\ \hline
X&Y&X&Z&Y(IYY)^{m-3}I&Y&X 
}
&\hspace{15pt}
\ba{ccccccc}{
Z&I&X&Z&Y(IYY)^{m-3}I&Y&X \\
Y&Y&&&&&\\ \hline
X&Y&X&Z&Y(IYY)^{m-3}I&Y&X 
} \nt \\
\ba{ccccccc}{
&X&X&Z&Y(IYY)^{m-3}I&Y&X \\
X&Z&& &&&\\ \hline
X&Y&X&Z&Y(IYY)^{m-3}I&Y&X 
},&
}
which leads to
\balign{
J^1_{XX}q_{XYXZY(IYY)^{m-3}IZ}+J^1_{XX}q_{ZXZY(IYY)^{m-3}IYX}-J^1_{XX}q_{X(YIY)^{m-1}X}& \nt \\
-J^1_{YY}q_{ZIXZ(YIY)^{m-2}X}-J^1_{XZ}q_{XXZ(YIY)^{m-2}X}&=0. \lb{rank1-XX-ZZ-gen-mid6}
}
The coefficients of $k-1$-support operators $X(YIY)^{m-1}X$ and $ZIXZ(YIY)^{m-2}X$ have already been computed by \lref{rank1-ZZ-k+1-k-1} and  \lref{rank1-ZZ-k+1-k-1-2nd}, respectively.
The remaining coefficients of $k-2$-support operators, ${ZXZY(IYY)^{m-3}IYX}$ and ${XXZ(YIY)^{m-2}X}$, are computed by considering commutators generating $k$-support operators $ZXZ(YIY)^{m-2}YIZ$ and $XXZ(YIY)^{m-2}YIZ$ as
\balign{
\ba{ccccccc}{
Z&X&Z&(YIY)^{m-2}&X&& \\
&&&&Z&I&Z \\ \hline
Z&X&Z&(YIY)^{m-2}&Y&I&Z
}
&\hspace{15pt}
\ba{ccccccc}{
Z&I&Z&(YIY)^{m-2}&Y&I&Z \\
&X&X&&&& \\ \hline
Z&X&Z&(YIY)^{m-2}&Y&I&Z
} \nt \\
\ba{ccccccc}{
&Z&Z&(YIY)^{m-2}&Y&I&Z \\
Z&Y&&&&& \\ \hline
Z&X&Z&(YIY)^{m-2}&Y&I&Z
}
&\hspace{15pt}
\ba{ccccccc}{
&Y&Z&(YIY)^{m-2}&Y&I&Z \\
Z&Z&&&&& \\ \hline
Z&X&Z&(YIY)^{m-2}&Y&I&Z
}
}
and
\eq{
\ba{cccccc}{
X&X&Z(YIY)^{m-2}&X&& \\
&&&Z&I&Z \\ \hline
X&X&Z(YIY)^{m-2}&Y&I&Z
}
\hspace{15pt}
\ba{cccccc}{
&Y&Z(YIY)^{m-2}&Y&I&Z \\
X&Z&&&& \\ \hline
X&X&Z(YIY)^{m-2}&Y&I&Z
},
}
which lead to
\eqa{
-J^2_Zq_{ZXZ(YIY)^{m-2}X}-J^1_{XX}q_{Z(IYY)^{m-1}IZ}-J^1_{YZ}q_{ZZ(YIY)^{m-2}YIZ}+J^1_{ZZ}q_{YZ(YIY)^{m-2}YIZ}=0
}{rank1-XX-ZZ-gen-mid7}
and
\eqa{
-J^2_Zq_{XXZ(YIY)^{m-2}X}+J^1_{XZ}q_{YZ(YIY)^{m-2}YIZ}=0,
}{rank1-XX-ZZ-gen-mid8}
respectively.
Here, the coefficients of $k-1$-support operators $ZZ(YIY)^{m-2}YIZ$, $YZ(YIY)^{m-2}YIZ$, and $YZ(YIY)^{m-2}YIZ$ have already been computed in \lref{rank1-ZZ-k+1-k-1-2nd}, and the coefficient of $k$-support operator ${Z(IYY)^{m-1}IZ}$ is what we aim to evaluate through this analysis.

Summing \eref{rank1-XX-ZZ-gen-mid5} from $n=0$ to $n=m-4$, and plugging Eqs.~\eqref{rank1-XX-ZZ-gen-mid1}, \eqref{rank1-XX-ZZ-gen-mid2}, \eqref{rank1-XX-ZZ-gen-mid6}, \eqref{rank1-XX-ZZ-gen-mid7}, and \eqref{rank1-XX-ZZ-gen-mid8}, we finally obtain
\eqa{
c_{XX-ZZ}^{1,k}\[ -(2m-1)(J^1_{XX})^2+(J^1_{YY})^2-(J^1_{YZ})^2\] (J^2_Z)^m (J^1_{XX})^{m-2}=0.
}{rank1-XX-ZZ-gen-fin1}
If $J^1_{YY}=0$, then the sum in the square bracket is nonzero (strictly negative), and we conclude $c_{XX-ZZ}^{1,k}=0$.
If $J^1_{YY}\neq 0$, we replace the role of $X$ and $Y$ in our whole argument, which yields
\eqa{
c_{XX-ZZ}^{1,k}\[ -(2m-1)(J^1_{YY})^2+(J^1_{XX})^2-(J^1_{XZ})^2\] (J^2_Z)^m (J^1_{XX})^{m-2}=0.
}{rank1-XX-ZZ-gen-fin2}
Since the sum of Eqs.~\eqref{rank1-XX-ZZ-gen-fin1} and \eqref{rank1-XX-ZZ-gen-fin2} is
\eqa{
c_{XX-ZZ}^{1,k}\[ -(2m-2)(J^1_{XX})^2-(2m-2)(J^1_{YY})^2-(J^1_{YZ})^2-(J^1_{YZ})^2\] (J^2_Z)^m (J^1_{XX})^{m-2}=0.
}{rank1-XX-ZZ-gen-fin}
The sum in the square bracket is nonzero (strictly negative) and $J^1_{XX}$ and $J^2_Z$ are assumed to be nonzero, we arrive at the desired result $c_{XX-ZZ}^{1,k}=0$.
In summary, we obtain $c_{XX-ZZ}^{1,k}=0$  regardless of the value of $J^1_{YY}$, which completes the proof for case A.

\bthm{
Consider a Hamiltonian \eqref{rank1-standard} with $J_Z^2\neq 0$ and $J^1_{XX}\neq 0$.
This Hamiltonian has no $k$-local conserved quantity with $4\leq k\leq L/2$.
}

\section{Rank 1: Case with $J^1_{XX}=J^1_{YY}=0$ and $J^1_{XZ}\neq 0$ (Case B1)}\lb{s:rank1-B1}


We next consider case B1, where our Hamiltonian in consideration is \eref{rank1-standard} with $J^1_{XX}=J^1_{YY}=0$ and at least one of $J^1_{XZ}$ or $J^1_{YZ}$ is nonzero.
In this case, by applying a proper local spin rotation in the $XY$ subspace, without loss of generality we suppose $J^1_{XZ}\neq 0$ and $J^1_{YZ}=0$.
The Hamiltonian considered in this section is expressed as
\balign{
H=&\sum_i \mx{X_{i+2}&Y_{i+2}&Z_{i+2}}\mx{0 && \\ &0& \\ &&J_Z^2}\mx{X_i\\ Y_i\\ Z_i}+\sum_i \mx{X_{i+1}&Y_{i+1}&Z_{i+1}}\mx{0 &0&J_{XZ}^1 \\ 0&0&0 \\ J_{XZ}^1&0&J_{ZZ}^1}\mx{X_i\\ Y_i\\ Z_i} \nt \\
&+\sum_i \mx{h_X&h_Y&h_Z}\mx{X_i\\ Y_i\\ Z_i}, \lb{rank1-B1}
}
where $J_Z^2$ and $J^1_{XZ}$ take nonzero values, and $J^1_{ZZ}$ and $h_P$ ($P\in \{X,Y,Z\}$) can take both zero or nonzero values.


\subsection{Restricting possible forms of $k$-local conserved quantity}

\subsubsection{Preliminary treatment}

Thanks to \lref{rank1-k+2}, in a candidate of $k$-local conserved quantity $Q$, a $k$-support operator which may have a nonzero coefficient takes the form of $ZI\cdots W$ (and its reflection) and $Z\cdots Z$.

\bigskip

We first show that the former $ZI\cdots W$ should accompany a zero coefficient.
This fact is easily demonstrated by considering commutators generating $k+1$-support operator $ZYI\cdots W$ as follows.

In the case of $W=Y$, this operator is generated only by
\eq{
\ba{ccccc}{
&Z&I&\cdots&Y \\
Z&X&&& \\ \hline
Z&Y&I&\cdots&Y
}
\hspace{15pt}
\ba{ccccc}{
&X&I&\cdots&Y \\
Z&Z&&& \\ \hline
Z&Y&I&\cdots&Y
}.
}
However, \lref{rank1-k+2} shows that $XI\cdots Y$ has zero coefficient, which directly implies $q_{ZI\cdots Y}=0$.

In the case of $W=X$, this operator is generated by
\eq{
\ba{cccccc}{
&Z&I&\cdots&*&X \\
Z&X&&&& \\ \hline
Z&Y&I&\cdots&*&X
}
\hspace{15pt}
\ba{cccccc}{
&X&I&\cdots&*&X \\
Z&Z&&&& \\ \hline
Z&Y&I&\cdots&*&X
}
\hspace{15pt}
\ba{cccccc}{
Z&Y&I&\cdots&W& \\
&&&&Z&X \\ \hline
Z&Y&I&\cdots&\Wc&X
}.
}
However, \lref{rank1-k+2} shows that both $XI\cdots *X$ and $ZYI\cdots W$ have zero coefficients, which directly implies $q_{ZI\cdots X}=0$.

\blm{\lb{l:rank1-B1-k-ZW-zero}
Consider a Hamiltonian \eqref{rank1-B1} with nonzero $J_Z^2$ and $J^1_{XZ}$.
In a candidate of a $k$-support conserved quantity $Q$, a $k$-support operator in the form of $Z\cdots W$ or $W\cdots Z$ ($W\in \{X,Y\}$) has zero coefficient.
}

\subsubsection{Analysis of $k$-support operator $Z\cdots Z$}

Thus, we move to specify the possible form of $k$-support operator $Z\cdots Z$ with a nonzero coefficient in detail.
To this end, we consider three types of operators, a $k$-support operator in the form of $Z\cdots Z$, a $k-1$-support operator in the form of $Z\cdots Y$, and a $k-1$-support operator in the form of $Z\cdots X$, with classifying by the second rightmost operator, and demonstrate that we can repeatedly add $ZX$ to left and remove some terms from right.
This procedure implies that any operator is reduced to an operator in the form of $ZYYYYY\cdots$.

\bigskip

\underline{Case of $Z\cdots XZ$}

We first show that a $k$-support operator in the form of  $Z\cdots XZ$ has a zero coefficient.
This can be easily shown to have a zero coefficient by considering commutators generating $k+1$-support operator $ZY\cdots XZ$ as
\eq{
\ba{ccccc}{
&Z&\cdots&X&Z \\
Z&X&&& \\ \hline
Z&Y&\cdots&X&Z
}
\hspace{15pt}
\ba{cclcc}{
&Y&W\cdots&X&Z \\
Z&I&Z&& \\ \hline
Z&Y&\cdots\cdots&X&Z
}
\hspace{15pt}
\ba{ccccc}{
Z&Y&\cdots&Y& \\
&&&Z&Z \\ \hline
Z&Y&\cdots&X&Z
}
\hspace{15pt}
\ba{ccrcc}{
Z&Y&\cdots W&Y& \\
&&Z&I&Z \\ \hline
Z&Y&\cdots\cdots&X&Z
},
}
while the latter three operators have already been shown to have zero coefficients by \lref{rank1-B1-k-ZW-zero}.
This fact means that $Z\cdots XZ$ also has a zero coefficient.

\blm{\lb{l:rank1-B1-k-ZXZ-zero}
Consider a Hamiltonian \eqref{rank1-B1} with nonzero $J_Z^2$ and $J^1_{XZ}$.
In a candidate of a $k$-support conserved quantity $Q$, a $k$-support operator in the form of $Z\cdots XZ$ or $ZX\cdots Z$ has zero coefficient.
}

\bigskip

\underline{Case of $Z\cdots YZ$}

We next consider the case of $Z\cdots YZ$.
In this case, by considering commutators generating $k+1$-support operator $ZY\cdots YZ$ as\fn{
Here, we need not consider
\eq{
\ba{cclcc}{
&Y&\cdots&Y&Z \\
Z&I&Z&& \\ \hline
Z&Y&\cdots &Y&Z
}
\hspace{15pt}
\ba{ccccc}{
Z&Y&\cdots&X& \\
&&&Z&Z \\ \hline
Z&Y&\cdots &Y&Z
}
}
since $k$-support operators $Y\cdots YZ$ and $ZY\cdots X$ have already been shown to have zero coefficient in \lref{rank1-k+2}.
}
\eq{
\ba{ccccc}{
&Z&\cdots&Y&Z \\
Z&X&&& \\ \hline
Z&Y&\cdots &Y&Z
}
\hspace{15pt}
\ba{ccccc}{
Z&Y&\cdots&Z& \\
&&&X&Z \\ \hline
Z&Y&\cdots &Y&Z
},
}
we find that $Z\cdots YZ$ forms a pair with $k$-support operator $ZY\cdots Z$.
We can continue the procedure by applying the analysis on $k$-support operator $Z\cdots Z$.

\bigskip

\underline{Case of $Z\cdots ZZ$}

We next consider the case of $Z\cdots ZZ$ by considering commutators generating $k+1$-support operator $ZY\cdots ZZ$ as\fn{
Here we need not consider the contribution of
\eq{
\ba{cccccc}{
Z&Y&\cdots&*&Y&  \\
&&&&X&Z \\ \hline
Z&Y&\cdots&W&Z&Z 
},
}
since this $k$-support operator $ZY\cdots *Y$ have already been shown to have zero coefficient in \lref{rank1-B1-k-ZW-zero}.
}
\eq{
\ba{cccccc}{
&Z&\cdots&*&Z&Z \\
Z&X&&&& \\ \hline
Z&Y&\cdots&*&Z&Z 
}
\hspace{15pt}
\ba{cccccc}{
Z&Y&\cdots&W^{\rm c}&Z&  \\
&&&Z&I&Z \\ \hline
Z&Y&\cdots&W&Z&Z 
}.
}
This relation implies that $*$ should be $X$ or $Y$ ($*=\Wc\in \{X,Y\}$), and $Z\cdots WZZ$ and $ZY\cdots \Wc Z$ form a pair.
We notice that the above $W$ should be $X$ for a nonzero coefficient for the following reason:
If $W=Y$, \lref{rank1-B1-k-ZXZ-zero} suggests that $ZY\cdots XZ$ has zero coefficient, implying that $Z\cdots YZZ$ also has zero coefficient.

\blm{\lb{l:rank1-B1-k-ZYZZ-zero}
Consider a Hamiltonian \eqref{rank1-B1} with nonzero $J_Z^2$ and $J^1_{XZ}$.
In a candidate of a $k$-support conserved quantity $Q$, a $k$-support operator in the form of $Z\cdots YZZ$ or $ZZY\cdots Z$ has zero coefficient.
}

If $W=X$, the operator $Z\cdots XZZ$ is connected to a $k$-support operator in the form of $ZY\cdots YZ$, from which we can continue the procedure.

\bigskip

\underline{Case of $Z\cdots IZ$}

We finally consider the case of $Z\cdots IZ$.
Obviously, the third rightmost operator should be $X$ or $Y$.

We first treat $Z\cdots XIZ$.
We consider commutators generating $k+1$-support operator $ZY\cdots XIZ$ as
\eq{
\ba{cccccc}{
&Z&\cdots&X&I&Z \\
Z&X&&&& \\ \hline
Z&Y&\cdots&X&I&Z
}
\hspace{15pt}
\ba{cccccc}{
Z&Y&\cdots&Y&& \\
&&&Z&I&Z \\ \hline
Z&Y&\cdots&X&I&Z
},
}
which leads to
\eq{
J^1_{XZ}q_{Z\cdots XIZ}+J^2_Zq_{ZY\cdots Y}=0.
}
In this case, $k$-support operator $Z\cdots XIZ$ is connected to $k-1$-support operator $ZY\cdots Y$.
This operator takes the form of $Z\cdots Y$, and thus the procedure continues.

We next treat $Z\cdots YIZ$.
We consider commutators generating $k+1$-support operator $ZY\cdots YIZ$ as
\eq{
\ba{cccccc}{
&Z&\cdots&Y&I&Z \\
Z&X&&&& \\ \hline
Z&Y&\cdots&Y&I&Z
}
\hspace{15pt}
\ba{cccccc}{
Z&Y&\cdots&X&& \\
&&&Z&I&Z \\ \hline
Z&Y&\cdots&Y&I&Z
},
}
which leads to
\eq{
J^1_{XZ}q_{Z\cdots YIZ}-J^2_Zq_{ZY\cdots X}=0.
}
In this case, $k$-support operator $Z\cdots YIZ$ is connected to $k-1$-support operator $ZY\cdots X$.
This operator takes the form of $Z\cdots X$, and thus the procedure continues.

\subsubsection{Analysis of $k-1$-support operator $Z\cdots Y$}

\underline{Case of $Z\dots XY$}

For our later use, we first show that a $k-1$-support operator in the form of $W\cdots Y$ ($W\in \{X,Y\}$) (and its reflection) has zero coefficient.
This fact is confirmed by considering commutators generating $k+1$-support operator $ZI\Wc\cdots Y$ as
\eq{
\ba{ccccc}{
&&W&\cdots&Y \\
Z&I&Z&& \\ \hline
Z&I&\Wc&\cdots&Y
},
}
which implies $q_{W\cdots Y}=0$.

\blm{\lb{l:rank1-B1-k-1-WY-zero}
Consider a Hamiltonian \eqref{rank1-B1} with nonzero $J_Z^2$ and $J^1_{XZ}$.
In a candidate of a $k$-support conserved quantity $Q$, a $k-1$-support operator in the form of $W\cdots Y$ or $Y\cdots W$ ($W\in \{X,Y\}$) has zero coefficient.
}

Using this, we shall also show that a $k-1$-support operator in the form of $Z\cdots XY$ (and $YX\cdots Z$) has zero coefficients.
This fact can be confirmed by considering commutators generating $k$-support operator $ZY\cdots XY$ as
\eq{
\ba{ccccc}{
&Z&\cdots&X&Y \\
Z&X&&& \\ \hline
Z&Y&\cdots&X&Y
}
\hspace{15pt}
\ba{ccccc}{
&X&\cdots&X&Y \\
Z&Z&&& \\ \hline
Z&Y&\cdots&X&Y
}
\hspace{15pt}
\ba{cccccc}{
&Y&*&\cdots&X&Y \\
Z&I&Z&&& \\ \hline
Z&Y&*'&\cdots&X&Y
}
\hspace{15pt}
\ba{ccccc}{
Z&Y&\cdots&X&Z \\
&&&&X \\ \hline
Z&Y&\cdots&X&Y
}.
}
However, \lref{rank1-B1-k-1-WY-zero} tells that two middle $k-1$-support operators $X\cdots XY$ $Y*\cdots XY$ have zero coefficients, and \lref{rank1-B1-k-ZXZ-zero} tells that the latter $k$-support operator $ZY\cdots XZ$ has zero coefficient.
This fact directly implies that $k-1$-support operator $Z\cdots XY$ also has a zero coefficient.

\blm{\lb{l:rank1-B1-k-1-ZWXY-zero}
Consider a Hamiltonian \eqref{rank1-B1} with nonzero $J_Z^2$ and $J^1_{XZ}$.
In a candidate of a $k$-support conserved quantity $Q$, a $k-1$-support operator in the form of $Z\cdots XY$ or $YX\cdots Z$ has zero coefficient.
}

\bigskip

\underline{Case of $Z\cdots YY$}

To analyze it, we consider commutators generating $k$-support operator $ZY\cdots YY$ as
\eq{
\ba{cccccc}{
&Z&\cdots &Y&Y \\
Z&X&&& \\ \hline
Z&Y&\cdots &Y&Y
}
\hspace{15pt}
\ba{ccccc}{
Z&Y&\cdots &Y&Z \\
&&&&X \\ \hline
Z&Y&\cdots &Y&Y
},
}
which leads to
\eq{
J^1_{XY}q_{Z\cdots YY}+h_X q_{ZY\cdots YZ}=0.
}
Thus, $k-1$-support operator $Z\cdots YY$ is connected to $k$-support operator $ZY\cdots YZ$, and thus our procedure can be continued.

\bigskip

\underline{Case of $Z\cdots ZY$}

The case of $Z\cdots ZY$ is a slightly complicated case, since three commutators generate $k$-support operator $ZY\cdots ZY$ as\fn{
Here we need not consider the contributions of
\eq{
\ba{cccccc}{
&X&*'&\cdots &Z&Y \\
Z&Z&&&& \\ \hline
Z&Y&*'&\cdots &Z&Y
}
\hspace{15pt}
\ba{cccccc}{
&Y&*&\cdots &Z&Y \\
Z&I&Z&&& \\ \hline
Z&Y&*'&\cdots &Z&Y
},
}
since these $k-1$-support operators have already been shown to have zero coefficients in \lref{rank1-B1-k-1-WY-zero}.
}
\eq{
\ba{ccccc}{
&Z&\cdots &Z&Y \\
Z&X&&& \\ \hline
Z&Y&\cdots &Z&Y
}
\hspace{15pt}
\ba{cccccc}{
Z&Y&\cdots &Z&Z \\
&&&&X \\ \hline
Z&Y&\cdots &Z&Y
}
\hspace{15pt}
\ba{cccccc}{
Z&Y&\cdots &I&Z \\
&&&Z&X \\ \hline
Z&Y&\cdots &Z&Y
}.
}
Owing to \lref{rank1-B1-k-ZYZZ-zero}, the $k$-support operator in the second commutator, $ZY\cdots ZZ$ may have nonzero coefficient only if the right end takes the form of $\cdots XZZ$.
In other words, except for the case of $Z\cdots XZY$, we need not consider the middle commutator, implying no bifurcation.
Thus, $k-1$-support operator $Z\cdots YZY$ forms a unique pair with $k$-support operator $ZY\cdots YIZ$, from which we can continue our procedure.

In the case of $Z\cdots XZY$, we encounter a bifurcation and this operator is connected to two $k$-support operators, $ZY\cdots XZZ$ and $ZY\cdots XIZ$.
Both operators take the form of $Z\cdots Z$, and thus we can continue our procedure for both operators.

\bigskip

\underline{Case of $Z\cdots IY$}

To analyze it, we consider commutators generating $k$-support operator $ZY\cdots IY$ as
\eq{
\ba{cccccc}{
&Z&\cdots &I&Y \\
Z&X&&& \\ \hline
Z&Y&\cdots &I&Y
}
\hspace{15pt}
\ba{ccccc}{
Z&Y&\cdots &I&Z \\
&&&&X \\ \hline
Z&Y&\cdots &I&Y
},
}
which leads to
\eq{
J^1_{XY}q_{Z\cdots IY}+h_X q_{ZY\cdots IZ}=0.
}
Thus, $k-1$-support operator $Z\cdots IY$ is connected to $k$-support operator $ZY\cdots IZ$, and thus our procedure can be continued.

\subsubsection{Analysis of $k-1$-support operator $Z\cdots X$}

\underline{Case of $Z\cdots XX$}

We first consider $Z\cdots XX$, which forms a unique pair with $k-1$-support operator $ZY\cdots Y$ as
\eq{
\ba{cccccc}{
&Z&\cdots &X&X \\
Z&X&&& \\ \hline
Z&Y&\cdots &X&X
}
\hspace{15pt}
\ba{cccccc}{
Z&Y&\cdots &Y& \\
&&&Z&X \\ \hline
Z&Y&\cdots &X&X
},
}
which leads to
\eq{
J^1_{XZ}q_{Z\cdots XX}+J^1_{XZ}q_{ZY\cdots Y}=0.
}
The connected $k-1$-support operator is in the form of $ZY\cdots Y$, and thus our analysis can be continued.

\bigskip

\underline{Case of $Z\cdots YX$}

In this case, we again encounter a bifurcation.
We consider commutators generating $k$-support operator $ZY\cdots YX$ as
\eq{
\ba{ccccc}{
&Z\cdots&Y&X \\
Z&X&& \\ \hline
Z&Y\cdots&Y&X
}
\hspace{15pt}
\ba{ccccc}{
Z&Y\cdots&X& \\
&&Z&X \\ \hline
Z&Y\cdots&Y&X
}
\hspace{15pt}
\ba{ccccc}{
Z&Y\cdots&Y&Z \\
&&&Y \\ \hline
Z&Y\cdots&Y&X
}.
}
The $k-1$-support operator $ZY\cdots X$ takes the form of $Z\cdots X$, from which we can continue our procedure.
The $k$-support operator $ZY\cdots YZ$ takes the form of $Z\cdots Z$, and thus we can also continue our procedure.

\bigskip

\underline{Case of $Z\cdots ZX$}

We notice that $k-1$-support operator $Z\cdots ZX$ forms a unique pair with $k$-support operator $ZYY\cdots ZZ$ as\fn{
Here, we need not consider the contribution of
\eq{
\ba{cccccc}{
&X&\cdots &A&X \\
Z&Z&&& \\ \hline
Z&Y&\cdots &A&X
}
}
with $A=Z$, because this $k-1$-support operator $X\cdots AX$ forms a pair with $k$-support operator $X\cdots AYIZ$ as
\eq{
\ba{ccccccc}{
X&\cdots&A&X&& \\
&&&Z&I&Z \\ \hline
X&\cdots&A&Y&I&Z
}
\hspace{15pt}
\ba{ccccccc}{
&X&\cdots&A&Y&I&Z\\
X&Z&&&&& \\ \hline
X&*&\cdots&A&Y&I&Z
},
}
while $X\cdots AYIZ$ has already been shown to have zero coefficient by \lref{rank1-B1-k-ZW-zero}.

The same argument holds for $A=I$.
}
\eq{
\ba{cccccc}{
&Z&\cdots &Z&X \\
Z&X&&& \\ \hline
Z&Y&\cdots &Z&X
}
\hspace{15pt}
\ba{cccccc}{
Z&Y&\cdots &Z&Z \\
&&&&Y \\ \hline
Z&Y&\cdots &Z&X
},
}
which leads to
\eq{
J^1_{XZ}q_{Z\cdots ZX}-h_Yq_{ZY\cdots ZZ}=0.
}
Since $k$-support operator $ZY\cdots ZZ$ takes the form of $Z\cdots Z$, we can continue our procedure.

\bigskip

\underline{Case of $Z\cdots IX$}

We notice that $k-1$-support operator $Z\cdots IX$ forms a unique pair with $k$-support operator $ZYY\cdots IZ$ as
\eq{
\ba{cccccc}{
&Z&\cdots &I&X \\
Z&X&&& \\ \hline
Z&Y&\cdots &I&X
}
\hspace{15pt}
\ba{cccccc}{
Z&Y&\cdots &I&Z \\
&&&&Y \\ \hline
Z&Y&\cdots &I&X
},
}
which leads to
\eq{
J^1_{XZ}q_{Z\cdots IX}-h_Yq_{ZY\cdots IZ}=0.
}
Since $k$-support operator $ZY\cdots IZ$ takes the form of $Z\cdots Z$, we can continue our procedure.

\subsubsection{Possible form of $k$-support operator}\lb{s:rank1-B1-kform-fin}

Through the presented procedure, an operator is connected to another operator, and we continue our procedure.
However, in some cases, this procedure does not provide a new operator but the original operator itself, and we can get no further information.
Before specification of the form of $k$-support operator, we here briefly comment on this situation.

One case is $k$-support operator $ZY^{k-2}Z$.
The procedure for $Z\cdots YZ$ provides the same operator.
However, by taking the coefficient and sign into account we find that the analysis tells
\eq{
2J^1_{XZ}q_{ZY^{k-2}Z}=0,
}
implying that $ZY^{k-2}Z$ has zero coefficient.

Another case is $k-1$-support operator $ZY^{k-3}X$.
The procedure for $Z\cdots X$ provides the same $k-1$-support operator $ZY^{k-3}X$ and $k$-support operator $ZY^{k-2}Z$, whose coefficient is shown to be zero in the above argument.
We, however, notice that $ZY^{k-3}X$ is what we reduce all operators to.
Thus, this circulation is harmless.

\bigskip

Through the above arguments, all the $k$-support operators are reduced to $k-1$-support operator $ZY^{k-3}X$, or equivalently, $k$-support operator $ZY^{k-3}IZ$, which is obtained by adding $ZIZ$ to the right end of  $ZY^{k-3}X$ and removing $ZX$ from left.
Now we provide the expression of coefficients of these $k$-support operators.

To treat the bifurcation transparently, we introduce the following symbols which have a similar role to a doubling operator:
\eq{
\lXi:=ZX, \hspace{15pt}
\rXi:=XZ, \hspace{15pt}
\lOmega:=\ba{ccc}{Z&I&Z \\ &Z&X \\ \hline \hline}, \hspace{15pt}
\rOmega:=\ba{ccc}{X&Z \\ Z&I&Z \\ \hline \hline}.
}
Aligned symbols (operators) represent the product of these operators with proper shift, as doubling-product operators and extended-doubling-product operators.
For example, $\tz \rXi Y\lOmega \lXi \lXi\tz$ represents the following operator
\eqa{
\tz \rXi Y\lOmega \lXi \lXi\tz=
\ba{cccccccccc}{
Z&I&Z&&&&&&& \\
&&X&Z&&&&&& \\
&&&Y&&&&&& \\
&&&Z&I&Z&&&& \\
&&&&Z&X&&&& \\
&&&&&Z&X&&& \\
&&&&&&Z&X&& \\
&&&&&&&Z&I&Z \\ \hline \hline
Z&I&Y&X&Z&X&Y&Y&I&Z
}.
}{rank1-XZ-column-ex}

The series of arguments presented above suggest that all the removed and added operators are expressed by a sequence of $\{\tz, \lXi, \rXi, \lOmega, \rOmega, X, Y\}$ (see also Table.~\ref{table}).
In addition, a possible sequence should satisfy several conditions.

\bdf{
Consider an operator expressed in the form of $\cdots \tz \cdots \tz\cdots $, where the second $\cdots$ does not contain $\tz$.
Then, the sequence between two $\tz$ should satisfy the following  {\it consistent condition}:
\bi{
\item Between two $\tz$'s, a single $X$ or $Y$ appears, which we denote by $W$.
\item Only $\rXi$ and $\rOmega$ appear between left $\tz$ and $W$,  and only $\lXi$ and $\lOmega$ appear between $W$ and right $\tz$.
\item Between the left end and the leftmost $\tz$, only $\lXi$ and $\lOmega$ appear. Between the right end and the rightmost $\tz$, only $\rXi$ and $\rOmega$ appear.
}
}

\begin{table}[t]
\centering
\begin{tabular}{c|c||c|c}
support & operator & removed symbol & next form \\ \hline \hline
$k$ & $Z\cdots YZ$ & $\rXi$ & $Z\cdots Z$ ($k$) \\ \hline
$k$ & $Z\cdots ZZ$ & $\rOmega$ & $Z\cdots Z$ ($k$) \\ \hline
$k$ & $Z\cdots IZ$ & $\tz$ & $Z\cdots X$  ($k-1$) or $Z\cdots Y$  ($k-1$) \\ \hline \hline
$k-1$ & $Z\cdots YY$ & $X$ & $Z\cdots Z$ ($k$) \\ \hline
$k-1$ & $Z\cdots ZY$(1) & $\lOmega$ & $Z\cdots X$  ($k-1$) \\ \hline
\multirow{2}{*}{$k-1$} & \multirow{2}{*}{$Z\cdots ZY$(2)} & $\lOmega$ & $Z\cdots Y$  ($k-1$)  \\
&& $X$ & $Z\cdots ZZ$ ($k$) \\ \hline
$k-1$ & $Z\cdots IY$ & $X$ & $Z\cdots IZ$ ($k$) \\ \hline \hline
$k-1$ & $Z\cdots XX$ & $\lXi$ & $Z\cdots Y$ ($k-1$) \\ \hline
\multirow{2}{*}{$k-1$} & \multirow{2}{*}{$Z\cdots YX$} & $\lXi$ & $Z\cdots X$  ($k-1$)  \\
&& $Y$ & $Z\cdots YZ$ ($k$) \\ \hline
$k-1$ & $Z\cdots ZX$ & $Y$ & $Z\cdots ZZ$ ($k$) \\ \hline
$k-1$ & $Z\cdots IX$ & $Y$ & $Z\cdots IZ$ ($k$) \\ \hline
\end{tabular}
\caption{We list our procedure for removing operators from the right end in terms of symbols $\Xi$ and $\Omega$.
Here, $Z\cdots ZY$(1) and $Z\cdots ZY$(2) represent two possible cases of $Z\cdots ZY$.}
\lb{table}
\end{table}

This means that the sequence takes the form of
\eqa{
\{ \lOmega , \lXi\} ^* \tz \{ \rOmega , \rXi\} ^* W \{ \lOmega , \lXi\} ^* \tz \{ \rOmega , \rXi\} ^* W \cdots W \{ \lOmega , \lXi\} ^* \tz \{ \rOmega , \rXi\} ^*,
}{rank1-B1-k-consistent}
where $\{ A, B\} ^*$ is a sequence of $A$ and $B$ with arbitrary length.
For example, $\lOmega\lXi\lOmega \tz Y\tz \lXi\lXi\lXi\lOmega\lXi X\rOmega \tz $ satisfies the consistent condition.
In contrast, $\lXi \tz \rXi \lOmega \tz \rXi$ does not satisfy the consistent condition, since there is no $X$ or $Y$ between two $\tz$.
Similarly, $\tz \rXi \lXi \rOmega X \lOmega \tz$ does not satisfy the consistent condition, since there is $\lXi$ between left $\tz$ and $X$.

We note that the expression of operators has some ambiguity owing to the aforementioned bifurcation.
For $W=Y$, the position of $Y$ in the sequence of $\Xi$'s is not fixed.
For example, the three sequences of symbols, $\tz \lOmega Y \rXi \rXi\tz$, $\tz \lOmega \lXi Y \rXi\tz$, $\tz \lOmega \lXi \lXi Y \tz$, represent the same operator:
\eqa{
\ba{ccccccccc}{
Z&I&Z&&&&&& \\
&&X&Z&&&&& \\
&&Z&I&Z&&&& \\
&&&&Y&&&& \\
&&&&Z&X&&& \\
&&&&&Z&X&& \\
&&&&&&Z&I&Z \\ \hline \hline
Z&I&X&Z&Y&Y&Y&I&Z
}
\hspace{15pt}
\ba{ccccccccc}{
Z&I&Z&&&&&& \\
&&X&Z&&&&& \\
&&Z&I&Z&&&& \\
&&&&X&Z&&& \\
&&&&&Y&&& \\
&&&&&Z&X&& \\
&&&&&&Z&I&Z \\ \hline \hline
Z&I&X&Z&Y&Y&Y&I&Z
}
\hspace{15pt}
\ba{ccccccccc}{
Z&I&Z&&&&&& \\
&&X&Z&&&&& \\
&&Z&I&Z&&&& \\
&&&&X&Z&&& \\
&&&&&X&Z&& \\
&&&&&&Y&& \\
&&&&&&Z&I&Z \\ \hline \hline
Z&I&X&Z&Y&Y&Y&I&Z
}.
}{rank1-XZ-ambiguity-ex}
For $W=X$, the position of $X$ in the sequence of $\Omega$'s is not fixed in a similar manner to above.

In spite of this ambiguity, the product of the interaction coefficients is always the same among these expressions (e.g., \eref{rank1-XZ-ambiguity-ex}).
In addition, the signs in multiplying these operators are also the same.
This fact can be confirmed by the fact that $\lXi ^l Y \rXi ^{m-l}$ provides the same products of interaction coefficients and the same sign regardless of $l$, and a similar fact holds for $\lOmega^l X\rOmega^{m-l}$.

Thus, by removing these symbols from right and adding $\lXi$ or $\lOmega$ (we choose a proper one\fn{
Note that if we add only $\lXi$ to the left, then in some cases we cannot derive a contradiction although it should have a zero coefficient.
An example is $\lXi\lXi\cdots \lXi Y \rXi\rXi\cdots \rXi=ZYY\cdots YYZ$, where removal of $\rXi$ and addition of $\lXi$ do not change the operator. 
To clarify the contradiction, we add $\lOmega$ to the left end, which results in $\lOmega \lXi\cdots \lXi Y \rXi\rXi\cdots \rXi\lr \cdots \lr \lOmega \lOmega \cdots \lOmega Y\rXi$.
The last operator $\lOmega \lOmega \cdots\lOmega Y\rXi=ZZXZX\cdots ZXZXZ$ does not form a pair since we cannot remove both $XZ$ and $ZIZ$ from the right end, which means that this operator has zero coefficient.
}) 
repeatedly, we find that a $k$-support operator which may have a nonzero coefficient is expressed in the form of a sequence of $\{\tz, \lXi, \rXi, \lOmega, \rOmega, X, Y\}$ with the consistent condition.
We also find that any $k$-support operator which may have a nonzero coefficient is connected to\fn{
We note that a sequence which may have a nonzero coefficient should accompany at least a single $\tz$ for the following reason.
If there is no $\tz$ in $\bsA$, in our procedure of removing operators from the right, only $Z\cdots YZ$ and $Z\cdots ZZ$ appear, implying that $\bsA$ is finally reduced to $ZYYY\cdots YYZ$.
However, this operator has already been shown to have a zero coefficient at the beginning of this subsection.
}
$\tz\rXi\rXi\cdots \rXi=ZIYYY\cdots YYZ$ and $\lXi\lXi\cdots \lXi\tz=ZYY\cdots YIZ$.
Summarizing these observations, we arrive at the following result:

\blm{\lb{l:rank1-B1-kform}
Consider a Hamiltonian \eqref{rank1-B1} with nonzero $J_Z^2$ and $J^1_{XZ}$.
In a candidate of a $k$-support conserved quantity $Q$, a $k$-support operator $\bsA$ may have nonzero coefficient only if $\bsA$ is obtained by a product of a sequence of $\{\tz, \lXi, \rXi, \lOmega, \rOmega, X, Y\}$ satisfying the consistent condition (i.e., satisfying the form of \eref{rank1-B1-k-consistent}).

In addition, let $M$ be the number of $\tz$ in this sequence, and $W_i$ ($1\leq i\leq M-1$) be the unique $W$ between $i$-th and $i+1$-th $\tz$.
We denote by $N_i$ the total number of $\lXi$ and $\rXi$ (resp. $\lOmega$ and $\rOmega$) touching $W_i$ in the case of $W_i=Y$ (resp. $W_i=X$).
We express $\bsA$ by a sequential product of $\Psi_j\in \{\tz, \lXi, \rXi, \lOmega, \rOmega\}$ and $W_i\in \{X,Y\}$ ($1\leq i\leq M-1$).
Then, the coefficient of $\bsA$ is computed as
\eq{
q_{\bsA}=\sigma c_{XZ}^{1,k}\prod_j J_{\Psi_j} \prod_i h_{W_i},
}
where we defined $J_{\tz}:=J^2_Z$, $J_{\Xi}:=J^1_{XZ}$, and $J_{\Omega}:=J^2_Z J^1_{XZ}$.
The sign $\sigma=\pm1$ is properly chosen.
}

The value of $\sigma$ is specified as follows:
We first write the sequence of symbols in the column expression as \eref{rank1-XZ-column-ex}.
We see operators in this expression from top to bottom.
Then, all operators except for the top one have a single overlap site to the product of the above operators.
In \eref{rank1-XZ-column-ex}, for example, $XZ$ in the second row has overlap on $X$ with the above $ZIZ$, and $ZIZ$ in the fourth line has overlap on the first $Z$ with $\abs{Z\cdot Y}=X$.
We multiply $\sigma$(above operator, present operator) on all overlaps, which is the desired $\sigma$.
For example, the sign of \eref{rank1-XZ-column-ex} is computed as
\eq{
\sigma=\sigma(Z,X)\sigma(Z,Y)\sigma(\abs{Z\cdot Y},Z)\sigma(Z,X)\sigma(\abs{Z\cdot X},Z)\sigma(X,Z)\sigma(X,Z)=+1.
}

In the derivation of \lref{rank1-B1-kform}, we have not shown that the coefficient $c_{XZ}^{1,k}$ are the same for all operators (i.e., the coefficient might be position-dependent).
The commonness of coefficients is confirmed in the next subsection.

\subsection{Commonness of coefficients}

We here demonstrate the commonness of coefficients.
If one of $h_X$ or $h_Y$ is nonzero, then the coefficient $c_{XZ}^{1,k}$ does not separate into sectors since we can add and remove $\tz$ and choose the position of $\tz$ properly with the help of a single $X$ or $Y$.
An example is
\balign{
&(\tz\rXi^m)_1\lr \cdots \lr (\lXi^{l} \tz \rXi^{m-l})_{L-l+1} \lr (\lXi^{l-1} \tz \rXi^{m-l} X)_{L-l+2} \lr (\lXi^{l-2} \tz \rXi^{m-l} X\tz)_{L-l+2} \lr \cdots  \nt \\
&\cdots \lr ( \tz \rXi^{m-l} X\tz \rXi^{l-2})_{1} \lr (\rXi^{m-l} X\tz \rXi^{l-1})_{3} \lr \cdots \lr (X\tz \rXi^{m-1})_{m-l+3} \lr (\tz \rXi^{m})_{m-l+3}.
}

On the other hand, if both $h_X$ and $h_Y$ are zero, then a $k$-support operator which may have a nonzero coefficient contains a single $\tz$, and the position of $\tz$ cannot move by adding and removing other operators ($\Xi$ and $\Omega$).
In this case, we need to construct a sequence connecting two operators whose $\tz$'s sit on different sites.

We take $(\lXi\lXi\tz)_1=(ZYYIZ)_1$ with $k=5$ as an example.
We first consider commutators generating $k+1$-support operator $(ZYYYIZ)_L$ as
\fn{
Here we do not consider the contribution of
\eq{
\ba{cccccc}{
&X&Y&Y&I&Z \\
Z&Z&&&& \\ \hline
Z&Y&Y&Y&I&Z
},
}
since this $k$-support operator does not satisfy the form of \lref{rank1-B1-kform}.
}
\eqa{
\ba{cccccc}{
&Z_1&Y&Y&I&Z \\
Z&X&&&& \\ \hline
Z&Y&Y&Y&I&Z
}
\hspace{15pt}
\ba{cccccc}{
Z_L&Y&Y&X&&\\
&&&Z&I&Z \\ \hline
Z&Y&Y&Y&I&Z
}.
}{rank1-XZ-common-mid1}

We next consider commutators generating $k$-support operator $(ZYYYX)_L$ as
\eq{
\ba{ccccc}{
Z_L&Y&Y&X& \\
&&&Z&X \\ \hline
Z&Y&Y&Y&X
}
\hspace{15pt}
\ba{ccccc}{
&Z_1&Y&Y&X \\
Z&X&&& \\ \hline
Z&Y&Y&Y&X
}.
}
Here we need not consider the contributions of
\eq{
\hspace{15pt}
\ba{ccccc}{
&X&Y&Y&X \\
Z&Z&&& \\ \hline
Z&Y&Y&Y&X
}
\hspace{15pt}
\ba{ccccc}{
&Y&X&Y&X \\
Z&I&Z&& \\ \hline
Z&Y&Y&Y&X
}.
}
The former $k-1$-support operator $XYYX$ forms a pair with $XYYIZ$ as
\eq{
\ba{cccccc}{
X&Y&Y&X&& \\
&&&Z&I&Z \\ \hline
X&Y&Y&Y&I&Z
}
\hspace{15pt}
\ba{cccccc}{
&X&Y&Y&I&Z \\
X&Z&&&& \\ \hline
X&Y&Y&Y&I&Z
},
}
while $k$-support operator $XYYIZ$ does not satisfy the form of \lref{rank1-B1-kform}, implying zero coefficient.
The latter $k-1$-support operator $YXYX$ is the unique operator generating $k+1$-support operator $YXYYIZ$ as
\eq{
\ba{cccccc}{
Y&X&Y&X&& \\
&&&Z&I&Z \\ \hline
Y&X&Y&Y&I&Z
},
}
which implies a zero coefficient.

We finally consider commutators generating $k+1$-support operator $(ZYYYIZ)_1$ as
\eq{
\ba{cccccc}{
&Z_2&Y&Y&I&Z \\
Z&X&&&& \\ \hline
Z&Y&Y&Y&I&Z
}
\hspace{15pt}
\ba{cccccc}{
Z_1&Y&Y&X&&\\
&&&Z&I&Z \\ \hline
Z&Y&Y&Y&I&Z
},
}
which is a one-site shift of \eref{rank1-XZ-common-mid1}.

Combining these relations, we have a sequence of pairs as
\eq{
(ZYYIZ)_1\lr (ZYYX)_L\lr (ZYYX)_1\lr (ZYYIZ)_2,
}
which clearly shows the one-site shift of the operator $ZYYIZ$.
Applying this shift process repeatedly, we can move the position of $\tz$ to any place, which confirms the commonness of coefficient $c_{XZ}^{1,k}$.

Similar arguments as above work for general $k$, leading to the following lemma.

\blm{
Consider a Hamiltonian \eqref{rank1-B1} with nonzero $J_Z^2$ and $J^1_{XZ}$.
The coefficient $c_{XZ}^{1,k}$ in \lref{rank1-B1-kform} is indeed position independent.
}

\subsection{Integrable case: $J^1_{ZZ}=h_X=h_Z=0$}

We notice that a Hamiltonian \eqref{rank1-B1} with nonzero is integrable if $J^1_{ZZ}=h_X=h_Z=0$ is satisfied ($J_Z^2$, $J^1_{XZ}$, and $h_Y$ can take a nonzero value).
The Hamiltonian reads
\eqa{
H=\sum_i \[ J^2_Z Z_iZ_{i+2}+ J^1_{XZ}(X_iZ_{i+1}+Z_iX_{i+1}) +h_Y Y_i\] .
}{rank1-B1-int-H}
This Hamiltonian can be mapped onto a well-known integrable model, XYZ model, through a modified Kramers-Wannier transformation~\cite{Katsura}.

Precisely, we apply the following (nonlocal) transformation, which keeps the rule of the Pauli products
\fn{
At the boundary, the simple periodic boundary condition is not kept after the transformation: $X_1\neq X_{L+1}$ and $Z_{2}\neq Z_{L+2}$.
However, its product still satisfies the periodic boundary condition $X_1Z_2=X_{L+1}Z_{L+2}$, and thus we safely transform our Hamiltonian $H$ to $H'$.
}:
\balign{
X_i &\to Z_1Z_2\cdots Z_{i-1}Y_iX_{i+1}, \\
Y_i &\to X_iX_{i+1}, \\
Z_i &\to Z_1Z_2\cdots Z_{i-1}Z_i.
}
Under this transformation, the Hamiltonian \eqref{rank1-B1-int-H} is mapped onto
\eqa{
H'=\sum_i \[ J^2_Z Z_iZ_{i+1}+ J^1_{XZ}(Y_iX_{i+1}+X_iY_{i+1}) +h_Y X_iX_{i+1}\] .
}{rank1-B1-int-H-mapped}
A proper global spin rotation diagonalizes this Hamiltonian, which reads the XYZ model.

\bthm{\lb{t:rank1-B1-int}
Consider a Hamiltonian \eqref{rank1-B1}.
This Hamiltonian is integrable if $J^1_{ZZ}=h_X=h_Z=0$ is satisfied.
}

We note that the four interaction terms in the original Hamiltonian \eqref{rank1-B1-int-H} and the four terms in its mapped one \eqref{rank1-B1-int-H-mapped} have one-to-one correspondence, which implies that the local conserved quantities in the mapped system (XYZ model) is also local conserved quantities in the original system.
In fact, we have the 4-local conserved quantity of \eref{rank1-B1-int-H}, for example, expressed as
\balign{
Q^4=\sum_i &\[ J^2_ZJ^1_{XZ}(Z_iY_{i+2}Z_{i+3}-Z_iY_{i+1}Z_{i+3}+Z_iZ_{i+1}Y_{i+2}-Y_iZ_{i+1}Z_{i+2}) \right. \nt \\
&\left.+(J^1_{XZ})^2(X_iY_{i+1}Z_{i+2}-Z_iY_{i+1}X_{i+2})+J^2_Zh_Y(X_iZ_{i+2}-Z_iX_{i+2})\] .
}

\subsection{Demonstrating that the remaining $k$-support operators have zero coefficients}
\subsubsection{Case with $J^1_{ZZ}\neq 0$}\lb{s:rank1-B1-ZZ}

We here consider a Hamiltonian \eqref{rank1-B1} with nonzero $J_Z^2$, $J^1_{XZ}$, and $J^1_{ZZ}$, where all magnetic field terms, $h_X$, $h_Y$, and $h_Z$ can take nonzero values.
Since all the remaining coefficients employ the common factor $c_{XZ}^{1,k}$, it suffices to prove that $k$-support operator $\tz\rXi\cdots \rXi =ZIY^{k-3}Z$ has zero coefficient.

We first consider commutators generating $k$-support operator $ZZXY^{k-4}Z$ as\fn{
Here we do not consider the contributions of
\eq{
\ba{ccccccc}{
Z&Z&X&Y^{k-5}&X&Y& \\
&&&&Z&I&Z \\ \hline
Z&Z&X&Y^{k-6}&Y&Y&Z
}
\hspace{15pt}
\ba{cccccc}{
&Y&X&Y^{k-5}&Y&Z \\
Z&X&&&& \\ \hline
Z&Z&X&Y^{k-5}&Y&Z
}
}
since these $k-1$-support operators have already been shown to have zero coefficients in \lref{rank1-B1-k-1-ZWXY-zero}.
}
\eqa{
\ba{cccccc}{
Z&I&Y&Y^{k-5}&Y&Z \\
&Z&Z&&& \\ \hline
Z&Z&X&Y^{k-5}&Y&Z
}
\hspace{15pt}
\ba{cccccc}{
Z&Z&X&Y^{k-5}&Z& \\
&&&&X&Z \\ \hline
Z&Z&X&Y^{k-5}&Y&Z
}
\hspace{15pt}
\ba{cccccc}{
&Z&Y&Y^{k-5}&Y&Z \\
Z&I&Z&&& \\ \hline
Z&Z&X&Y^{k-5}&Y&Z
},
}{rank1-B1-ZZ-com1}
which leads
to
\eqa{
J^1_{ZZ}q_{ZIY^{k-3}Z}+J^1_{XZ}q_{ZZXY^{k-5}Z}+J^2_Zq_{ZY^{k-3}Z}=0.
}{rank1-B1-ZZ-mid1}

We first treat the last $k-1$-support operator $ZY^{k-3}Z$, by considering commutators generating $k$-support operator $ZY^{k-2}Z$ as
\fn{
Here we need not consider the contributions of
\eq{
\ba{ccccccc}{
&Y&X&Y^{k-5}&Y&Y&Z \\
Z&I&Z&&&& \\ \hline
Z&Y&Y&Y^{k-5}&Y&Y&Z
}
\hspace{15pt}
\ba{ccccccc}{
Z&Y&Y&Y^{k-5}&X&Y& \\
&&&&Z&I&Z \\ \hline
Z&Y&Y&Y^{k-5}&Y&Y&Z
},
}
since these $k-1$-support operators have already been shown to have zero coefficients in  \lref{rank1-B1-k-1-ZWXY-zero}.
}
\eq{
\ba{ccccccc}{
&Z&Y&Y^{k-6}&Y&Y&Z \\
Z&X&&&&& \\ \hline
Z&Y&Y&Y^{k-6}&Y&Y&Z
}
\hspace{15pt}
\ba{ccccccc}{
Z&Y&Y&Y^{k-6}&Y&Z& \\
&&&&&X&Z \\ \hline
Z&Y&Y&Y^{k-6}&Y&Y&Z
}
\hspace{15pt}
\ba{ccccccc}{
&X&Y&Y^{k-6}&Y&Y&Z \\
Z&Z&&&&& \\ \hline
Z&Y&Y&Y^{k-6}&Y&Y&Z
}
\hspace{15pt}
\ba{ccccccc}{
Z&Y&Y&Y^{k-6}&Y&X& \\
&&&&&Z&Z \\ \hline
Z&Y&Y&Y^{k-6}&Y&Y&Z
},
}
which leads to
\eq{
2J^1_{XZ}q_{ZY^{k-3}Z}-J^1_{ZZ}q_{XY^{k-3}Z}-J^1_{ZZ}q_{ZY^{k-3}X}=0.
}
Both $XY^{k-3}Z$ and $ZY^{k-3}X$ form a pair with $k$-support operators $ZIY^{k-3}Z$ and $ZY^{k-3}IZ$ respectively as
\eq{
\ba{cccccc}{
&&X&Y^{k-4}&Y&Z \\
Z&I&Z&&& \\ \hline
Z&I&Y&Y^{k-4}&Y&Z
}
\hspace{15pt}
\ba{cccccc}{
Z&I&Y&Y^{k-4}&Z& \\
&&&&X&Z \\ \hline
Z&I&Y&Y^{k-4}&Y&Z
}
}
and
\eq{
\ba{cccccc}{
Z&Y&Y^{k-4}&X&& \\
&&&Z&I&Z \\ \hline
Z&Y&Y^{k-4}&Y&I&Z
}
\hspace{15pt}
\ba{cccccc}{
&Z&Y^{k-4}&Y&I&Z \\
Z&X&&&& \\ \hline
Z&Y&Y^{k-4}&Y&I&Z
},
}
which respectively lead to
\eq{
-J^2_Zq_{XY^{k-3}Z}+J^1_{XZ}q_{ZIY^{k-3}Z}=0
}
and
\eq{
-J^2_Zq_{ZY^{k-3}X}+J^1_{XZ}q_{ZY^{k-3}IZ}=0.
}
Since \lref{rank1-B1-kform} suggests $q_{ZY^{k-3}IZ}=(-1)^{k+1}q_{ZIY^{k-3}Z}$, we find
\eqa{
q_{ZY^{k-3}Z}=\bcases{
\frac{J^1_{ZZ}}{J^2_Z}q_{ZIY^{k-3}Z} & k \ {\rm is \ odd} \\
0& k \ {\rm is \ even}
}.
}{rank1-B1-ZZ-mid2}

We next treat the middle $k-1$-support operator in \eref{rank1-B1-ZZ-com1}, $ZZXY^{k-5}Z$.
To this end, we examine a more general relation.
We consider commutators generating $k$-support operator $ZY^{n+1}ZXY^{k-5-n}Z$ as\fn{
Here we need not consider the contribution of
\eq{
\ba{ccccccccc}{
&Y&X&Y^{n-1}&Z&X&Y^{k-6-n}&Y&Z \\
Z&I&Z&&&&& \\ \hline
Z&Y&Y&Y^{n-1}&Z&X&Y^{k-6-n}&Y&Z 
},
}
since this $k-1$-support operator has already been shown to have zero coefficient in \lref{rank1-B1-k-1-ZWXY-zero}.

We also do not consider the contribution of
\eq{
\ba{cccccccc}{
&X&Y^{n}&Z&X&Y^{k-6-n}&Y&Z \\
Z&Z&&&&&& \\ \hline
Z&Y&Y^{n}&Z&X&Y^{k-6-n}&Y&Z 
},
}
since this $k-1$-support operator $XY^{n}ZXY^{k-5-n}Z$ is shown to have zero coefficient as follows:
This operator forms a pair with $k$-support operator  $ZIY^{n+1}ZXY^{k-6-n}Z$ as
\eq{
\ba{cccccc}{
&&X&Y^{n}ZXY^{k-6-n}&Y&Z \\
Z&I&Z&&& \\ \hline
Z&I&Y&Y^{n}ZXY^{k-6-n}&Y&Z
}
\hspace{15pt}
\ba{cccccc}{
Z&I&Y&Y^{n}ZXY^{k-6-n}&Z& \\
&&&&X&Z \\ \hline
Z&I&Y&Y^{n}ZXY^{k-6-n}&Y&Z
}.
}
However, this $k$-support operator $ZIY^{n+1}ZXY^{k-6-n}Z$ does not satisfy the form in \lref{rank1-B1-kform} and thus has zero coefficient.
}
\balign{
\ba{cccccccc}{
&Z&Y^{n}&Z&X&Y^{k-6-n}&Y&Z \\
Z&X&&&&&& \\ \hline
Z&Y&Y^{n}&Z&X&Y^{k-6-n}&Y&Z 
}
&\hspace{15pt}
\ba{cccccccc}{
Z&Y&Y^{n}&Z&X&Y^{k-6-n}&Z&  \\
&&&&&&X&Z \\ \hline
Z&Y&Y^{n}&Z&X&Y^{k-6-n}&Y&Z 
} \nt \\
\ba{cccccccc}{
Z&Y&Y^{n}&I&Y&Y^{k-6-n}&Y&Z  \\
&&&Z&Z&&& \\ \hline
Z&Y&Y^{n}&Z&X&Y^{k-6-n}&Y&Z 
}
&\hspace{15pt}
\ba{cccccccc}{
Z&Y&Y^{n}&Z&X&Y^{k-6-n}&X&  \\
&&&&&&Z&Z \\ \hline
Z&Y&Y^{n}&Z&X&Y^{k-6-n}&Y&Z 
}, \lb{rank1-B1-ZZ-com2}
}
which leads to
\eqa{
J^1_{XZ}q_{ZY^{n}ZXY^{k-5-n}Z}+J^1_{XZ}q_{ZY^{n+1}ZXY^{k-6-n}Z}+J^1_{ZZ}q_{ZY^{n+1}IY^{k-4-n}Z}-J^1_{ZZ}q_{ZY^{n+1}ZXY^{k-6-n}X}=0.
}{rank1-B1-ZZ-mid3}

Here, the last $k-1$-support operator $ZY^{n+1}ZXY^{k-6-n}X$ forms a pair with $k$-support operator $ZY^{n}ZXY^{k-5-n}IZ$ as
\eqa{
\ba{cccccc}{
Z&Y&Y^{n}ZXY^{k-6-n}&X&& \\
&&&Z&I&Z \\ \hline
Z&Y&Y^{n}ZXY^{k-6-n}&Y&I&Z
}
\hspace{15pt}
\ba{cccccc}{
&Z&Y^{n}ZXY^{k-6-n}&Y&I&Z \\
Z&X&&&& \\ \hline
Z&Y&Y^{n}ZXY^{k-6-n}&Y&I&Z
},
}{rank1-B1-ZZ-com3}
which leads to
\eqa{
-J^2_Z q_{ZY^{n+1}ZXY^{k-6-n}X}-J^1_{XZ}q_{ZY^nZXY^{k-5-n}IZ}=0.
}{rank1-B1-ZZ-mid4}
This $k$-support operator $ZY^{n}ZXY^{k-5-n}IZ$ satisfies the form in \lref{rank1-B1-kform}.
Plugging \eref{rank1-B1-ZZ-mid4} into \eref{rank1-B1-ZZ-mid3}, we have
\eqa{
(-1)^n \( q_{ZY^{n}ZXY^{k-5-n}Z}+q_{ZY^{n+1}ZXY^{k-6-n}Z}+\frac{J^1_{ZZ}}{J^1_{XZ}}q_{ZY^{n+1}IY^{k-4-n}Z}+\frac{J^1_{ZZ}}{J^2_Z}q_{ZY^nZXY^{k-5-n}IZ}\) =0.
}{rank1-B1-ZZ-mid7}
This relation holds for $0\leq n\leq k-6$.

For $n=k-5$, the $k-1$-support operator $ZY^{k-5}ZXZ$ is connected to another $k-1$-support operator as\fn{
Here we need not consider the contribution of
\eq{
\ba{cccccc}{
&X&Y^{k-5}&Z&X&Z \\
Z&Z&&&& \\ \hline
Z&Y&Y^{k-5}&Z&X&Z
}
}
since this $k-1$-support operator has already been shown to have zero coefficient in \lref{rank1-B1-k-1-ZWXY-zero}.
}
\eq{
\ba{cccccc}{
&Z&Y^{k-5}&Z&X&Z \\
Z&X&&&& \\ \hline
Z&Y&Y^{k-5}&Z&X&Z
}
\hspace{15pt}
\ba{cccccc}{
Z&Y&Y^{k-5}&Z&Y& \\
&&&&Z&Z \\ \hline
Z&Y&Y^{k-5}&Z&X&Z
}
\hspace{15pt}
\ba{cccccc}{
Z&Y&Y^{k-5}&I&Y&Z \\
&&&Z&Z& \\ \hline
Z&Y&Y^{k-5}&Z&X&Z
},
}
which leads to
\eqa{
J^1_{XZ}q_{ZY^{k-5}ZXZ}+J^1_{ZZ}q_{ZY^{k-4}ZY}+J^1_{ZZ}q_{ZY^{k-4}IYZ}=0.
}{rank1-B1-ZZ-mid5}
The $k-1$-support operator $ZY^{k-4}ZY$ is further connected as\fn{
Here we need not consider the contributions of
\eq{
\ba{ccccc}{
&X&Y^{k-4}&Z&Y \\
Z&Z&&& \\ \hline
Z&Y&Y^{k-4}&Z&Y
}
\hspace{15pt}
\ba{cccccc}{
&Y&X&Y^{k-5}&Z&Y \\
Z&I&Z&&& \\ \hline
Z&Y&Y&Y^{k-5}&Z&Y
}
}
since these $k-1$-support operators have already shown to have zero coefficients in \lref{rank1-B1-k-1-WY-zero}.

We also need not consider the contribution of
\eq{
\ba{ccccc}{
Z&Y&Y^{k-4}&Z&Z \\
&&&&X \\ \hline
Z&Y&Y^{k-4}&Z&Y
},
}
since this $k$-support operator has already been shown to have a zero coefficient in \lref{rank1-B1-k-ZYZZ-zero}.
}
\eq{
\ba{ccccc}{
&Z&Y^{k-4}&Z&Y \\
Z&X&&& \\ \hline
Z&Y&Y^{k-4}&Z&Y
}
\hspace{15pt}
\ba{ccccc}{
Z&Y&Y^{k-4}&I&Z \\
&&&Z&X \\ \hline
Z&Y&Y^{k-4}&Z&Y
},
}
which leads to
\eqa{
J^1_{XZ}q_{ZY^{k-4}ZY}+J^1_{XZ}q_{ZY^{k-3}IZ}=0.
}{rank1-B1-ZZ-mid6}

Now we combine the obtained relations.
From \lref{rank1-B1-kform}, we notice that
\eqa{
q_{ZY^{n+1}IY^{k-4-n}Z}=(-1)^{n+1}q_{ZIY^{k-3}Z}
}{ZYIYZ-parity}
and
\eq{
q_{ZY^nZXY^{k-5-n}IZ}=q_{ZZXY^{k-5}IZ}.
}
Summing \eref{rank1-B1-ZZ-mid7} from $n=0$ to $n=k-6$, combining Eqs.~\eqref{rank1-B1-ZZ-mid5} and \eqref{rank1-B1-ZZ-mid6}, with plugging these relations, we have
\eqa{
q_{ZZXY^{k-5}Z}-(k-3)\frac{J^1_{ZZ}}{J^1_{XZ}}q_{ZIY^{k-3}Z}=0
}{rank1-B1-ZZ-mid7}
for odd $k$, and
\eqa{
q_{ZZXY^{k-5}Z}+\frac{J^1_{ZZ}}{J^2_Z}q_{ZZXY^{k-5}IZ}-(k-3)\frac{J^1_{ZZ}}{J^1_{XZ}}q_{ZIY^{k-3}Z}=0
}{rank1-B1-ZZ-mid8}
for even $k$.
In the case of even $k$, we further use the following relation suggested by \lref{rank1-B1-kform}:
\eq{
q_{ZZXY^{k-5}IZ}=-\frac{J^2_Z}{J^1_{XZ}}q_{ZIY^{k-3}Z}.
}

Finally, combining Eqs.~\eqref{rank1-B1-ZZ-mid1}, \eqref{rank1-B1-ZZ-mid2}, with \eref{rank1-B1-ZZ-mid7} (odd $k$) or \eref{rank1-B1-ZZ-mid8} (even $k$), we arrive at
\eq{
(k-1)J^1_{ZZ}q_{ZIY^{k-3}Z}=0
}
regardless of the parity of $k$.
This directly implies that the coefficient of $ZIY^{k-3}Z$ is zero, which is the desired result.

\bthm{\lb{t:rank1-B1-ZZ}
Consider a Hamiltonian \eqref{rank1-B1} with nonzero $J_Z^2$, $J^1_{XZ}$ and $J^1_{ZZ}$.
This Hamiltonian has no $k$-local conserved quantity with $4\leq k\leq L/2$.
}

\subsubsection{Case with $h_X\neq 0$ and $J^1_{ZZ}=0$}

We next consider the Hamiltonian in case B1 with $h_X\neq 0$ and $J^1_{ZZ}=0$:
\balign{
H=&\sum_i \mx{X_{i+2}&Y_{i+2}&Z_{i+2}}\mx{0 && \\ &0& \\ &&J_Z^2}\mx{X_i\\ Y_i\\ Z_i}+\sum_i \mx{X_{i+1}&Y_{i+1}&Z_{i+1}}\mx{0 &0&J_{XZ}^1 \\ 0&0&0 \\ J_{XZ}^1&0&0}\mx{X_i\\ Y_i\\ Z_i} \nt \\
&+\sum_i \mx{h_X&h_Y&h_Z}\mx{X_i\\ Y_i\\ Z_i}, \lb{rank1-B1-X-standard}
}
where $h_Y$ and $h_Z$ can take both zero and nonzero values.
Thanks to \lref{rank1-B1-kform}, all the remaining coefficients employ the common factor $c_{XZ}^{1,k}$, implying that it suffices to prove that $k$-support operator $\tz\rXi\cdots \rXi =ZIY^{k-3}Z$ has zero coefficient.

First we consider commutators generating $k$-support operator $ZIZY^{k-4}Z$ as\fn{
Here we need not consider the contribution of
\eq{
\ba{ccccccc}{
Z&I&Y&Y^{k-6}&X&Y& \\
&&&&Z&I&Z \\ \hline
Z&I&Z&Y^{k-6}&Y&Y&Z
},
}
since this $k-1$-support operator has already been shown to have zero coefficient in \lref{rank1-B1-k-1-ZWXY-zero}.
}
\eqa{
\ba{ccccccc}{
Z&I&Y&Y^{k-6}&Y&Y&Z \\
&&X&&&& \\ \hline
Z&I&Z&Y^{k-6}&Y&Y&Z
}
\hspace{15pt}
\ba{ccccccc}{
Z&I&Z&Y^{k-6}&Y&Z& \\
&&&&&X&Z \\ \hline
Z&I&Z&Y^{k-6}&Y&Y&Z
},
}{rank1-B1-X-com1}
which leads to
\eqa{
-h_X q_{ZIY^{k-3}Z}+J^1_{XZ}q_{ZIZY^{k-5}Z}=0.
}{rank1-B1-X-mid1}

We next consider commutators generating $k$-support operator $ZY^{n+1}IZY^{k-5-n}Z$ as\fn{
Here we need not consider the contributions of
\eq{
\ba{cccccccc}{
&Y&X&Y^{n-1}I&Z&Y^{k-6-n}&Y&Z \\
Z&I&Z&&&&& \\ \hline
Z&Y&Y&Y^{n-1}I&Z&Y^{k-6-n}&Y&Z
}
\hspace{15pt}
\ba{ccccccc}{
Z&Y^{n+1}I&Z&Y^{k-7-n}&X&Y& \\
&&&&Z&I&Z \\ \hline
Z&Y^{n+1}I&Z&Y^{k-7-n}&Y&Y&Z
},
}
since these $k-1$-support operators have already been shown to have zero coefficients in \lref{rank1-B1-k-1-ZWXY-zero}.
Similar arguments hold for Eqs.~\eqref{rank1-B1-X-midcom1} and \eqref{rank1-B1-X-midcom2}.
}
\eq{
\ba{ccccccc}{
&Z&Y^nI&Z&Y^{k-6-n}&Y&Z \\
Z&X&&&&& \\ \hline
Z&Y&Y^nI&Z&Y^{k-6-n}&Y&Z
}
\hspace{15pt}
\ba{ccccccc}{
Z&Y&Y^nI&Z&Y^{k-6-n}&Z& \\
&&&&&X&Z \\ \hline
Z&Y&Y^nI&Z&Y^{k-6-n}&Y&Z
}
\hspace{15pt}
\ba{ccccccc}{
Z&Y&Y^nI&Y&Y^{k-6-n}&Y&Z \\
&&&X&&& \\ \hline
Z&Y&Y^nI&Z&Y^{k-6-n}&Y&Z
},
}
which leads to
\eqa{
J^1_{XZ}q_{ZY^nIZY^{k-5-n}Z}+J^1_{XZ}q_{ZY^{n+1}IZY^{k-6-n}Z}-h_Xq_{ZY^{n+1}IY^{k-4-n}Z}=0.
}{rank1-B1-X-mid2}
This relation holds for $0\leq n\leq k-6$.

For $n=k-5$, we consider commutators generating $k$-support operator $ZY^{k-4}IZZ$ as
\eqa{
\ba{cccccc}{
&Z&Y^{k-5}&I&Z&Z \\
Z&X&&&& \\ \hline
Z&Y&Y^{k-5}&I&Z&Z
}
\hspace{15pt}
\ba{cccccc}{
Z&Y&Y^{k-5}&I&Y& \\
&&&&X&Z \\ \hline
Z&Y&Y^{k-5}&I&Z&Z
}
\hspace{15pt}
\ba{cccccc}{
Z&Y&Y^{k-5}&I&Y&Z\\
&&&&X& \\ \hline
Z&Y&Y^{k-5}&I&Z&Z
},
}{rank1-B1-X-midcom1}
which leads to
\eqa{
J^1_{XZ}q_{ZY^{k-5}IZZ}-J^1_{XZ}q_{ZY^{k-4}IY}-h_Xq_{ZY^{k-4}IYZ}=0.
}{rank1-B1-X-mid3}
Finally, we consider commutators generating $k$-support operator $ZY^{k-3}IY$ as
\eqa{
\ba{ccccc}{
&Z&Y^{k-4}&I&Y \\
Z&X&&& \\ \hline
Z&Y&Y^{k-4}&I&Y
}
\hspace{15pt}
\ba{ccccc}{
Z&Y&Y^{k-4}&I&Z \\
&&&&X \\ \hline
Z&Y&Y^{k-4}&I&Y
},
}{rank1-B1-X-midcom2}
which leads to
\eqa{
J^1_{XZ}q_{ZY^{k-4}IY}-h_Xq_{ZY^{k-3}IZ}=0.
}{rank1-B1-X-mid4}

Summing \eref{rank1-B1-X-mid2} from $n=0$ to $n=k-6$ with multiplying $(-1)^{n+1}$, and plugging Eqs.~\eqref{rank1-B1-X-mid1}, \eqref{rank1-B1-X-mid3}, and \eqref{rank1-B1-X-mid4}, we arrive at
\eq{
-h_X(k-2)q_{ZIY^{k-3}Z}=0,
}
where we used \eref{ZYIYZ-parity}.
This relation directly implies the desired relation, $q_{ZIY^{k-3}Z}=0$.

Combining our finding with \tref{rank1-B1-ZZ}, we have the following theorem:

\bthm{\lb{t:rank1-B1-X}
Consider a Hamiltonian \eqref{rank1-B1} with nonzero $J_Z^2$, $J^1_{XZ}$.
We assume that one of $J^1_{ZZ}$ and $h_X$ is nonzero.
Then, this Hamiltonian has no $k$-local conserved quantity with $4\leq k\leq L/2$.
}

\subsubsection{Case with $h_Z\neq 0$ and $J^1_{ZZ}=h_X=0$}

We finally consider the Hamiltonian in case B1 with $h_Z\neq 0$ and $J^1_{ZZ}=h_X=0$:
\balign{
H=&\sum_i \mx{X_{i+2}&Y_{i+2}&Z_{i+2}}\mx{0 && \\ &0& \\ &&J_Z^2}\mx{X_i\\ Y_i\\ Z_i}+\sum_i \mx{X_{i+1}&Y_{i+1}&Z_{i+1}}\mx{0 &0&J_{XZ}^1 \\ 0&0&0 \\ J_{XZ}^1&0&0}\mx{X_i\\ Y_i\\ Z_i} \nt \\
&+\sum_i \mx{0&h_Y&h_Z}\mx{X_i\\ Y_i\\ Z_i}, \lb{rank1-B1-X-standard}
}
where $h_Y$ can take both zero and nonzero values.
Thanks to \lref{rank1-B1-kform}, all the remaining coefficients employ the common factor $c_{XZ}^{1,k}$, implying that it suffices to prove that $k$-support operator $\tz\rXi\cdots \rXi =ZIY^{k-3}Z$ has zero coefficient.

First we consider commutators generating $k$-support operator $ZIXY^{k-4}Z$ as\fn{
Here we need not consider the contribution of 
\eq{
\ba{ccccccc}{
Z&I&X&Y^{k-6}&X&Y& \\
&&&&Z&I&Z \\ \hline
Z&I&X&Y^{k-6}&Y&Y&Z
},
}
since this $k-1$-support operator has already been shown to have zero coefficient in \lref{rank1-B1-k-1-ZWXY-zero}.
}
\eq{
\ba{cccccc}{
Z&I&Y&Y^{k-5}&Y&Z \\
&&Z&&& \\ \hline
Z&I&X&Y^{k-5}&Y&Z
}
\hspace{15pt}
\ba{cccccc}{
&&Y&Y^{k-5}&Y&Z \\
Z&I&Z&&& \\ \hline
Z&I&X&Y^{k-5}&Y&Z
}
\hspace{15pt}
\ba{cccccc}{
Z&I&X&Y^{k-5}&Z& \\
&&&&X&Z \\ \hline
Z&I&X&Y^{k-5}&Y&Z
},
}
which leads to
\eqa{
h_Zq_{ZIY^{k-3}Z}+J^2_Zq_{Y^{k-3}Z}+J^1_{XZ}q_{ZIXY^{k-5}Z}=0.
}{rank1-B1-Z-mid1}

We first examine the $k-2$-support operator $Y^{k-3}Z$, by considering commutators generating $k-1$-support operator $Y^{k-2}Z$ as\fn{
Here we need not consider the contribution of
\eq{
\ba{ccc}{
Y^m&X&Y^{k-3-m}Z \\
&Z& \\ \hline
Y^m&Y&Y^{k-3-m}Z
}
}
with $1\leq m\leq k-3$, since $k-1$-support operator $Y^mXY^{k-3-m}Z$ forms a pair with $k$-support operator $ZIXY^{m-1}XY^{k-3-m}Z$, which does not satisfy \lref{rank1-B1-kform} and thus has zero coefficient.

We also need not consider the contribution of
\eq{
\ba{cc}{
Y^{k-2}&X \\
&Y \\ \hline
Y^{k-2}&Z
},
}
since $k-1$-support operator $Y^{k-2}X$ has already shown to be zero in \lref{rank1-B1-k-1-WY-zero}.

Moreover, we need not consider the contribution of
\eq{
\ba{ccccc}{
Y&Y^{k-5}&X&Y& \\
&&Z&I&Z \\ \hline
Y&Y^{k-5}&Y&Y&Z
},
}
since this $k$-support operator $ZIXY^{k-5}XY$ is generated only by
\eq{
\ba{cccc}{
&&Y&Y^{k-5}XY \\
Z&I&Z& \\ \hline
Z&I&X&Y^{k-5}XY
},
}
implying $q_{ZIXY^{k-5}XY}=0$.
Note that no commutator of $k$-support operator satisfying the form in \lref{rank1-B1-kform} and 1-support operator $Y$ or $Z$ generates $ZIXY^{k-5}XY$.
}
\eqa{
\ba{ccccc}{
Y&Y^{k-5}&Y&Z& \\
&&&X&Z \\ \hline
Y&Y^{k-5}&Y&Y&Z
}
\hspace{15pt}
\ba{ccccc}{
X&Y^{k-5}&Y&Y&Z \\
Z&&&& \\ \hline
Y&Y^{k-5}&Y&Y&Z
},
}{rank1-B1-Z-com1}
which leads to
\eqa{
J^1_{XZ}q_{Y^{k-3}Z}-h_Zq_{XY^{k-3}Z}=0.
}{rank1-B1-Z-mid2pre1}
The $k-1$-support operator $XY^{k-3}Z$ forms a pair with $k$-support operator $ZIY^{k-3}Z$ as
\eq{
\ba{cccccc}{
&&X&Y^{k-4}&Y&Z \\
Z&I&Z&&& \\ \hline
Z&I&Y&Y^{k-4}&Y&Z
}
\hspace{15pt}
\ba{cccccc}{
Z&I&Y&Y^{k-4}&Z& \\
&&&&X&Z \\ \hline
Z&I&Y&Y^{k-4}&Y&Z
},
}
which leads to
\eqa{
-J^2_Zq_{XY^{k-3}Z}+J^1_{XZ}q_{ZIY^{k-3}Z}=0.
}{rank1-B1-Z-mid2pre2}
Plugging \eref{rank1-B1-Z-mid2pre2} into \eref{rank1-B1-Z-mid2pre1}, we find
\eqa{
q_{Y^{k-3}Z}=\frac{h_Z}{J^2_Z}q_{ZIY^{k-3}Z}.
}{rank1-B1-Z-mid2}

We next treat the $k-1$-support operator in \eref{rank1-B1-Z-mid1}, $ZIXY^{k-5}Z$.
To this end, we consider commutators generating $k$-support operator $ZY^{n+1}IXY^{k-5-n}Z$ as\fn{
Here we need not consider the contributions of
\eq{
\ba{ccccccccc}{
&Y&X&Y^{n-1}&I&X&Y^{k-6-n}&Y&Z \\
Z&I&Z&&&&&& \\ \hline
Z&Y&Y&Y^{n-1}&I&X&Y^{k-6-n}&Y&Z
}
\hspace{15pt}
\ba{ccccccccc}{
Z&Y&Y^n&I&X&Y^{k-7-n}&X&Y& \\
&&&&&&Z&I&Z \\ \hline
Z&Y&Y^n&I&X&Y^{k-7-n}&Y&Y&Z
},
}
since these $k-1$-support operators have already been shown to have zero coefficients in \lref{rank1-B1-k-1-ZWXY-zero}.

A similar argument is employed in \eref{rank1-B1-Z-midcom4}.
}
\eq{
\ba{cccccccc}{
&Z&Y^n&I&X&Y^{k-6-n}&Y&Z \\
Z&X&&&&&& \\ \hline
Z&Y&Y^n&I&X&Y^{k-6-n}&Y&Z
}
\hspace{15pt}
\ba{cccccccc}{
Z&Y&Y^n&I&X&Y^{k-6-n}&Z& \\
&&&&&&X&Z \\ \hline
Z&Y&Y^n&I&X&Y^{k-6-n}&Y&Z
}
\hspace{15pt}
\ba{cccccccc}{
Z&Y&Y^n&I&Y&Y^{k-6-n}&Y&Z \\
&&&&Z&&& \\ \hline
Z&Y&Y^n&I&X&Y^{k-6-n}&Y&Z
},
}
which leads to
\eqa{
J^1_{XZ}q_{ZY^nIXY^{k-5-n}Z}+J^1_{XZ}q_{ZY^{n+1}IXY^{k-6-n}Z}+h_Zq_{ZY^{n+1}IY^{k-3-n}Z}=0.
}{rank1-B1-Z-mid3}
This relation holds for $0\leq n\leq k-6$.

For $n=k-5$, we consider commutators generating $k$-support operator $ZY^{k-4}IXZ$ as
\eqa{
\ba{cccccc}{
&Z&Y^{k-5}&I&X&Z \\
Z&X&&&& \\ \hline
Z&Y&Y^{k-5}&I&X&Z
}
\hspace{15pt}
\ba{cccccc}{
Z&Y&Y^{k-5}&I&Y&Z \\
&&&&Z& \\ \hline
Z&Y&Y^{k-5}&I&X&Z
},
}{rank1-B1-Z-midcom4}
which leads to
\eqa{
J^1_{XZ}q_{ZY^{k-5}IXZ}+h_Zq_{ZY^{k-4}IYZ}=0.
}{rank1-B1-Z-mid4}

Summing \eref{rank1-B1-Z-mid3} with multiplying $(-1)^n$ from $n=0$ to $n=k-6$, and plugging the sum and Eqs.~\eqref{rank1-B1-Z-mid2} and \eqref{rank1-B1-Z-mid4} into \eref{rank1-B1-Z-mid1}, we arrive at
\eq{
(k-2)h_Zq_{ZIY^{k-3}Z}=0,
}
where we used \eref{ZYIYZ-parity}.
This relation directly implies the desired relation, $q_{ZIY^{k-3}Z}=0$.

Combining our finding with \tref{rank1-B1-X}, we have the following theorem:

\bthm{\lb{t:rank1-B1}
Consider a Hamiltonian \eqref{rank1-B1} with nonzero $J_Z^2$, $J^1_{XZ}$.
We assume that one of $J^1_{ZZ}$, $h_X$, and $h_Z$ is nonzero.
Then, this Hamiltonian has no $k$-local conserved quantity with $4\leq k\leq L/2$.
}

Since a Hamiltonian \eqref{rank1-B1} with $J^1_{ZZ}=h_X=h_Z=0$ has already been shown to be integrable in \tref{rank1-B1-int}, we complete the classification of Hamiltonians in case B1.

\section{Rank 1: Case with $J^1_{XX}=J^1_{YY}=J^1_{XZ}=J^1_{YZ}=0$ and $J^1_{ZZ}\neq 0$ (Case B2)}\lb{s:rank1-B2}

We finally consider case B2, where all the matrix elements of $J^1$ except for $J^1_{ZZ}$ are zero, and $J^1_{ZZ}$ is nonzero.
In this case, by applying a proper global spin rotation in the $XY$ subspace, we can set $h_Y=0$.
Since the case with $h_X=0$ is classical and thus integrable, we here suppose $h_X\neq 0$.
The Hamiltonian considered in this section is expressed as
\eqa{
H=\sum_i J^2_Z Z_{I+2}Z_i +J^1_{ZZ}Z_{i+1}Z_i + h_XX_i +h_Z Z_i,
}{rank1-B2}
where $J^2_Z$, $J^1_{ZZ}$, and $h_X$ take nonzero values, and $h_Z$ can take both zero and nonzero values.

We note that our proof follows a similar strategy to that for the mixed-field Ising chain with nearest-neighbor interaction ($H=\sum_i J Z_{i+1}Z_i+h_X X_i+h_Z Z_i$)~\cite{Chi24-1}, though some additional cares are required in our setup.

\bigskip

We consider a candidate of $k$-local conserved quantity $Q$ and show that all the coefficients of $k$-support operators are zero.

\lref{rank1-k+2} tells that a $k$-support operator both of whose ends are not $Z$ (i.e., $W\cdots W'$ with $W,W'\in \{X,Y\}$) has zero coefficient.
For later use, we here present a similar result that a $k-1$-support operator in the form of $W\cdots W'$ ($W,W'\in \{X,Y\}$) has zero coefficient.
This claim is confirmed by the fact that $k+1$-support operator $ZI\Wc\cdots W'$ is generated only by
\eq{
\ba{ccccc}{
&&W&\cdots&W' \\
Z&I&Z&& \\ \hline
Z&I&\Wc&\cdots&W'
}.
}

\blm{\lb{l:rank1-B2-WW-k-1}
Consider a Hamiltonian \eqref{rank1-B2} with nonzero $J_Z^2$, $J^1_{ZZ}$, and $h_X$.
In a candidate of a $k$-support conserved quantity $Q$, a $k-1$-support operator  in the form of $W\cdots W'$ ($W,W'\in \{X,Y\}$) has zero coefficient.
}

From \lref{rank1-k+2}, we find that a $k$-support operator may have a nonzero coefficient only if one of its ends is $Z$.
We first treat operators in the form of $Z\cdots W$ and $W\cdots Z$ with $W\in \{X,Y\}$ in \sref{rank1-B2-WZ}, and then treat operators in the form of $Z\cdots Z$ in \sref{rank1-B2-ZZ}.

\subsection{Analysis of $W\cdots Z$}\lb{s:rank1-B2-WZ}

\subsubsection{Restricting possible forms of $k$-support operators}

In the first step, we specify the possible operator form of $W\cdots Z$ (and $Z\cdots W$) which may have a nonzero coefficient.

We first notice that an operator $W\cdots Z$ with a nonzero coefficient should take the form of $W\cdots W'IZ$ ($W'\in \{X,Y\}$).
In other words, $W\cdots ZIZ$ and $W\cdots IIZ$ have zero coefficients.
This fact can be seen by considering commutators generating $ZI\Wc \cdots AIZ$ as
\eq{
\ba{ccccccc}{
&&W&\cdots&A&I&Z \\
Z&I&Z&&&& \\ \hline
Z&I&\Wc&\cdots&A&I&Z
}
\hspace{15pt}
\ba{ccccccc}{
Z&I&\Wc&\cdots&*&& \\
&&&&Z&I&Z \\ \hline
Z&I&\Wc&\cdots&A&I&Z
}.
}
For the existence of the latter commutator, operator $A$ should be $X$ or $Y$.

We next consider commutators generating $k$-support operator $YI\Wc \cdots \Wpc$ as\fn{
Here we need not consider the contribution of
\eq{
\ba{cccc}{
Y&I&\Wc\cdots &W' \\
&&&Z \\ \hline
Y&I&\Wc\cdots &\Wpc 
},
}
since the $k$-support operator $YI\Wc \cdots W'$ has already been shown to have zero coefficient in \lref{rank1-k+2}.
}
\eq{
\ba{cccc}{
Z&I&\Wc\cdots &\Wpc \\
X&&& \\ \hline
Y&I&\Wc\cdots &\Wpc 
}
\hspace{15pt}
\ba{cccc}{
Y&I&\Wc\cdots &Z \\
&&&X \\ \hline
Y&I&\Wc\cdots &\Wpc 
}.
}
For the existence of the latter commutator, $W'$ should be $X$.
Notice that the latter $k$-support operator $YI\Wc\cdots Z$ also takes the form of $W\cdots Z$, and thus we can apply the above argument repeatedly.
Through this procedure, we specify the possible form of an operator in the form of $W\cdots Z$:

\blm{\lb{l:rank1-B2-WZ-kform}
Consider a Hamiltonian \eqref{rank1-B2} with nonzero $J_Z^2$, $J^1_{ZZ}$, and $h_X$.
In a candidate of a $k$-support conserved quantity $Q$, a $k$-support operator where one of the ends is not $Z$ may have a nonzero coefficient only if it is $Y(IX)^mIZ$ or $ZI(XI)^mY$ with $k=2m+3$.

In addition, these two coefficients are related as $q_{Y(IX)^mIZ}=-q_{ZI(XI)^mY}$.
}

Note in passing that the remaining operators in \lref{rank1-B2-WZ-kform} can be expressed as
\eq{
\ba{ccccccccccccc}{
X&&&&&&&&&&&& \\
Z&I&Z&&&&&&&&&& \\
&&X&&&&&&&&&& \\
&&Z&I&Z&&&&&&&& \\
&&&&X&&&&&&&& \\
&&&&Z&I&Z&&&&&& \\
&&&&&&&\ddots&&&&& \\
&&&&&&&&Z&I&Z&& \\
&&&&&&&&&&X&& \\
&&&&&&&&&&Z&I&Z \\ \hline
Y&I&X&I&X&I&\cdots&\cdots&\cdots&I&X&I&Z
}\hspace{15pt}
\ba{ccccccccccccc}{
Z&I&Z&&&&&&&&&& \\
&&X&&&&&&&&&& \\
&&Z&I&Z&&&&&&&& \\
&&&&X&&&&&&&& \\
&&&&Z&I&Z&&&&&& \\
&&&&&&&\ddots&&&&& \\
&&&&&&&&Z&I&Z&& \\
&&&&&&&&&&X&& \\
&&&&&&&&&&Z&I&Z \\
&&&&&&&&&&&&X \\ \hline
Z&I&X&I&X&I&\cdots&\cdots&\cdots&I&X&I&Y
},
}
which is helpful in understanding procedures in this and the next subsection.

We remark that we do not discuss the commonnness of coefficients of $q$, since our proof presented below works as the proof that $(Y(IX)^mIZ)_i$ and $(ZI(XI)^mY)_i$ have zero coefficients for all $i$.

\subsubsection{Demonstrating that the remaining $k$-support operators have zero coefficients}

Our remaining task for $W\cdots Z$ is to show that $k$-support operator $Y(IX)^mIZ$ has zero coefficient.
To this end, we first consider commutators generating $k+1$-support operator $ZX(IX)^mIZ$ as
\eq{
\ba{ccccccc}{
&Y&(IX)^{m-1}&I&X&I&Z \\
Z&Z&&&&& \\ \hline
Z&X&(IX)^{m-1}&I&X&I&Z
}
\hspace{15pt}
\ba{ccccccc}{
Z&X&(IX)^{m-1}&I&Y&& \\
&&&&Z&I&Z \\ \hline
Z&X&(IX)^{m-1}&I&X&I&Z
},
}
which leads to
\eqa{
J^1_{ZZ}q_{Y(IX)^mIZ}+J^2_Z q_{ZX(IX)^{m-1}IY}=0.
}{rank1-B2-WZ-mid1}
The $k-1$-support operator $ZX(IX)^{m-1}IY$ is connected to other operators by considering commutators generating $k$-support operator $ZZY(IX)^{m-1}IY$ as\fn{
Note that $k$-support operators one of whose end is not $Z$ should take the form shown in \lref{rank1-B2-WZ-kform}, and thus we need not consider, for example, the contributions of
\eq{
\ba{cccccc}{
Z&Z&Y&(IX)^{m-1}&I&X \\
&&&&&Z \\ \hline
Z&Z&Y&(IX)^{m-1}&I&Y
}
\hspace{15pt}
\ba{cccccc}{
Z&Z&Z&(IX)^{m-1}&I&Y \\
&&X&&& \\ \hline
Z&Z&Y&(IX)^{m-1}&I&Y
}
}
and other similar commutators between a $k$-support operator and a 1-support operator.
}
\eq{
\ba{cccccc}{
&Z&X&(IX)^{m-1}&I&Y \\
Z&I&Z&&& \\ \hline
Z&Z&Y&(IX)^{m-1}&I&Y
}
\hspace{15pt}
\ba{cccccc}{
Z&Z&Y&(IX)^{m-1}&I&Z \\
&&&&&X \\ \hline
Z&Z&Y&(IX)^{m-1}&I&Y
}
\hspace{15pt}
\ba{cccccc}{
Z&I&X&(IX)^{m-1}&I&Y \\
&Z&Z&&& \\ \hline
Z&Z&Y&(IX)^{m-1}&I&Y
},
}
which leads to
\eqa{
-J^2_Zq_{ZX(IX)^{m-1}IY}+h_Xq_{ZZY(IX)^{m-1}IZ}+J^1_{ZZ}q_{Z(IX)^mIY}=0.
}{rank1-B2-WZ-mid2}
Note that at present we have no information on $q_{ZZY(IX)^{m-1}IZ}$ since both ends of this $k$-support operator $ZZY(IX)^{m-1}IZ$ are $Z$ and therefore this operator is out of the scope of \lref{rank1-B2-WZ-kform}.

To treat $k$-support operator $ZZY(IX)^{m-1}IZ$ and similar operators, we further consider commutators generating $k$-support operator $Y(IX)^n ZY (IX)^{m-1-n}IZ$ as
\balign{
\ba{ccccccccc}{
Z&(IX)^n&Z&Y&(IX)^{m-2-n}&I&X&I&Z \\
X&&&&&&&& \\ \hline
Y&(IX)^n&Z&Y&(IX)^{m-2-n}&I&X&I&Z
}&
\hspace{15pt}
\ba{ccccccccc}{
Y&(IX)^n&Z&Y&(IX)^{m-2-n}&I&Y&& \\
&&&&&&Z&I&Z \\ \hline
Y&(IX)^n&Z&Y&(IX)^{m-2-n}&I&X&I&Z
}
\nt \\
\ba{ccccccccc}{
Y&(IX)^n&I&X&(IX)^{m-2-n}&I&X&I&Z \\
&&Z&Z&&&&& \\ \hline
Y&(IX)^n&Z&Y&(IX)^{m-2-n}&I&X&I&Z
},&
}
which leads to
\eqa{
h_Xq_{Z(IX)^nZY(IX)^{m-1-n}IZ}+J^2_Zq_{Y(IX)^nZY(IX)^{m-2-n}IY}-J^1_{ZZ}q_{Y(IX)^mIZ}=0.
}{rank1-B2-WZ-mid3}
In addition, we consider commutators generating $k$-support operator $Z(IX)^{n-1} ZY (IX)^{m-2-n}IZ$ as
\balign{
\ba{ccccccccc}{
&&Y&(IX)^n&Z&Y&(IX)^{m-2-n}&I&Y \\
Z&I&Z&&&&&& \\ \hline
Z&I&X&(IX)^n&Z&Y&(IX)^{m-2-n}&I&Y
}&
\hspace{15pt}
\ba{ccccccccc}{
Z&I&X&(IX)^n&Z&Y&(IX)^{m-2-n}&I&Z \\
&&&&&&&&X \\ \hline
Z&I&X&(IX)^n&Z&Y&(IX)^{m-2-n}&I&Y
}
\nt \\
\ba{ccccccccc}{
Z&I&X&(IX)^n&I&X&(IX)^{m-2-n}&I&Y \\
&&&&Z&Z&&& \\ \hline
Z&I&X&(IX)^n&Z&Y&(IX)^{m-2-n}&I&Y
},&
}
which leads to
\eqa{
J^2_Zq_{Y(IX)^nZY(IX)^{m-2-n}IY}+h_Xq_{Z(IX)^{n+1}ZY(IX)^{m-2-n}IZ}-J^1_{ZZ}q_{Z(IX)^mIY}=0.
}{rank1-B2-WZ-mid4}
Subtracting \eref{rank1-B2-WZ-mid4} from \eref{rank1-B2-WZ-mid3} with inserting $q_{Y(IX)^mIZ}=-q_{ZI(XI)^mY}$, we find
\eqa{
h_Xq_{Z(IX)^nZY(IX)^{m-1-n}IZ}-h_Xq_{Z(IX)^{n+1}ZY(IX)^{m-2-n}IZ}-2J^1_{ZZ}q_{Y(IX)^mIZ}=0.
}{rank1-B2-WZ-mid5}
This relation holds for $0\leq n\leq m-2$.

For $n=m-1$, we consider commutators generating $k$-support operator $Y(IX)^{m-1}ZYIZ$ as
\eq{
\ba{cccccc}{
Z&(IX)^{m-1}&Z&Y&I&Z \\
X&&&&& \\ \hline
Y&(IX)^{m-1}&Z&Y&I&Z 
}
\hspace{15pt}
\ba{cccccc}{
Y&(IX)^{m-1}&Z&X&& \\
&&&Z&I&Z \\ \hline
Y&(IX)^{m-1}&Z&Y&I&Z 
}
\hspace{15pt}
\ba{cccccc}{
Y&(IX)^{m-1}&I&X&I&Z \\
&&Z&Z&& \\ \hline
Y&(IX)^{m-1}&Z&Y&I&Z 
},
}
which leads to
\eqa{
h_Xq_{Z(IX)^{m-1}ZYIZ}-J^2_Z q_{Y(IX)^{m-1}ZX}-J^1_{ZZ}q_{Y(IX)^mIZ}=0.
}{rank1-B2-WZ-mid6}
Finally, we consider commutators generating $k$-support operator $Z(IX)^mZX$ as
\eq{
\ba{cccccc}{
&&Y&(IX)^{m-1}&Z&X \\
Z&I&Z&&& \\ \hline
Z&I&X&(IX)^{m-1}&Z&X
}
\hspace{15pt}
\ba{cccccc}{
Z&I&X&(IX)^{m-1}&I&Y \\
&&&&Z&Z \\ \hline
Z&I&X&(IX)^{m-1}&Z&X
},
}
which leads to
\eqa{
J^2_Zq_{Y(IX)^{m-1}ZX}+J^1_{ZZ}q_{Z(IX)^mIY}=0.
}{rank1-B2-WZ-mid7}

Summing \eref{rank1-B2-WZ-mid5} from $n=0$ to $n=m-2$, and plugging Eqs.~\eqref{rank1-B2-WZ-mid1}, \eqref{rank1-B2-WZ-mid2}, \eqref{rank1-B2-WZ-mid6}, and \eqref{rank1-B2-WZ-mid7}, we arrive at
\eq{
2mJ^1_{ZZ}q_{Z(IX)^mIY}=0.
}
Since $J^1_{ZZ}\neq 0$, this relation directly implies $q_{Z(IX)^mIY}=0$, which is the desired relation.

\blm{\lb{l:rank1-B2-WZ-fin}
Consider a Hamiltonian \eqref{rank1-B2} with nonzero $J_Z^2$, $J^1_{ZZ}$, and $h_X$.
In a candidate of a $k$-support conserved quantity $Q$, a $k$-support operator where one of the ends is not $Z$ has a zero coefficient.
}

\subsection{Analysis of $Z\cdots Z$}\lb{s:rank1-B2-ZZ}

The remaining operators with nonzero coefficients take the form of $Z\cdots Z$.
In the following, we shall show that these operators also have zero coefficients.
Our approach is similar to that presented in the previous subsection for $W\cdots Z$, though some additional cares are required.

\subsubsection{Restricting possible forms of $k$-support operators}

Our first goal is to specify the possible operator form of $Z\cdots Z$ which may have a nonzero coefficient.
To this end, we first consider commutators generating $k$-support operator $Y\cdots Z$ as\fn{
Here we need not consider the contribution of
\eq{
\ba{ccccc}{
Y&\cdots&*&W& \\
&&&Z&Z \\ \hline
Y&\cdots&*&*&Z
},
}
since $k-1$-support operator $Y\cdots *W$ with $W\in \{X,Y\}$ have already shown to have zero coefficient in \lref{rank1-B2-WW-k-1}.
}
\eq{
\ba{ccccc}{
Z&\cdots&*&*&Z \\
X&&&& \\ \hline
Y&\cdots&*&*&Z
}
\hspace{15pt}
\ba{ccccc}{
Y&\cdots&W&& \\
&&Z&I&Z \\ \hline
Y&\cdots&*&*&Z
}.
}
The latter $k-2$-support operator $Y\cdots W$ forms a pair with $k$-support operator $ZIX\cdots Z$ as
\eq{
\ba{ccccc}{
&&Y&\cdots&W \\
Z&I&Z&& \\ \hline
Z&I&X&\cdots&W
}
\hspace{15pt}
\ba{ccccc}{
Z&I&X&\cdots&Z \\
&&&&X \\ \hline
Z&I&X&\cdots&W
}.
}
For the existence of the latter commutator, $W=Y$ is imposed.
The $k$-support operator $ZIX\cdots Z$ takes the form of $Z\cdots Z$, and thus we can apply this argument repeatedly.
Through this, we arrive at the following fact:

\blm{\lb{l:rank1-B2-ZZ-kform}
Consider a Hamiltonian \eqref{rank1-B2} with nonzero $J_Z^2$, $J^1_{ZZ}$, and $h_X$.
In a candidate of a $k$-support conserved quantity $Q$, a $k$-support operator has a nonzero coefficient only if the operator is $Z(IX)^mIZ$ with $k=2m+3$.
}

Note in passing that the remaining operators in \lref{rank1-B2-ZZ-kform} can be expressed as
\eq{
\ba{ccccccccccccc}{
Z&I&Z&&&&&&&&&& \\
&&X&&&&&&&&&& \\
&&Z&I&Z&&&&&&&& \\
&&&&X&&&&&&&& \\
&&&&Z&I&Z&&&&&& \\
&&&&&&&\ddots&&&&& \\
&&&&&&&&Z&I&Z&& \\
&&&&&&&&&&X&& \\
&&&&&&&&&&Z&I&Z \\ \hline
Z&I&X&I&X&I&\cdots&\cdots&\cdots&I&X&I&Z
}.
}

We remark that we do not discuss the commonness of coefficients of $q$, since our proof presented below works as the proof that $(Z(IX)^mIZ)_i$ has zero coefficient for all $i$.

\subsubsection{Restricting possible forms of $k-1$-support operators}

In this subsection, we shall examine possible forms of some $k-1$-support operators.

Since a $k-1$-support operator in the form of $W\cdots W'$ ($W, W'\in \{X,Y\}$) has already been shown to have zero coefficient in \lref{rank1-B2-WW-k-1}, we here examine a $k-1$-support operator in the form of $W\cdots Z$ and $Z\cdots W$ ($W\in \{X,Y\}$).
To treat them, we consider commutators generating $k+1$-support operator $ZI\Wc\cdots Z$, where we have two cases depending on the form of the operator.

First, if a $k-1$-support operator takes the form of $W\cdots IZ$, then we have
\eq{
\ba{ccccccc}{
&&W&\cdots&W'&I&Z \\
Z&I&Z&&&& \\ \hline
Z&I&\Wc&\cdots&W'&I&Z
}
\hspace{15pt}
\ba{ccccccc}{
Z&I&\Wc&\cdots&\Wpc&& \\
&&&&Z&I&Z \\ \hline
Z&I&\Wc&\cdots&W'&I&Z
}.
}
The latter $k-1$-support operator $ZI\Wc\cdots \Wpc$ forms a pair with another $k-1$-support operator as\fn{
Here we need not consider the contribution of
\eq{
\ba{cccccc}{
Y&I&\Wc&\cdots&Z&W \\
&&&&Z&Z \\ \hline
Y&I&\Wc&\cdots&I&\Wpc
}
\hspace{15pt}
\ba{cccccc}{
Y&I&\Wc&\cdots&I&W \\
&&&&Z&Z \\ \hline
Y&I&\Wc&\cdots&Z&\Wpc
},
}
since these $k-1$-support operators have already been shown to have zero coefficients in \lref{rank1-B2-WW-k-1}.
}
\eq{
\ba{ccccc}{
Z&I&\Wc&\cdots&\Wpc \\
X&&&& \\ \hline
Y&I&\Wc&\cdots&\Wpc
}
\hspace{15pt}
\ba{ccccc}{
Y&I&\Wc&\cdots&* \\
&&&&*' \\ \hline
Y&I&\Wc&\cdots&\Wpc
}.
}
For the existence of the latter commutator, $*$ should be $Z$ due to \lref{rank1-B2-WW-k-1}, which implies $*'=X$ and $\Wpc=Y$ (i.e., $W'=X$).
Now the latter commutator $YI\Wc\cdots Z$ takes the form of $W\cdots Z$, and thus we can repeatedly apply this procedure.

Second, if a $k-1$-support operator takes the form of $W\cdots PZ$ with $P\in \{X,Y,Z\}$, then we have
\eq{
\ba{ccccccc}{
&&W&\cdots&A&P&Z \\
Z&I&Z&&&& \\ \hline
Z&I&\Wc&\cdots&A&P&Z
}
\hspace{15pt}
\ba{ccccccc}{
Z&I&\Wc&\cdots&A&P^{\rm c}& \\
&&&&&Z&Z \\ \hline
Z&I&\Wc&\cdots&A&P&Z
}
\hspace{15pt}
\ba{ccccccc}{
Z&I&\Wc&\cdots&*&P& \\
&&&&Z&I&Z \\ \hline
Z&I&\Wc&\cdots&A&P&Z
}.
}
However, we notice that both $k$-support operators in the latter two commutators cannot be the form of $Z(IX)^mIZ$ from the following observations.
In the former one, $P^{\rm c}$ cannot be $Z$.
In the latter one, $*$ cannot be $I$.
Hence, the coefficient of $k-1$-support operator $W\cdots PZ$ turns out to be zero.

From this assertion, we find that only the first case, $W\cdots IZ$, is possible, and the first procedure should be applied ad infinitum.
To realize this, the $k-1$-support operator $W\cdots Z$ should take the form of $XIXI\cdots XIZ$.
This suggests that $k-1$ is odd.
On the other hand, we have already clarified that the remaining $k$-support operator is $Z(IX)^mIZ$, and thus $k$ should be odd, which is a contradiction.
This fact means that all the $k-1$-support operators in the form of $W\cdots Z$ have zero coefficients.

\blm{\lb{l:rank1-B2-WZ-k-1}
Consider a Hamiltonian \eqref{rank1-B2} with nonzero $J_Z^2$, $J^1_{ZZ}$, and $h_X$.
In a candidate of a $k$-support conserved quantity $Q$, a $k-1$-support operator in the form of $W\cdots Z$ or $Z\cdots W$ ($W\in \{X,Y\}$) has zero coefficient.
}

\subsubsection{Demonstrating that the remaining $k$-support operators have zero coefficients}

We finally show that $k$-support operator $Z(IX)^mIZ$ has zero coefficient.
This completes the proof for case B2, and thus the proof of our main theorem, \tref{main}, is also accomplished.

\bigskip

We first observe that a $k-2$-support operator in the form of $W\cdots W'$ ($W,W'\in \{X,Y\}$) may have a nonzero coefficient only if it takes the form of $Y(IX)^{m-1}IY$.
To see this fact, we consider commutators generating $k$-support operator $ZI\Wc\cdots W'$ as
\eq{
\ba{ccccc}{
&&W&\cdots&W' \\
Z&I&Z&& \\ \hline
Z&I&\Wc&\cdots&W'
}
\hspace{15pt}
\ba{ccccc}{
Z&I&\Wc&\cdots&Z \\
&&&&X \\ \hline
Z&I&\Wc&\cdots&W'
}.
}
Here, to restrict possible commutators, we used the fact that a $k$-support operator which may have a nonzero coefficient is only $Z(IX)^mIZ$.
For the existence of the latter commutator, $W'=Y$ and $\Wc \cdots =(XI)^m$ are imposed.

\blm{\lb{l:rank1-B2-WW-k-2}
Consider a Hamiltonian \eqref{rank1-B2} with nonzero $J_Z^2$, $J^1_{ZZ}$, and $h_X$.
In a candidate of a $k$-support conserved quantity $Q$, a $k-2$-support operator in the form of $W\cdots W'$ ($W, W'\in \{X,Y\}$) may have a nonzero coefficient only if this operator takes the form of $Y(IX)^{m-1}IY$ with $k=2m+3$.
}

We start with the fact that $k$-support operator $Z(IX)^mIZ$ forms a pair with $k-2$-support operator $Y(IX)^{m-1}IY$ as
\eq{
\ba{cccccc}{
Z&(IX)^{m-1}&I&X&I&Z \\
X&&&&& \\ \hline
Y&(IX)^{m-1}&I&X&I&Z
}
\hspace{15pt}
\ba{cccccc}{
Y&(IX)^{m-1}&I&Y&& \\
&&&Z&I&Z \\ \hline
Y&(IX)^{m-1}&I&X&I&Z
},
}
which leads to
\eqa{
h_Xq_{Z(IX)^mIZ}+J^2_Zq_{Y(IX)^{m-1}IY}=0.
}{rank1-B2-ZZ-mid1}

We next consider commutators generating $k-2$-support operator $XZ(XI)^{m-1}Y$ as
\eq{
\ba{cccc}{
Y&I&(XI)^{m-1}&Y \\
Z&Z&& \\ \hline
X&Z&(XI)^{m-1}&Y
}
\hspace{15pt}
\ba{cccc}{
X&Z&(XI)^{m-1}&Z \\
&&&X \\ \hline
X&Z&(XI)^{m-1}&Y
},
}
which leads to
\eqa{
J^1_{ZZ}q_{Y(IX)^{m-1}IY}+h_Xq_{XZ(XI)^{m-1}Z}=0.
}{rank1-B2-ZZ-mid2}
Here we need not consider contributions from operators with longer supports ($k-1$-support operators or $k$-support operators) as
\eqa{
\ba{ccccc}{
X&Z&(XI)^{m-1}&X&Z \\
&&&Z&Z \\ \hline
X&Z&(XI)^{m-1}&Y&
}
\hspace{15pt}
\ba{cccccc}{
X&Z&(XI)^{m-1}&X&I&Z \\
&&&Z&I&Z \\ \hline
X&Z&(XI)^{m-1}&Y&&
},
}{rank1-B2-com-shrink}
since these $k-1$-support operator and $k$-support operator have already been shown to have zero coefficients in \lref{rank1-B2-WZ-k-1} and \lref{rank1-B2-WZ-fin}.

We then consider commutators generating $k$-support operators $ZIYZ(XI)^{m-1}Z$ as
\eq{
\ba{cccccccc}{
&&X&Z&(XI)^{m-2}&X&I&Z \\
Z&I&Z&&&&& \\ \hline
Z&I&Y&Z&(XI)^{m-2}&X&I&Z
}
\hspace{15pt}
\ba{cccccccc}{
Z&I&Y&Z&(XI)^{m-2}&Y&& \\
&&&&&Z&I&Z \\ \hline
Z&I&Y&Z&(XI)^{m-2}&X&I&Z
}
\hspace{15pt}
\ba{cccccccc}{
Z&I&X&I&(XI)^{m-2}&X&I&Z \\
&&Z&Z&&&& \\ \hline
Z&I&Y&Z&(XI)^{m-2}&X&I&Z
}
}
which leads to
\eqa{
-J^2_Z q_{XZ(XI)^{m-1}Z}+J^2_Zq_{ZIYZ(XI)^{m-2}Y}-J^1_{ZZ}q_{ZI(XI)^{m}Z}=0.
}{rank1-B2-ZZ-mid3}

We further consider commutators generating $k-2$-support operator $Y(IX)^nIYZ(XI)^{m-2-n}Y$ as\fn{
Here we need not consider contributions from commutators with shrinking the size of support, for the same reason as \eref{rank1-B2-com-shrink}.
}
\eq{
\ba{ccccccc}{
Z&(IX)^n&I&Y&Z&(XI)^{m-2-n}&Y \\
X&&&&&& \\ \hline
Y&(IX)^n&I&Y&Z&(XI)^{m-2-n}&Y
}
\hspace{15pt}
\ba{ccccccc}{
Y&(IX)^n&I&Y&Z&(XI)^{m-2-n}&Z \\
&&&&&&X \\ \hline
Y&(IX)^n&I&Y&Z&(XI)^{m-2-n}&Y
}
\hspace{15pt}
\ba{ccccccc}{
Y&(IX)^n&I&X&I&(XI)^{m-2-n}&Y \\
&&&Z&Z&& \\ \hline
Y&(IX)^n&I&Y&Z&(XI)^{m-2-n}&Y
},
}
which leads to
\eqa{
h_Xq_{Z(IX)^nIYZ(XI)^{m-2-n}Y}+h_Xq_{Y(IX)^nIYZ(XI)^{m-2-n}Z}+J^1_{ZZ}q_{Y(IX)^{m-1}IY}=0.
}{rank1-B2-ZZ-mid4}
In addition, we consider commutators generating $k$-support operator $Z(IX)^{n+1}IYZ(XI)^{m-2-n}Z$ as
\balign{
\ba{ccccccccccc}{
&&Y&(IX)^n&I&Y&Z&(XI)^{m-3-n}&X&I&Z \\
Z&I&Z&&&&&&&& \\ \hline
Z&I&X&(IX)^n&I&Y&Z&(XI)^{m-3-n}&X&I&Z
}&
\hspace{15pt}
\ba{ccccccccccc}{
Z&I&X&(IX)^n&I&Y&Z&(XI)^{m-3-n}&Y&& \\
&&&&&&&&Z&I&Z \\ \hline
Z&I&X&(IX)^n&I&Y&Z&(XI)^{m-3-n}&X&I&Z
} \nt \\
\ba{ccccccccccc}{
Z&I&X&(IX)^n&I&X&I&(XI)^{m-3-n}&X&I&Z \\
&&&&&Z&Z&&&& \\ \hline
Z&I&X&(IX)^n&I&Y&Z&(XI)^{m-3-n}&X&I&Z
},&
}
which leads to
\eqa{
J^2_Zq_{Y(IX)^nIYZ(XI)^{m-2-n}Z}+J^2_Zq_{Z(IX)^{n+1}IYZ(XI)^{m-3-n}Y}-J^1_{ZZ}q_{Z(IX)^{m}IZ}=0.
}{rank1-B2-ZZ-mid5}
Combining Eqs.~\eqref{rank1-B2-ZZ-mid4} and \eqref{rank1-B2-ZZ-mid5} with the help of \eref{rank1-B2-ZZ-mid1}, we find
\eqa{
q_{Z(IX)^nIYZ(XI)^{m-2-n}Y}-q_{Z(IX)^{n+1}IYZ(XI)^{m-3-n}Y}+2\frac{J^1_{ZZ}}{h_X}q_{Y(IX)^{m-1}IY}=0.
}{rank1-B2-ZZ-mid6}
This relation holds for $0\leq n\leq m-3$.

For $n=m-2$, \eref{rank1-B2-ZZ-mid4} holds as it is, while \eref{rank1-B2-ZZ-mid5} should be modified.
We consider commutators generating $k$-support operator $Z(IX)^{m-1}IYZZ$ as
\eq{
\ba{cccccccc}{
&&Y&(IX)^{m-2}&I&Y&Z&Z \\
Z&I&Z&&&&& \\ \hline
Z&I&X&(IX)^{m-2}&I&Y&Z&Z
}
\hspace{15pt}
\ba{cccccccc}{
Z&I&X&(IX)^{m-2}&I&X&Z& \\
&&&&&Z&I&Z \\ \hline
Z&I&X&(IX)^{m-2}&I&Y&Z&Z
}
\hspace{15pt}
\ba{cccccccc}{
Z&I&X&(IX)^{m-2}&I&X&I&Z \\
&&&&&Z&Z& \\ \hline
Z&I&X&(IX)^{m-2}&I&Y&Z&Z
},
}
which leads to
\eqa{
J^2_Zq_{Y(IX)^{m-2}IYZZ}-J^2_Zq_{Z(IX)^mZ}-J^1_{ZZ}q_{Z(IX)^mIZ}=0.
}{rank1-B2-ZZ-mid7}
The obtained $k-1$-support operator $Z(IX)^mZ$ forms a pair with $k-2$-support operator $Y(IX)^{m-1}IY$ as\fn{
Here we need not consider the shrinking case (i.e., a commutator of $k$-support operator and $2$-support operator generates  $k-1$-support operator $Z(IX)^mZ$).
}
\eq{
\ba{ccccccc}{
Z&I&X&(IX)^{m-2}&I&X&Z \\
X&&&&&& \\ \hline
Y&I&X&(IX)^{m-2}&I&X&Z
}
\hspace{15pt}
\ba{ccccccc}{
Y&I&X&(IX)^{m-2}&I&Y& \\
&&&&&Z&Z \\ \hline
Y&I&X&(IX)^{m-2}&I&X&Z
},
}
which leads to
\eqa{
h_Xq_{Z(IX)^mZ}+J^1_{ZZ}q_{Y(IX)^{m-1}IY}=0.
}{rank1-B2-ZZ-mid8}

Summing \eref{rank1-B2-ZZ-mid6} from $n=0$ to $n=m-3$ and plugging Eqs.~\eqref{rank1-B2-ZZ-mid1}, \eqref{rank1-B2-ZZ-mid2}, \eqref{rank1-B2-ZZ-mid3}, \eref{rank1-B2-ZZ-mid4} with $n=m-2$, and Eqs.~\eqref{rank1-B2-ZZ-mid7} and \eqref{rank1-B2-ZZ-mid8}, we arrive at
\eq{
2m J^1_{ZZ}q_{Z(IX)^mIZ}=0.
}
Since $J^1_{ZZ}\neq 0$, we conclude $q_{Z(IX)^mIZ}=0$, which completes the proof.

\bthm{
Consider a Hamiltonian \eqref{rank1-standard} with $J_Z^2\neq 0$, $J^1_{XX}=J^1_{YY}=J^1_{XZ}=J^1_{YZ}=0$, and $J^1_{ZZ}\neq 0$.
This Hamiltonian has no $k$-local conserved quantity with $4\leq k\leq L/2$.
}

\section{Discussion}

We have classified the integrability and non-integrability of all $S=1/2$ chains with shift-invariant and inversion-symmetric next-nearest-neighbor interaction (i.e., Hamiltonians given by \eref{gen-form}).
We rigorously establish that there are only two integrable models (a classical model and a Bethe-solvable model) in this class and all other models are non-integrable.
This classification theorem confirms that there are no missing integrable models waiting to be discovered.
In the context of condensed matter physics, using our result we can safely employ a spin system on the zigzag ladder (except for the above two models) as a non-integrable system.
This theorem also tells that there is no intermediate system with a finite number of nontrivial local conserved quantities, which solves the Grabowski-Mathieu conjecture~\cite{GM95-2} and the Gombor-Pozsgay conjecture~\cite{GP21} in the affirmative within this class.

Towards a general theory of non-integrability, systems with next-nearest-neighbor interaction provide many insightful suggestions which are not seen in $S=1/2$ spin chains with nearest-neighbor interaction~\cite{YCS24-1, YCS24-2}.
First, after step 1 a coefficient of a remaining $k$-support operator is usually expressed by a product of interaction coefficients (or a local magnetic field) in the Hamiltonian.
This is true for the nearest-neighbor interaction case~\cite{Chi24-1, YCS24-1, YCS24-2}.
However, in case B1 of rank 1 (\lref{rank1-B1-kform} in \sref{rank1-B1-kform-fin}) the coefficient is expressed by not a single product but a sum of products.
We consider that the nearest-neighbor case is exceptional and a sum of products generally appears in the expression of a coefficient if off-diagonal interaction coefficients remain nonzero values.

Second, a naive guess from the Grabowski-Mathieu conjecture and the Gombor-Pozsgay conjecture (in particular, Conjecture 3 in Ref.~\cite{GP21}) lead to an anticipation that in step 2 it suffices to examine the absence of 5-local conserved quantities to confirm the non-integrability, since a 3-local operator on the zigzag ladder is at most 5-support in the one-dimensional spin chain.
However, as seen in the analysis of $Z\cdots Z$ in case A of rank 1 (\sref{rank1-A-ZZ}), the minimum candidate of local conserved quantities has support 6, not 5 or less.
Hence, observing only 5-local conserved quantities is insufficient to demonstrate the non-integrability.

Third, in previous literature the analysis of the minimum candidate of local conserved quantities turns out to be sufficient, and the case of general $k$ (in step 2) is a straightforward extension of the minimum nontrivial $k$, which is consistent with the philosophy of the Grabowski-Mathieu conjecture.
On the other hand, in case B1 of rank 1 (\sref{rank1-B1-ZZ}), although the minimum nontrivial candidate is $k=4$ (e.g., $\tz\rXi=ZIYZ$), some contributions appear only for $k\geq 5$, and thus observation on $k=4$ is insufficient to examine all possible types of commutators.
This discrepancy comes from the fact that the unit $\rOmega$ is 3-support and thus the minimum nontrivial candidate with $\rOmega$ is a 5-support operator $\tz\rOmega=ZIXZZ$.

Fourth, in previous literature, the finally obtained relation in step 2 from a sequence of sets of commutators takes the form of ``a product of (coefficients explicitly assumed to be nonzero)$\cdot$ (common coefficient) $=0$".
For example, in $S=1/2$ XYZ chain with $z$ magnetic field~\cite{Shi19}, the finally obtained relation takes the form of $h_Z (1-J_X/J_Y)c=0$ (besides a constant factor).
In this case, we explicitly assume $h_Z\neq 0$ and $J_X\neq J_Y$ (i.e., $1-J_X/J_Y\neq 0$), and hence $c=0$ is concluded.
On the other hand, in the analysis of $Z\cdots Z$ in case A of rank 1, the obtained relations \eqref{rank1-XX-ZZ-gen-fin1} and \eqref{rank1-XX-ZZ-gen-fin2} do not take this form in that we do not assume $-(2m-1)(J^1_{XX})^2+(J^1_{YY})^2-(J^1_{YZ})^2\neq 0$.
In fact, this quantity can be zero under the assumption of case A of rank 1.
To obtain the desired relation $c^{1,k}_{XX-ZZ}=0$, we need to add two obtained relations  \eqref{rank1-XX-ZZ-gen-fin1} and \eqref{rank1-XX-ZZ-gen-fin2} and further resort the fact that the sum of squares with the same sign is nonzero if one of the squares is nonzero.
As far as the author investigated, no proof without resorting to such a technical trick is discovered.

Fifth, in previous studies on nearest-neighbor interacting systems, if the possible form of $k$-support operators is restricted to (generalized) doubling-product operators, the non-integrability proof is completed at the analysis on commutators generating $k$-support operators.
On the other hand, in the case of rank 2 with $J^1_{ZZ}\neq 0$ and other matrix elements of $J^1$ are zero (\sref{rank2-ZZ}), although the possible form of $k$-support operators is restricted to extended doubling-product operators, to prove the absence of $k$-local conserved quantity we need to analyze commutators generating $k-1$-support operators.
This implies that general non-integrability proof might be more complicated than suggested in the previous literature even if we only consider the case that the possible form of $k$-support operators is restricted to (generalized) doubling-product operators.

\bigskip

{\it Acknowledgement}||
It is a pleasure to thank Hosho Katsura, for noticing that the Hamiltonian \eqref{rank1-B1-int-H} can be mapped onto the XYZ model.
The author is also grateful to Akihiro Hokkyo for careful reading of this manuscript and providing helpful comments.
The author also thanks Yuuya Chiba, Mizuki Yamaguchi, and Akihiro Hokkyo for fruitful discussions.

\bigskip

{\it Conflict of interest}||
The author declares no competing interest.

\bigskip

{\it Data availability}||
 No datasets were generated or analyzed during the
current study.

\end{document}